\documentclass[12pt]{article}
\usepackage{graphics}
\usepackage[dvips]{graphicx}
\usepackage{mathrsfs}
\usepackage{amssymb}
\usepackage{verbatim}
\usepackage{float}
\newcommand{\be}{\begin{equation}}
\newcommand{\ee}{\end{equation}}
\newcommand{\bea}{\begin{eqnarray}}
\newcommand{\eea}{\end{eqnarray}}

\def\4vol{{\int d^4x \sqrt{-g}}}

\def\simlt{\stackrel{<}{{}_\sim}}
\def\simgt{\stackrel{>}{{}_\sim}}
\newcommand{\nc}{\newcommand}

\nc{\nt}{\tilde{N}}
\nc{\ra}{\rightarrow}
\nc{\lsim}{\begin{array}{c}\,\sim\vspace{-21pt}\\< \end{array}}
\nc{\gsim}{\begin{array}{c}\sim\vspace{-21pt}\\> \end{array}}
\nc{\tnt}{\tilde{N}}
\nc{\tst}{\tilde{t}}

\nc{\LL}{L}
\nc{\vv}{\tilde{v}}

\title{
\vspace*{-1.3cm}
\begin{flushright}
\normalsize{
ANL-HEP-PR-06-25\\
EFI-06-04\\
FERMILAB-PUB-06-044-T
}
\end{flushright}
\vspace{0.5cm}
\Large
\textbf{Constraints on B and Higgs Physics \\
 in Minimal Low Energy Supersymmetric Models} 
\vspace*{0.5cm}
\author{\textbf{M.~Carena$^a$, A.~Menon$^{b,c}$, R.~Noriega-Papaqui$^{a,e}$,} 
\\
\textbf{A.~Szynkman$^{a,d,f}$ and C.E.M.~Wagner$^{b,c}$}\\ 
\\[0.5cm]
$^a$\normalsize\emph{Theoretical Physics Dept., Fermi National Laboratory,
Batavia, IL 60510} \\
$^b$\normalsize\emph{HEP Division, Argonne National Laboratory,
9700 Cass Ave.,
Argonne, IL 60439, USA} \\
$^c$\normalsize\emph{Enrico Fermi Inst., Univ. of Chicago,
5640 S. Ellis Ave., Chicago, IL 60637, USA} \\
$^d$\normalsize\emph{Lab. de Phys. Nucleaire, Univ. de Montreal,
C.P. 6128, Montreal, Canada H3C 3J7} \\
$^e$\normalsize\emph{Inst. de F\'{\i}sica, Univ.Aut\'onoma de Puebla. A. P. J-48 Puebla, M\'{e}xico} \\
$^f$\normalsize\emph{IFLP, Dept. de F\'{\i}sica, Univ. Nacional de
La Plata, C.C. 67, 1900 La Plata Argentina}}}

\begin{document}
\setcounter{page}{0}
\maketitle
\vspace{0.5cm}
\begin{abstract}
We study the implications of minimal  
flavor violating low energy supersymmetry 
scenarios for the search of new physics in the B and Higgs sectors at the 
Tevatron collider and the LHC. 
We show that the already stringent Tevatron bound on the 
decay rate $B_s \to \mu^+\mu^-$ sets strong constraints on the possibility of 
generating large corrections to the mass difference $\Delta M_s$ of the $B_s$ 
eigenstates. We also show that the $B_s \to \mu^+\mu^-$ bound together 
with the constraint on the branching ratio of the rare decay $b \to s \gamma$ 
has strong implications for the 
search of light, non-standard Higgs bosons at hadron colliders. 
In doing this, we demonstrate that 
the former expressions derived for the analysis of the double penguin 
contributions in the Kaon sector need to be corrected by additional terms 
for a realistic analysis of these effects. We also study a specific
non-minimal flavor violating scenario, where there are flavor changing
gluino-squark-quark interactions, governed by the CKM matrix elements,
and show that the $B$ and Higgs physics constraints are similar to
the ones in the minimal flavor violating case.
Finally we show that, in scenarios like electroweak baryogenesis which 
have light stops and charginos, there may be enhanced effects on the 
$B$ and $K$ mixing parameters, without any significant effect on 
the rate of $B_s \to \mu^+\mu^-$. 
\end{abstract}
\thispagestyle{empty}

\newpage

\setcounter{page}{1}
\section{Introduction}
The standard model (SM) provides an accurate description of all the
results from high energy physics experiments, in particular
precision electroweak measurements and flavor physics observables.
These experiments put strong constraints
on extensions of the SM that have tree-level flavor changing 
neutral current effects or large custodial symmetry breaking
effects. For renormalizable, 
weakly interacting theories, where the new exotic particles 
acquire large gauge invariant masses so that they decouple from the 
low energy effective theory, these constraints can be avoided. 
Low energy supersymmetry~\cite{Nilles:1983ge,Haber:1984rc}
is a particularly attractive example of this kind of theory. 
The minimal supersymmetric extension of the Standard Model or MSSM (with 
gauge invariant SUSY breaking masses of the order of 1 TeV) predicts an 
extended Higgs sector with a light SM-like Higgs boson of mass lower than 
135 GeV~\cite{mhiggsRG1}--\cite{mhiggsEP5} 
that agrees well with precision electroweak measurements.

However the structure of supersymmetry breaking parameters is
not well defined.  If there are no tree-level flavor changing transitions
in any gauge or super-gauge interaction, then the deviations from SM 
predictions are naturally small. Such small deviations can be achieved if the 
quark and squark mass matrices are block diagonalizable in the same basis.
For instance, this happens when the squark and slepton  
supersymmetry breaking masses are flavor independent. For these kinds of 
models, all flavor violating effects are induced at the loop-level and are 
governed by the CKM matrix elements, as in the SM. Many studies 
have concentrated on the properties of these minimal flavor violating 
scenarios 
(see, for example, Refs.~\cite{Bertolini:1990if}--\cite{Foster:2005wb}). 

In this article we shall analyze their flavor violating effects
in two quite generic cases. In the first case, we consider a low energy 
effective theory in which the quark and squark mass matrices are aligned
in flavor space and can be simultaneously diagonalized in blocks, as
described in the next section. We will remain agnostic about how this
effective low energy theory is UV completed. However, since the
Yukawa-induced
radiative corrections to the soft supersymmetry breaking parameters
tend to destroy the alignment of the squark and quark mass
matrices,
this situation may  be only naturally realized in models of low energy
supersymmetry breaking, where these radiative corrections 
are  small.  We call this low energy scenario Minimal 
Flavor Violation. 

In order to study the possible effect of Yukawa dependent radiative 
corrections we study a second case, in which we assume a departure from
the alignment condition by the presence of flavor violating effects 
proportional  to the CKM matrix elements. These effects are 
induced by corrections to the 
left-handed down squark mass matrices proportional to the product
of the up-quark Yukawa matrix and its hermitian conjugate (or, in 
general, powers of this product). We furthermore assume that
the right-handed down squark masses are flavor independent. 
As we will discuss in more detail in the next section, these
conditions at low energies are achieved, for instance, by Yukawa dependent 
radiative corrections,  if one starts from flavor independent
squark 
masses at a high energy scale at moderate values of $\tan\beta$.
One characteristics of this second scenario is that there are flavor 
violating down-squark-gluino vertices at tree-level. Since all 
flavor violating effects are governed by the CKM matrix elements,
this scenario would also enter within the general definition of minimal
flavor violating models given in Ref.~\cite{giudice}. However, due
to the presence of flavor violating couplings at tree level, we will 
denote it as non-minimal flavor violation in order 
distinguish it from the first scenario of flavor alignment at the weak 
scale, in which such tree-level effects are absent. As we will show,
the phenomenological predictions in this scenario 
are similar to those of the flavor alignment case, 
unless the  left-handed squarks and the gluino are very light.

Apart from the structure of supersymmetry breaking parameters,
the phases associated with them are also important. In minimal flavor 
violating schemes there are at least two phases that cannot be absorbed by 
redefining the low energy fields. For real values of the $\mu$ parameter,
these phases can be associated with a universal phase
for the gaugino masses and the trilinear mass parameter. In general,
however, one can choose independent phases for the different gaugino masses 
and trilinear mass parameters. CP-violating phases beyond the CKM one are 
required, for instance, in models of electroweak 
baryogenesis~\cite{CQW}--\cite{Balazs:2004ae}. In this scenario, there could be
 significant effects on $\Delta M_s$,  $\mathcal{BR}(B_s \rightarrow \mu^+ 
\mu ^-)$ and $\epsilon_K$ because of the presence of a light stop and extra 
phases in the chargino, neutralino and gluino sectors. We shall comment on 
the effects of these new CP violating phases below.

In this paper we attempt to develop a systematic method of treating the extra 
sources of flavor violation in the minimal and non-minimal flavor violating 
models described above. We show that the usual approach of calculating $\tan 
\beta$ enhanced FCNC (Flavor Changing Neutral Currents) effects in the Kaon 
sector does not agree with the exact results one finds in the limit of 
flavor independent masses. Thus, we develop a perturbative  approach that leads
to agreement with the exact result in this limit.

We shall emphasize the implications of the present bounds on  
$\mathcal{BR}(B_s \to \mu^+\mu^-)$
for future measurements at the Tevatron collider, both in Higgs as
well as in B-physics. In particular, we shall show 
that the present bound on $\mathcal{BR}(B_s \to \mu^+\mu^-)$ leads to strong
constraints on possible corrections to both $\Delta M_s$ and the Kaon mixing 
parameters in minimal flavor violating schemes. Moreover, 
we shall show that this bound, together with the constraint implied by the 
measurement of $\mathcal{BR}(b \to s \gamma)$ leads to limits on the
possibility detecting light, non-standard Higgs bosons in the
MSSM at the Tevatron collider. Throughout the paper we always 
take real values of $\mu A_t$, and therefore the Higgs sector  is 
approximately CP-invariant~\cite{Pilaftsis:1999qt,Carena:2002bb}, 
and will be treated as such. 

This article is organized as follows. In section 2, we define our
theoretical setup, giving the basic expressions necessary for
the analysis of the flavor violating effects at large values
of $\tan\beta$. In particular, we show how the first order perturbative
expressions in the CKM matrix elements are inappropriate to 
define the corrections in the Kaon
sector where higher order effects need to be considered. In section 3 we show 
the implications of the 
constraint on $\mathcal{BR}(B_s \to \mu^+\mu^-)$ for the mixing
parameters of the Kaon and B sectors in the large $\tan\beta$ regime.
In section 4, we explain the implications for Higgs searches
at the Tevatron. We reserve section 5 for our conclusions and some
technical details for the appendices.

\section{Theoretical Setup} 

\subsection{The resummed effective Lagrangian and the sparticle spectrum}

The importance of large $\tan \beta$ FCNC effects in supersymmetry 
has been known for sometime. The finite pieces of the one-loop self energy 
diagrams lead to an effective lagrangian for the quark-Higgs sector, valid at 
energy scales lower than the heavy squark masses, which has 
the generic form~\cite{Babu:1999hn}--\cite{Dedes:2002er},
\cite{deltamb2b,deltamb2}
\begin{eqnarray}
- \mathcal{L}_{eff} = \bar{d}_R^0 \mathbf{\hat{Y}_d} [\Phi_d^{0*}+\Phi_u^{*0} 
\left(\mathbf{\hat{\epsilon}_0+\hat{\epsilon}_Y \hat{Y}_u^{\dagger} \hat{Y}_u} 
\right)]d_L^0+\Phi_u^0 \bar{u}_R^0 \mathbf{\hat{Y}_u} u_L^0 + h.c. 
\label{Leff:eq} \\
- \mathcal{L}_{mass} = \frac{v_d}{\sqrt{2}} \bar{d}_R^0 \mathbf{\hat{Y}_d} [1+
\tan \beta \left(\mathbf{\hat{\epsilon}_0+\hat{\epsilon}_Y \hat{Y}_u^{\dagger} 
\hat{Y}_u} \right)]d_L^0 + \frac{v_u}{\sqrt{2}} \bar{u}_R^0 \mathbf{\hat{Y}_u} 
u_L^0 + h.c. \label{Lmass:eq}
\end{eqnarray}
in an arbitrary basis. The $\mathbf{\hat{\epsilon}_0}$ and $\mathbf{
\hat{\epsilon}_Y}$ matrices correspond to radiative 
contributions~\cite{deltamb1} coming from the loops shown in 
Fig.~\ref{effdiag:fig}. Their exact dependence on the supersymmetric mass 
parameters is given in Appendix~\ref{A2:sec}.

\begin{figure}
\begin{center}
\resizebox{10cm}{!}{\includegraphics{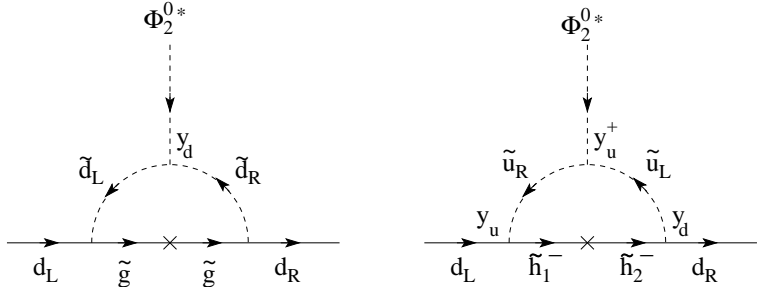}}
\end{center}
\caption{SUSY radiative corrections to the self-energies of the d-quarks
in the mass insertion approximation}
\label{effdiag:fig}
\end{figure} 

The flavor structure of the loop correction factors are independent of 
their momentum integrations. Therefore, in an arbitrary basis, the flavor 
dependence of the loop correction parameters are the same as that of the 
mass matrices and Yukawa couplings. Thus, the loop correction factors have the 
following flavor structure
\begin{eqnarray}
\mathbf{\hat{Y}_d} \hat{\epsilon}_0 &\propto& \mathbf{\hat{M}}_{\tilde{d}_R}^{
-2} \mathbf{\hat{Y}_d} \mathbf{\hat{M}}_{\tilde{d}_L}^{-2} 
\label{e0flavstruct:eq}\\
\mathbf{\hat{Y}_d} \hat{\epsilon}_Y \mathbf{\hat{Y}_u^{\dagger}\hat{Y}_u} 
&\propto& \mathbf{\hat{Y}_d \hat{M}_{\tilde{u}_L}^{-2} \hat{Y}_u^{\dagger} 
\hat{M}_{\tilde{u}_R}^{-2} \hat{Y}_u}
\end{eqnarray}
where $\mathbf{\hat{M}}^{-2}$ matrices are the non-diagonal inverse squark 
mass 
squared matrices. Thus the sparticle spectrum is intimately connected to 
the $\epsilon$ parameters which in turn affect the FCNC's. We look at two 
possible choices that connect the quark mass eigenstate basis to that of the squarks.

\subsubsection{Minimal Flavor Violation} \label{mvf:sec}

This scenario is similar to that discussed  
in Refs.~\cite{Isidori:2001fv,Buras:2002vd,Babu:1999hn,Hamzaoui:1998nu}, where 
one assumes an alignment of the quark and squark mass matrices in flavor
space. Therefore, in the low energy effective theory, the
diagonalization of the quark mass matrices
leads to squark mass matrices that are block 
diagonal. 
Using the following transformation matrices
\begin{eqnarray}
u_L^0 = \mathbf{U_L^Q} u_L, \;\; d_L^0 = \mathbf{U_L^Q V_0} d_L, \;\; 
u_R^0 = \mathbf{U_R^u} u_R, \;\; d_R^0 = \mathbf{U_R^d} d_R
\end{eqnarray}
to rotate the original quark supermultiplets into a basis where the tree 
level Yukawa couplings are diagonal, we get 
\begin{eqnarray}
\mathbf{Y_d} &=& \mathbf{U_R^{d\dagger} \hat{Y}_d U_L^Q V_0};  \nonumber \\
\mathbf{Y_u} &=& \mathbf{U_R^{u\dagger} \hat{Y}_u U_L^Q};  \nonumber \\
\mathbf{M_{\tilde{d}_R}^{-2}} &=& \mathbf{U_R^{d\dagger} \hat{M}_{\tilde{d}_R
}^{-2} U_R^d};  \; \; \;\;\;\;\;
\mathbf{M_{\tilde{d}_L}^{-2}} = \mathbf{V_0^{\dagger} U_L^{Q\dagger} \hat{M
}_{\tilde{d}_L}^{-2} U_L^Q V_0}; \nonumber \\
\mathbf{M_{\tilde{u}_R}^{-2}} &=& \mathbf{U_R^{u\dagger} \hat{M}_{\tilde{u}_R
}^{-2} U_R^u};  \; \; \;\;\;\;\;
\mathbf{M_{\tilde{u}_L}^{-2}} = \mathbf{U_L^{Q\dagger} \hat{M}_{\tilde{u}_L
}^{-2} U_L^Q}; \nonumber \\
\hat{\epsilon}_0 &\propto&\mathbf{U_L^Q V_0 M_{\tilde{d}_R}^{-2} M_{\tilde{d}_L
}^{-2} V_0^{\dagger} U_L^{Q\dagger}}; 
\;\;\;\;\;\;\;\; \hat{\epsilon}_0 = \mathbf{U_L^{Q}V_0} \epsilon_0 
\mathbf{V_0^{\dagger}U_L^{Q\dagger}};
\nonumber \\
\hat{\epsilon}_Y &\propto& \mathbf{U_L^Q M_{\tilde{u}_L}^{-2} 
M_{\tilde{u}_R}^{-2} U_L^{Q\dagger}} ;
\;\;\;\;\;\;\;\;\;\;\;\;\;\;\;\; \;\;
\hat{\epsilon}_Y = \mathbf{U_L^{Q}} \epsilon_Y 
\mathbf{U_L^{Q\dagger}};
\label{minflavor:eq}
\end{eqnarray}
where the un-hatted mass and Yukawa matrices are diagonal and $\mathbf{V}_0$ is
the tree level CKM matrix. Under this transformation the effective mass 
lagrangian becomes 
\begin{eqnarray}
- \mathcal{L}_{\rm mass} = \frac{v_d}{\sqrt{2}} \bar{d}_R \mathbf{Y_d} [1+\tan 
\beta \left(\mathbf{\epsilon_0+ V_0^{\dagger} \epsilon_Y |Y_u|^2 V_0}
\right)]d_L + \frac{v_u}{\sqrt{2}} \bar{u}_R \mathbf{Y_u} u_L + h.c. 
\label{LMFVmass:eq}
\end{eqnarray}
where the $\mathbf{\epsilon_0}$ and $\mathbf{\epsilon_Y}$ terms, defined in
Eq.~(\ref{minflavor:eq}) (see also Appendix A, 
Eq.~(\ref{e0form:eq}) and Eq.~(\ref{eyform:eq})),  are diagonal. 
Therefore the quark mass matrices receive off-diagonal terms proportional
to $\epsilon_Y$ at the 1-loop level and so need to be rediagonalized 
perturbatively. This procedure has been performed in 
Refs.~\cite{Isidori:2001fv, Buras:2002vd}. 
However, the calculation of the $(2,1)$ and $(1,2)$ components
of the neutral-Higgs-quark-quark coupling are affected by additional 
corrections not included in Refs.~\cite{Isidori:2001fv, Buras:2002vd}. In 
Appendix~\ref{A1:sec} we calculate the corrected couplings which we present 
here. Defining the down-quark neutral Higgs interaction Lagrangian to be
\begin{equation}
-\mathcal{L} = \bar{d}_R^J \left(X_{RL}^S \right)^{JI} d_L^I \phi_S + h.c.,
\end{equation}
we find that the neutral Higgs flavor changing coupling, with $I \neq J$, 
takes the form
\begin{eqnarray}
(X_{RL}^S)^{JI}=\frac{\bar{m}_{d_J} y_t^2 
\Gamma^{JI} (x_u^S - x_d^S \tan \beta)}{v_d (1+
\epsilon_0^J \tan \beta) (1+\epsilon_3 \tan \beta)} V_{eff}^{3J*} 
V_{eff}^{3I} \label{xrl:eq}
\end{eqnarray}
where we have ignored the small effects proportional to the first and second 
generation Yukawa couplings to find $\epsilon_J = \epsilon_0^J + \delta_{3J} 
\epsilon_Y y_t^2$, $x_u^S$ and 
$x_d^S$ are the Higgs scalar components on the neutral $\Phi_u^{0*}$ and 
$\Phi_d^{0*}$ fields  (see Appendix A, Eq.(\ref{neutHmix:eq})) and
\begin{eqnarray}
\Gamma^{3I} &=& \epsilon_Y \\
\Gamma^{J3} &=& \frac{\epsilon_Y(1+\epsilon_3^* \tan \beta)-
\epsilon_Y^*(\epsilon_3-
\epsilon_J) \tan \beta}{1+\epsilon_0^{3*} \tan \beta} \\
\Gamma^{21} &=& \frac{\epsilon_Y}{(1+\epsilon_2 \tan \beta)
|1+\epsilon_0^3 \tan 
\beta|^2} \left[(1+\epsilon_0^3\tan \beta)|1+\epsilon_3 \tan 
\beta|^2 - \right. \nonumber\\
& & \left.\epsilon_Y y_t^2 \tan \beta (1+\epsilon_3^*\tan \beta)(1+\epsilon_2
\tan \beta)- \epsilon_Y^* y_t^2 \tan \beta(1+\epsilon_2 \tan \beta)^2 
\right] 
\label{Gamma21:eq} \\
\Gamma^{12} &=& \frac{\epsilon_Y}{(1+\epsilon_2 \tan \beta) |1+\epsilon_0^3 
\tan \beta|^2}\left\{(1+\epsilon_0^3\tan \beta)|1+\epsilon_3 \tan \beta|^2 -
\right.\nonumber\\
& & \left. \epsilon_Y y_t^2 \tan \beta (1+\epsilon_3^*\tan \beta)(1+\epsilon_2\tan 
\beta)-\epsilon_Y^* y_t^2 \tan \beta(1+\epsilon_2 \tan \beta)(1+\epsilon_1 \tan 
\beta) \right. \nonumber \\ 
& & \left. + \frac{\epsilon_1-\epsilon_2}{\epsilon_Y} \left[ \frac{\epsilon_Y^*
\tan \beta}{1+\epsilon_2^* \tan \beta} - \frac{(\epsilon_Y^*)^2 y_t^2 \tan^2 
\beta}{(1+\epsilon_2^* \tan \beta)(1+\epsilon_3^* \tan \beta)} - \frac{|
\epsilon_Y|^2 y_t^2 \tan^2 \beta}{|1+\epsilon_3 \tan \beta|^2} - \right] \right
\}. \label{Gamma12:eq}
\end{eqnarray}

Here $V_{eff}$ is the CKM matrix obtained after diagonalization of
the one-loop mass matrix in Eq.~(\ref{LMFVmass:eq}). The relation between
this matrix and $V_0$ is given in the Appendix~\ref{A1:sec}.
Observe that in the limit of universal squark soft SUSY breaking masses the 
$\epsilon_0$ diagonal matrix is proportional to the identity and, in spite of 
their complicated form, all the $\Gamma^{IJ}$ become equal to $\epsilon_Y$. 
The difference between the above expressions and those obtained before in the 
literature will be discussed in more detail below.

\subsubsection{Non-minimal Flavor Violation using the CKM matrix
}\label{nmvf:sec}

As explained in the introduction, we shall discuss a second scenario
in which all flavor violating effects are proportional to CKM matrix
elements, and there are tree-level down-squark-gluino flavor violating
vertices in the low energy effective theory. This scenario is similar
to that discussed in  Ref.~\cite{Dedes:2002er}.  For the present discussion,
let us assume that we perform the diagonalization 
procedure in a single step under the transformation
\begin{eqnarray}
u_L^0 = \mathbf{U_L^Q} u_L, \;\; d_L^0 = \mathbf{U_L^Q V_{eff}} d_L, \;\; 
u_R^0 = \mathbf{U_R^u} u_R, \;\; d_R^0 = \mathbf{U_R^d} d_R
\label{diagquark:eq}
\end{eqnarray}
where instead of $\mathbf{V_0}$ the tree level CKM matrix we have 
$\mathbf{V}_{eff}$ the 
effective CKM matrix. This transformation leads to a diagonal quark mass 
matrix and a mass lagrangian of the form 
\begin{eqnarray}
- \mathcal{L}_{\rm mass} = \frac{v_d}{\sqrt{2}} \bar{d}_R \mathbf{U}_R^{d
\dagger} \mathbf{\hat{Y}_d} \mathbf{U}_L^{Q}[1+\tan \beta 
\left(\mathbf{\epsilon_0+\epsilon_Y |Y_u|^2}
\right)]\mathbf{V}_{eff} d_L + \frac{v_u}{\sqrt{2}} \bar{u}_R \mathbf{Y_u} u_L 
+ h.c. ,
\end{eqnarray}
under the assumption that the matrices 
\begin{eqnarray}
\mathbf{\epsilon_0} &=& \mathbf{U_L^{Q\dagger} \hat{\epsilon_0} U_L^Q} 
\nonumber \\
\mathbf{\epsilon_Y} &=& \mathbf{U_L^{Q\dagger} \hat{\epsilon_Y} U_L^Q} 
\label{e0eYnmfv:eq}
\end{eqnarray}
are diagonal~\cite{Dedes:2002er}. 
The condition that $\mathbf{U_L^{Q\dagger} \hat{\epsilon_Y} 
U_L^Q}$ is diagonal is the same as Eq.~(\ref{minflavor:eq}) in Minimal 
Flavor Violation. Thus we again need the u-squark mass matrix to be block 
diagonal in the u-quark eigenbasis. Therefore there are no flavor 
changing effects in the neutral up supergauge currents. 
 
However the assumption that $\mathbf{U_L^{Q\dagger} \hat{\epsilon_0} U_L^Q}$ 
is diagonal differs from Eq.~(\ref{minflavor:eq}) in MFV. 
From the flavor structure of $\hat{\mathbf{\epsilon}}_0$ in 
Eq.(\ref{e0flavstruct:eq}), we see that this 
can only be naturally fulfilled if 
\begin{equation}
\mathbf{M_{\tilde{d}_L}^{-2}} = \mathbf{U_L^{Q\dagger}  \hat{M}_{\tilde{d
}_L}^{-2} U_L^Q}, \;\;\;\; {\rm and} 
\;\;\;\;
\mathbf{M_{\tilde{d}_R}^{-2}} = \mathbf{U_R^{d\dagger}}
\mathbf{\hat{M}_{\tilde{d}_R}^{-2} U_R^d}
\label{diagsquark:eq}
\end{equation}
are diagonal and $[\mathbf{M_{\tilde{d}_R}^{-2}, Y_d V_{eff}^{\dagger}}]=0$.
The obvious way of satisfying this commutation relation is to require the
right-handed d-squark mass matrix to be flavor independent
or $\mathbf{M_{\tilde{d}_R}^{2}} \propto \mathbf{I}$. 
Observe that this analysis 
was not performed in Ref.~\cite{Dedes:2002er} and 
hence the above conditions were not required in that work. As stressed
in the introduction, the above flavor structure of mass matrices may
be achieved by Yukawa induced radiative corrections to universal, flavor
independent squark masses at high energy scales, at moderate values of 
$\tan\beta$. Assuming the squark masses are flavor independent at high 
energies, the only one-loop corrections that violate flavor are induced
by the up and down Yukawa matrices because the gauge interactions are flavor
blind. These corrections are given by~\cite{Dugan:1984qf}
\begin{equation}
\Delta M_{\tilde{Q}}^2 \simeq - \frac{1}{8\pi^2}\left[
\left(2 m_0^2 
+ M_{H_u}^2(0) + A_0^2 \right) Y_u^{\dagger} Y_u 
+
\left(2 m_0^2 
+ M_{H_d}^2(0) + A_0^2 \right) Y_d^{\dagger} Y_d \right] 
\log\left(\frac{M}{M_{SUSY}}\right),
\end{equation}
where $\tilde{Q}$ denote the left-handed squarks, 
$m_0$ is the common squark mass at the scale $M$ at which 
supersymmetry breaking is transmitted to the observable sector,
$M_{H_{u,d}}^2(0)$ and $A_0$
are the Higgs soft supersymmetry breaking masses and squark-Higgs 
trilinear mass parameters at that scale, and $M_{SUSY}$ is the
characteristic low energy squark mass scale.

Similarly,
the right-handed up and down squark mass matrices, receive
one-loop Yukawa-induced corrections proportional to
\begin{equation}
\Delta M_{\tilde{u}_R}^2 = - \frac{2}{8\pi^2} \left( 2 m_0^2 
+ M_{H_u}^2(0) + A_0^2 \right) Y_u Y_u^{\dagger}
\log\left(\frac{M}{M_{\rm SUSY}}\right) ,
\end{equation}
and
\begin{equation}
\Delta M_{\tilde{d}_R}^2 = - \frac{2}{8\pi^2} \left( 2 m_0^2 
+ M_{H_d}^2(0) + A_0^2 \right) Y_d Y_d^{\dagger}
\log\left(\frac{M}{M_{\rm SUSY}}\right) ,
\end{equation}
respectively. 

Therefore, while the Yukawa induced radiative corrections to the
right-handed squark mass matrices mantain the alignment of these matrices with
their corresponding Yukawa matrices, the corrections to the 
left-handed squark masses induced a misalignment 
between the quark and squark mass matrices governed by CKM matrix
elements. Since the dominant effects are governed by the third
generation Yukawa eigenvalues, the down-quark Yukawa effects
may be neglected at small or moderate values of $\tan\beta$ where
the bottom Yukawa coupling is much smaller than the top quark one.
In this case, one arrives at the properties of the squark mass
matrices specified in the non-minimal flavor violating scenario
defined above.  

In general, even at larger values of $\tan\beta$, the only flavor
violating squark-gluino vertices will be in the left-handed couplings
(and the Higgs-squark-squark vertices) and they will be governed by
CKM matrix elements.
The only difference between the large $\tan\beta$ case with respect to 
the non-minimal flavor violating model
defined above is that the masses of the right-handed down
squarks will no longer be flavor independent at low energies and 
therefore the  $\hat{\epsilon}_0$ matrix
will not be aligned with the $\hat{\epsilon}_Y$ one. However, the
flavor properties of the large $\tan\beta$ scenario
are quite similar the non-minimal flavor violating scenario 
specified above and therefore this scenario will allow us to 
study the possible effects 
of the Yukawa induced radiative corrections to the squark mass matrices,
in particular the ones associated with the flavor violating 
down-squark-gluino couplings at tree-level.

Following the argument in Ref.~\cite{Dedes:2002er} we can rewrite the effective
lagrangian in terms of the mass eigenstates as
\begin{eqnarray}
- \mathcal{L}_{eff}&=&\frac{\sqrt{2}}{v_u} (\Phi_d^{0*} - \Phi_u^{0*} \tan 
\beta) \bar{d}_R \mathbf{\bar{m}_d} V_{eff}^{\dagger} \mathbf{R}^{-1} V_{eff} 
d_L + \frac{\sqrt{2}}{v_u} \Phi_u^{0*} \bar{d}_R \mathbf{\bar{m}_d} d_L \nonumber \\
& & +\Phi_u^0 \bar{u}_R \mathbf{Y_u} u_L + h.c. 
\end{eqnarray}
where $V_{eff}$ is the effective CKM matrix, $\mathbf{Y_u}$ is the 
diagonal up Yukawa matrix, $\mathbf{\bar{m}_d}$ is the diagonal
down-quark running mass matrix, and 
\begin{eqnarray}
\mathbf{R} = \mathbf{1} + \mathbf{\epsilon_0} \tan \beta +  \mathbf{
\epsilon_Y} |\mathbf{Y_u}|^2 \tan \beta.
\end{eqnarray}
Therefore, neglecting\footnote{This approximation breaks down in the limit $1+
\epsilon_0 \tan \beta \rightarrow 0$, the singularity in $[X_{RL}^{dS}]$ 
proportional to $y_t$ cancels against those coming from $y_c$ and $y_u$ as 
discussed in Ref.~\cite{Dedes:2002er}} $y_u$ and $y_c$ as compared to $y_t$, 
and defining
\begin{eqnarray}
\epsilon_J=\epsilon_0^J+\epsilon_Y y_t^2 \delta^{J3} \label{epsilonJ:eq}
\end{eqnarray}
for all $J$, we find
\begin{eqnarray}
(\mathbf{R}^{-1})^{JI} &=& \frac{1}{1+\epsilon_J \tan \beta} \delta^{JI}
\end{eqnarray}

If we assume a generational mass splitting so that the first two generations 
are equally massive and 
heavier than the third generation we find $\epsilon_0^1=\epsilon_0^2 =
\epsilon_0$. In this case the flavor changing effects are not solely 
dependent on $\epsilon_Y$, but they also depend on the difference between the 
loop factors $(\epsilon_3-\epsilon_0)$:
\begin{eqnarray}
(X_{RL}^S)^{JI}=\frac{\bar{m}_{d_J} (\epsilon_3-\epsilon_0) (x_u^S - x_d^S 
\tan \beta)}{v_d (1+\epsilon_0 \tan \beta) (1+\epsilon_3 \tan \beta)} 
V_{eff}^{3J*} V_{eff}^{3I} \label{xrlnmfv:eq}.
\end{eqnarray}

The reason we call this scenario non-minimal flavor violation is that the 
diagonalization procedure induces 
flavor changing effects in the gluino-quark-squark couplings which lead to 
additional contributions to flavor changing processes. Indeed,
the assumption that $\mathbf{\epsilon_0}$ and $\mathbf{\epsilon_Y}$ in 
Eq.(\ref{e0eYnmfv:eq}) are diagonal leads to the appearance of CKM elements 
in the down quark-squark-gluino coupling, as it is clear from 
Eqs.~(\ref{diagquark:eq}) and (\ref{diagsquark:eq}).
Because the left-handed squarks 
are not diagonalized by the same rotation as the left-handed quarks, 
the effective gluino Lagrangian becomes
\begin{eqnarray}
\mathcal{L}_{\tilde{g}} &=& - \sqrt{2} g_s \left[ \bar{u}_{L}\
 \tilde{g}^{a} \ T^a \tilde{u}_{L} - \bar{u}_{R}\
 \tilde{g}^{a} \ T^a \tilde{u}_{R} \right]    \nonumber \\
& & +\sqrt{2} g_s \left[ \bar{d}_{L}\
 \tilde{g}^{a} \ T^a\ \mathbf{V} \tilde{d}_{L} - \bar{d}_{R}\
 \tilde{g}^{a} \ T^a \tilde{d}_{R}  \right].
\label{lag-gluinos:eq}
\end{eqnarray}
The appearance of the CKM matrix in the gluino couplings induces flavor changing
box diagrams that can in principle produce large effects. 

\subsubsection{The uniform squark mass limit}

The two flavor changing scenarios discussed above coincide for the case of 
uniform squark masses. Since, in this limit, the transformation performed in 
Section~\ref{nmvf:sec} requires no approximations or assumptions the expression
for the FCNC's are exact. However, the perturbative approach in 
Section~\ref{mvf:sec} provides expressions for the FCNCs that are only valid 
up to a certain order in the off-diagonal CKM matrix elements. For the 
perturbative approach in Section~\ref{mvf:sec} to be valid we need the two 
expression for the FCNC's to be equal to at least quadratic order in the 
off-diagonal CKM matrix elements. However, as discussed above, comparing the 
results of Ref.~\cite{Buras:2002vd} and Ref.~\cite{Dedes:2002er} this is 
clearly not true for the $(2,1)$ and $(1,2)$ components of the down quark-Higgs
couplings $X_{RL}$. 

In the uniform squark limit, the flavor violating coupling given in 
Eq.~(\ref{xrlnmfv:eq}) has the form
\begin{eqnarray}
(X_{RL}^{S})^{JI} = \frac{\bar{m}_{d_J} \epsilon_Y y_t^2 (x_u^S - 
x_d^S \tan \beta)}{v_d(1+\epsilon_3 \tan \beta)(1+\epsilon_0 \tan 
\beta)} V_{eff}^{3J*} V_{eff}^{3I} \label{uniformXRL:eq}.
\end{eqnarray}
which does not agree with the results in Ref.~\cite{Buras:2002vd}, where 
they find the corrected coupling to be
\begin{eqnarray}
(X_{RL}^{S})^{21} &=& \frac{\bar{m}_{d_J} \epsilon_Y y_t^2}{v_d}
\frac{|1+\epsilon_3 \tan \beta|^2}{|1+\epsilon_0 \tan \beta|^2(1+
\epsilon_0 \tan \beta)^2} V_{eff}^{3J*} V_{eff}^{3I5} 
(x_u^S-x_d^S\tan \beta).
\end{eqnarray} 

To understand this difference between the results of 
Ref.~\cite{Buras:2002vd} and Ref.~\cite{
Dedes:2002er} we need to look at the approximations made in Ref.~\cite{
Buras:2002vd}. Diagonalizing the tree level quark mass matrices in 
Eq.~(\ref{LMFVmass:eq}) leads to uncorrected diagonal masses 
$\mathbf{m_d}$ and a CKM matrix $\mathbf{V_0}$. However the large $\tan \beta$ 
enhanced radiative corrections lead to off-diagonal terms in the mass 
matrix, which have the form
\begin{eqnarray}
(\mathbf{m_d + \Delta m_d})^{JI} &=& 
 m_{d_J} \left((1+\epsilon_J \tan \beta) 
\delta^{JI} + \epsilon_Y y_t^2 \tan \beta \lambda_0^{JI} \right)
\end{eqnarray} 
where $\lambda_0^{JI} = V_0^{3J*} V_0^{3I}$ for $J \neq I$ and $\epsilon_J$ is
defined in eqn.~(\ref{epsilonJ:eq}). We have also neglected contributions to 
the diagonal elements of the form $|V_0^{3J}|^2$ for $J \neq 3$ as they are 
subdominant. Hence, to go to 
the physical quark basis we need to further diagonalize 
this effective mass matrix by unitary matrices 
$\mathbf{D_{L,R}}$ so that
\begin{eqnarray}
e^{-i\theta_J} (\mathbf{D_R^{\dagger}(m_d+\Delta m_d)D_L})^{JI} = 
\bar{m}_{d_J} \delta^{JI}  
\label{massdiag:eq}
\end{eqnarray}
where $\theta_J = \arg(1+\epsilon_J \tan \beta)$. The approach taken in 
Ref.~\cite{Buras:2002vd} is to perturbatively expand the diagonalization 
matrices $\mathbf{D_L}$ and $\mathbf{D_R}$ so that
\begin{eqnarray}
\mathbf{D_L} &=& \mathbf{1 + \Delta D_L} \\
\mathbf{D_R} &=& \mathbf{1 + \Delta D_R}
\end{eqnarray}
where the unitarity of $\mathbf{D_{L,R}}$ to linear order in $\Delta$ leads to 
conditions $(\mathbf{\Delta D_{L,R}})^{\dagger}=-\mathbf{\Delta D_{L,R}}$, so
that when $J \neq I$ in Eq.(\ref{massdiag:eq}) we have the condition 
\begin{eqnarray}
e^{-i\theta_J}(\mathbf{-(\Delta D_R) \bar{m}_d+\Delta m_d+ \bar{m}_d 
(\Delta D_L)})^{JI} = 0 \label{massoffdiag:eq},
\end{eqnarray}
where the $\bar{m}_d$ includes higher order terms and higher 
orders in $\mathbf{
\Delta}$ have been neglected. Using Eq.\ (\ref{massoffdiag:eq}) and its 
dagger along with the hierarchy in quark masses gives us
\begin{eqnarray}
(\mathbf{\Delta D_L})^{JI} = \left\{ \begin{array}{ll}
- \frac{\epsilon_Y y_t^2 \tan \beta}{1+\epsilon_J \tan \beta} \lambda_0^{JI} 
& J > I \\
\frac{\epsilon_Y^* y_t^2 \tan \beta}{1+\epsilon_I^* \tan \beta} \lambda_0^{JI} 
& J < I
\end{array} \right.
\end{eqnarray}
and
\begin{eqnarray}
(\mathbf{\Delta D_R})^{JI} = \left\{ \begin{array}{ll}
- \frac{\bar{m}_{d_I}}{\bar{m}_{d_J}} \left(\frac{\epsilon_Y y_t^2 \tan \beta}{
1+\epsilon_J \tan \beta} + \frac{\epsilon_Y^* y_t^2 \tan \beta}{1+\epsilon_I^* 
\tan \beta}\right) e^{i(\theta_J-\theta_I)} \lambda_0^{JI} & J > I \\
\frac{\bar{m}_{d_J}}{\bar{m}_{d_I}} \left(\frac{\epsilon_Y y_t^2 \tan \beta}{
1+\epsilon_I \tan \beta} + \frac{\epsilon_Y^* y_t^2 \tan \beta}{1+\epsilon_J^* 
\tan \beta}\right) e^{i(\theta_J-\theta_I)} \lambda_0^{JI}  & J < I
\end{array} \right.
\end{eqnarray}
Putting these matrices back into Eq.\ (\ref{massoffdiag:eq}) with $(J,I)=(2,1)$ 
the dominant terms have the form
\begin{eqnarray}
e^{-i\theta_2}(\mathbf{\bar{m}_d \Delta D_L})^{21} =-\frac{\bar{m}_s \epsilon_Y 
y_t^2 \tan \beta}{1+\epsilon_2 \tan \beta} \lambda_0^{21},
\end{eqnarray}
which are comparable to the terms that were neglected in 
Eq.~(\ref{massoffdiag:eq}) like
\begin{eqnarray}
e^{-i\theta_2}(\mathbf{(\Delta m_d) (\Delta D_L)})^{21} = - \frac{\bar{m}_s 
\epsilon_Y^2 y_t^4 \tan^2 \beta}{(1+\epsilon_2 \tan \beta)(1+\epsilon_3 \tan 
\beta)} \lambda_0^{21}.
\end{eqnarray}
This is particularly true for values of $\epsilon_3 < 0$ and large values 
of $\tan\beta$. Therefore the deviation between Ref.~\cite{Dedes:2002er} 
and Ref.~\cite{Buras:2002vd} in the Kaon sector is due to a breakdown in the 
perturbative series leading to first and second order contributions being 
comparable. The expansion shown in Ref.~\cite{Buras:2002vd} works for the $(1,3),
(2,3),(3,1)\mbox{ and }(3,2)$ components as they expanded the mass matrices only
to first order. As mentioned above, an analysis of the second order corrections,
together with a derivation of Eqs.~(\ref{Gamma21:eq})--(\ref{Gamma12:eq}) is
presented in Appendix \ref{A1:sec}.

\subsubsection{Flavor changing in the Charged Higgs Coupling}

The process of calculating the flavor changing couplings for the charged
goldstone modes is exactly the same as in Ref.~\cite{Buras:2002vd}. As the couplings 
of the goldstone has to match those of the W-bosons at tree level, so as to form
its longitudinal mode, the flavor changing effects have to be
\begin{eqnarray}
(P_{LR}^{G+})^{JI} &=& - \frac{\sqrt{2}}{v} V_{eff}^{JI} \bar{m}_{d_I}\\
(P_{RL}^{G+})^{JI} &=& \frac{\sqrt{2}}{v} \bar{m}_{u_I} V_{eff}^{JI} 
\end{eqnarray}
The charged Higgs has the effective lagrangian~\cite{deltamb2} 
\begin{eqnarray}
\mathcal{L}_{eff}^{H+} &=& \frac{\sqrt{2}}{v} \bar{u}_R 
\left[ \cot \beta \mathbf{m}_u - \frac{v_d}{\sqrt{2}}\tan \beta \mathbf{
\Delta Y}_u \right]  V_{eff} d_L H^+  +  \nonumber \\
& & \frac{\sqrt{2}}{v} \bar{u}_L V_{eff} \mathbf{D}_{L}^{\dagger} \left[
\tan \beta \mathbf{m}_d - \frac{v_u}{\sqrt{2}} \cot \beta \mathbf{\Delta Y}_d 
\right] \mathbf{D}_R d_R H^+  
\end{eqnarray}
where 
\begin{eqnarray}
(\mathbf{\Delta Y}_u)^{JI} &=&   y_{u_J} (\epsilon_0^{'J} \delta^{JI} + 
\epsilon_Y^{'} y_b^2 V_0^{J3} V_0^{I3*}) \\
(\mathbf{\Delta Y}_d)^{JI} &=& - y_{d_J} (\epsilon_0^J \delta^{JI} + \epsilon_Y 
y_t^2 V_0^{3J*} V_0^{3I})
\end{eqnarray}
are the generic form of corrections to the down(up) Yukawas after neglecting the
Yukawas of the first two generations. The matrices $\epsilon_0'$ and 
$\epsilon_Y'$ are closely related to $\epsilon_0$ and $\epsilon_Y$ and their
form is given in Appendix~\ref{A1:sec}. Hence, we find for $I=1,2,3$
\begin{eqnarray}
(P_{RL}^{H+})^{3I} &=& \frac{\sqrt{2}}{v} m_t \cot \beta \;V_{eff}^{3I} \left( 1
- \tan \beta \left(\epsilon_0^{'3} \right.\right. \nonumber \\
& & \left. \left. + \epsilon_Y^{'} y_b^2 \left[\frac{1+\epsilon_3 
\tan \beta}{1+\epsilon_3^0 \tan \beta} \delta^{3I} - \frac{\epsilon_Y y_t^2 
\tan 
\beta}{1+\epsilon_3^0 \tan \beta}\right]\right)\right), \label{H+RL3i:eq}
\end{eqnarray}
for $J\neq3$
\begin{eqnarray}
(P_{RL}^{H+})^{J3} &=& \frac{\sqrt{2}}{v} m_{u_J} \cot \beta \;V_{eff}^{J3} 
\left(1 - \tan \beta \left (\epsilon_0^{'J} + \epsilon_Y^{'} y_b^2 \frac{1+
\epsilon_3^* \tan \beta}{1+\epsilon_3^{0*} \tan \beta} \right) \right) 
\label{H+RLj3:eq}
\end{eqnarray}
and finally for $(J,I)=(2,1),(1,2),(1,1) \mbox{ and } (2,2)$
\begin{eqnarray}
(P_{RL}^{H+})^{JI} &=& \frac{\sqrt{2}}{v}  m_{u_J} \cot \beta \; V_{eff}^{JI} 
\left(1 - \tan \beta \epsilon_0^{'J} \right) \label{H+RL21:eq}
\end{eqnarray}
which agrees with Ref.~\cite{Buras:2002vd} if the phases are neglected. To find the 
left-right coupling we neglect the $(\mathbf{\Delta Y}_d)$ as it is suppressed 
by $\cot \beta$ so that we have for $I\neq3$
\begin{eqnarray}
(P_{LR}^{H+})^{3I} &=& \frac{\sqrt{2}}{v} \frac{\bar{m}_{d_I} \tan \beta(1+
\epsilon_3 \tan \beta)}{(1+\epsilon_0^3 \tan \beta)(1+\epsilon_3^* \tan \beta)} 
V_{eff}^{3I} \left(\frac{1+\epsilon_0^{3*} \tan \beta}{1+\epsilon_I^* \tan 
\beta} - \frac{\epsilon_Y y_t^2 \tan \beta}{1+\epsilon_3 \tan \beta}\right),
\label{H+LR3i:eq}
\end{eqnarray}
for $J\neq3$
\begin{eqnarray}
(P_{LR}^{H+})^{J3} &=& \frac{\sqrt{2}}{v} \frac{\bar{m}_b \tan \beta}{1+
\epsilon_0^{3*} \tan \beta} V_{eff}^{J3} \label{H+LRj3:eq}
\end{eqnarray}
and for $(J,I)=(3,3) \mbox{ and } J \neq 3 \neq I$
\begin{eqnarray}
(P_{LR}^{H+})^{33} &=& \frac{\sqrt{2}}{v} \frac{\bar{m}_b \tan \beta}{1+
\epsilon_3^* \tan \beta} V_{eff}^{33} \label{H+LR33:eq}\\
(P_{LR}^{H+})^{JI} &=& \frac{\sqrt{2}}{v} \frac{\bar{m}_{d_I} \tan \beta}{1+
\epsilon_I^* \tan \beta} V_{eff}^{JI} \label{H+LR21:eq}
\end{eqnarray}

\section{Flavor changing processes in the Kaon and $B_s$-Meson systems}

\subsection{$\Delta F=2$ processes}

The effective Hamiltonian that contributes to $\Delta F=2$ processes in the Kaon
and $B_s$ meson systems have the generic form
\begin{eqnarray}
\mathcal{H}_{eff}^{\Delta F=2} = \frac{G_f^2 M_W^2}{16 \pi^2} \sum_i C_i(\mu) 
Q_i(\mu) \label{heff2:eq}
\end{eqnarray}
where $C_i(\mu)$ are the Wilson coefficients evaluted at the scale $\mu$. The 
$\Delta F$ operators for a meson of the form $(\bar{q}^J q^I)$ are
\begin{eqnarray}
& & Q^{VLL} = (\bar{q}_L^J \gamma_\mu q_L^I)(\bar{q}_L^J \gamma^\mu q_L^I) 
\nonumber\\
& &Q_1^{SLL} = (\bar{q}_R^J q_L^I)(\bar{q}_R^J q_L^I) \nonumber \\
& &Q_2^{SLL} = (\bar{q}_R^J \sigma_{\mu \nu} q_L^I)(\bar{q}_R^J \sigma^{\mu \nu}
q_L^I ) \nonumber \\
& & Q^{VRR} = (\bar{q}_R^J \gamma_\mu q_R^I)(\bar{q}_R^J \gamma^\mu q_R^I) 
\nonumber\\
& &Q_1^{SRR} = (\bar{q}_L^J q_R^I)(\bar{q}_L^J q_R^I). \\
& &Q_2^{SRR} = (\bar{q}_L^J \sigma_{\mu \nu} q_R^I)(\bar{q}_L^J \sigma^{\mu \nu}
q_R^I) \nonumber \\
& &Q_1^{LR} = (\bar{q}_L^J \gamma_\mu q_L^I)(\bar{q}_R^J \gamma^\mu q_R^I) \nonumber \\
& &Q_2^{LR} = (\bar{q}_R^J q_L^I)(\bar{q}_L^J q_R^I) \nonumber 
\end{eqnarray}

So for the $K^0-\bar{K}^0$ system the quantities of interest to us 
are $\epsilon_K$ and the eigenstate mass difference $\Delta M_K$, 
which to a good approximation have the form
\begin{eqnarray}
\Delta M_K =  2 {\rm Re}(\langle \bar{K}^0|
H_{eff}^{\Delta S=2}| K^0\rangle)  \;\;\;\;\;\;\;\;\;
\epsilon_K = \frac{e^{i\pi/4}}{\sqrt{2} \Delta M_K} {\rm Im}(\langle \bar{K}^0|
H_{eff}^{\Delta S=2}| K^0\rangle).
\end{eqnarray} 
The SUSY contribution to the  matrix element for the meson $M$ may be
written down as
\begin{eqnarray}
\langle \bar{M}|H_{eff}^{\Delta S=2}| M\rangle^{SUSY} &=& \frac{G_f^2 M_W^2}{
12 \pi^2} m_M F_{M}^2 \eta_2 \hat{B}_M \left[\bar{P}^{VLL}(C^{VLL}
(\mu_{SUSY})+C^{VRR}(\mu_{SUSY})) + \right. \nonumber \\ 
& & \bar{P}_1^{SLL}(C_1^{SLL}(\mu_{SUSY})+C_1^{SRR}(\mu_{SUSY})) + \nonumber \\
& & \bar{P}_2^{SLL}(C_2^{SLL}(\mu_{SUSY})+C_2^{SRR}(\mu_{SUSY})) \nonumber \\ 
& & +\bar{P}_1^{LR} C_1^{LR}(\mu_{SUSY}) + \left. \bar{P}_2^{LR} C_2^{LR}(
\mu_{SUSY}) \right] . \label{m12:eq}
\end{eqnarray}
For the Kaon system $m_K =0.498 \mbox{ GeV}, F_K = 0.16 \mbox{ GeV}$, the 
values of the NLO QCD factors from Ref.~\cite{Dedes:2002er} are
\begin{eqnarray}
\bar{P}_1^{VLL} = 0.25,\; \bar{P}_1^{LR} = -18.6,\; \bar{P}_2^{LR} = 30.6,\; 
\bar{P}_1^{SLL} = -9.3,\; \bar{P}_2^{SLL} = -16.6
\end{eqnarray}
for which the values $\eta_2=0.57, \hat{B}_K=0.85$ have been used. 
The dominant contributions 
as shown in Ref.~\cite{Buras:2002vd,Dedes:2002er} come from the double penguin 
diagrams which on matching give contributions to the Wilson coefficients
\begin{eqnarray}
C_2^{LR} = - \frac{16 \pi^2}{G_f^2 (V_{eff}^{21})^2 M_W^2} \sum_{S=1}^3 \frac{1
}{M_S^2} (X_{RL}^S)^{21} (X_{LR}^S)^{21} \nonumber \\
C_1^{SLL} = - \frac{8 \pi^2}{G_f^2 (V_{eff}^{21})^2 M_W^2} \sum_{S=1}^3 \frac{1
}{M_S^2} (X_{RL}^S)^{21} (X_{RL}^S)^{21} \label{dp:eq}\\
C_1^{SRR} = - \frac{8 \pi^2}{G_f^2 (V_{eff}^{21})^2 M_W^2} \sum_{S=1}^3 \frac{1
}{M_S^2} (X_{LR}^S)^{21} (X_{LR}^S)^{21}. \nonumber
\end{eqnarray}
Additional subleading contributions at large $\tan \beta$ come from the 
charged-Higgs boson and chargino box-diagram contributions to $\epsilon_K$, and
their form are given in the Appendix A.4 of Ref.~\cite{Buras:2002vd}. 

Similarly, for the $B_s$  eigenstate mass  differences
$\Delta M_s$, using again 
Eq.~(\ref{heff2:eq}) for $\Delta B=2$ processes, we get, approximately,
\begin{eqnarray}
\Delta M_s &=& 2 |\langle \bar{B}_s | H_{eff}^{\Delta B=2} | B_s \rangle| 
\end{eqnarray}
Therefore, using 
Eq.~(\ref{m12:eq}), the mass difference in the $\bar{B}_s-B_s$ meson system can 
be found using $m_{B_s} =5.37 \mbox{ GeV}, F_{B_s} = 0.230 \mbox{ GeV}$ and the 
values of NLO QCD factors from Ref.~\cite{Buras:2002vd} being
\begin{eqnarray}
\bar{P}_1^{VLL} =0.254,\; \bar{P}_1^{LR} = -0.71,\; \bar{P}_2^{LR} = 0.90,\; 
\bar{P}_1^{SLL} = -0.37,\; \bar{P}_2^{SLL} = -0.72
\end{eqnarray}
for which  the values $\eta_B=0.55, \hat{B}_{B_S}=1.3$
have been used. Again, the dominant 
contributions come from double-penguin diagrams which have the same form as 
Eq.~(\ref{dp:eq}) with the indices $(2,1) \rightarrow (3,2)$ and there are 
subdominant contributions from the box diagrams with charged Higgs bosons and 
stop-charginos.

\subsection{$\Delta F=1$ processes contributing to $B_s\rightarrow \mu^+ \mu^-$}

The effective Hamiltonian that contributes to $\Delta F=1$ processes in the 
$B_s$ meson system has the form
\begin{eqnarray}
\mathcal{H}_{eff}^{\Delta B=1} = \frac{G_f \alpha_{em}}{\sqrt{2}\pi s_w^2} 
V_{eff}^{tb} V_{eff}^{ts} \sum_i c_i(\mu) \mathcal{O}_i(\mu) \label{heff1:eq}
\end{eqnarray}
where the operators $\mathcal{O}$ are
\begin{eqnarray}
& &\mathcal{O}_A = (\bar{b}_L \gamma^\mu s_L)(\bar{l} \gamma_\mu \gamma_5 l) \\
& &\mathcal{O'}_A = (\bar{b}_R \gamma^\mu s_R)(\bar{l} \gamma_\mu \gamma_5 l)
\nonumber \\
& &\mathcal{O}_S = m_b (\bar{b}_R s_L)(\bar{l} l)\nonumber \\
& &\mathcal{O'}_S = m_s (\bar{b}_L s_R)(\bar{l} l)\nonumber \\
& &\mathcal{O}_P = m_b(\bar{b}_R s_L)(\bar{l} \gamma_5 l)\nonumber \\
& &\mathcal{O'}_S = m_s(\bar{b}_L s_R)(\bar{l} \gamma_5 l).\nonumber
\end{eqnarray}
The operators $\mathcal{O}_A \; \mbox{and} \; \mathcal{O'}_A$ 
can be dropped as $c_A \; 
\mbox{and} \; c'_A$ are proportional to the 
muon mass and so are small at large 
$\tan \beta$. Also the other primed operators are suppressed with respect to 
the unprimed ones due to the hierarchy of quark masses. So the dominant 
contributions at large $\tan \beta$ come from the penguin diagrams leading to 
the contributions 
\begin{eqnarray}
c_S &=& - \frac{4 \pi^2 m_{\mu} \tan \beta}{\bar{m}_b M_W^2 2^{7/4} G^{3/2} 
V_{eff}^{ts} \sin \beta} \sum_{I=1}^3 \frac{1}{M_I^2} (X_{RL}^I)^{32} O^{1I} 
\nonumber \\
c_P &=& i \frac{4 \pi^2 m_{\mu} \tan \beta}{\bar{m}_b M_W^2 2^{7/4} G^{3/2} 
V_{eff}^{ts}} \sum_{I=1}^3 \frac{1}{M_I^2} (X_{RL}^I)^{32} O^{3I}. 
\label{cscp:eq}
\end{eqnarray}
where $O^{IJ}$ is the neutral Higgs diagonalization matrix and related to
$x_u^S$ and $x_d^S$ through Eq.(\ref{neutHmix:eq}). 
Hence, in the large $\tan \beta$ limit we find~\cite{Buras:2002vd}
\begin{eqnarray}
\mathcal{BR}(B_s \rightarrow \mu^+ \mu^-) = 2.32 \times 10^{-6} M_{B_s}^2 (
|c_S|^2 + |c_P|^2) \label{bsmumu:eq}
\end{eqnarray}

\section{Numerical Results: Minimal Flavor Violation}

In this section we will study some of the phenomenological implications of the 
scenarios of minimal flavor violation. The quantities of 
interest in the following section are $\Delta M_K$, $\epsilon_K$, and in 
particular the observables in the $B$ sector, $\Delta M_s$ and $\mathcal{BR}(
B_s\rightarrow \mu^+ \mu^-)$. The standard model theoretical prediction of 
$\Delta M_s$ has errors associated with the quantities $\bar{m}_t, V_{ts},$ and
$F_{B_s}\sqrt{B_{B_s}}$ that lead to large theoretical 
uncertainties~\cite{Buras:2002yj,Battaglia:2003in}.
There is good agreement between the central values for the SM prediction for
$\Delta M_s$ obtained by the  CKMfitter and UTFit groups. Their evaluation
of the uncertainties is somewhat different. 
The UTFit group finds the $2\sigma$ range~\cite{utfit}
\begin{eqnarray}
16.7 \; {\rm ps}^{-1} \leq (\Delta M_s)^{SM} \leq 26.9 \; {\rm ps}^{-1} 
\label{deltamsbound:eq}
\end{eqnarray}
with central value $21.5\; {\rm ps}^{-1}$, which is consistent with the 
CKMfitter groups' $2\sigma$ range~\cite{ckmfitter}
\begin{eqnarray}
14.9 \; {\rm ps}^{-1} \leq (\Delta M_s)^{SM} \leq 31.4 \; {\rm ps}^{-1} 
\label{deltamsbounda:eq}
\end{eqnarray}
and central value $21.7 \; {\rm ps}^{-1}$. Additionally, the D0 collaboration 
has reported a signal consistent with values of $\Delta M_s$ in the range
\begin{eqnarray}
21 \; (\mbox{ps})^{-1} > \Delta M_s > 19 \; \mbox{ps}^{-1}
\end{eqnarray} 
at the 90 \% confidence level~\cite{Abazov:2006dm}. 
More 
recently, the CDF collaboration has made a measurement of 
$\Delta M_s$~\cite{CDFdeltams},  
\begin{eqnarray}
\Delta M_s = (17.33^{+0.42}_{-0.21} \pm 0.07 \mbox{(syst)}) \mbox{ps}^{-1} .
\end{eqnarray}

The experimental bound~\cite{Bernhard:2005yn,CDFmumu} 
\begin{eqnarray}
\mathcal{BR}(B_s\rightarrow \mu^+ \mu^-) \leq 1 \times 10^{-7} 
\label{bsmumubound:eq}
\end{eqnarray}
puts strong restrictions on possible flavor violating effects induced by
the double penguin contributions in the large $\tan\beta$ regime. 
The dominant contributions for large $\tan \beta$ to $\Delta M_s$ and 
$\mathcal{BR}(B_s\rightarrow \mu^+ \mu^-)$ come from the same penguin diagrams.
The dominant contributions to $\epsilon_0^J$ and $\epsilon_Y$  come from the 
gluino d-squark loop and  the chargino u-squark loop, respectively. Hence, for
heavy squarks, the form of these loop corrections can be written approximately 
as
\begin{eqnarray}
|\epsilon_0^3| &\approx& \frac{2 \alpha_s}{3 \pi} |M_3| |\mu| C_0(
m_{\tilde{b}_1}^2, m_{\tilde{b}_2}^2,|M_3|^2) \label{e0j:eq}\\
|\epsilon_Y| &\approx& \frac{1}{16 \pi^2} |A_t| |\mu| C_0(m_{\tilde{t}_1}^2,
m_{\tilde{t}_2}^2,|\mu|^2) \label{eY:eq},
\end{eqnarray}
where $C_0$ is the standard Passarino-Veltman function.

\subsection{Phenomenological constraints on Double Penguin Contributions in the 
MFV scenario}

\subsubsection{The effect of $\mathcal{BR}(B_s\rightarrow \mu^+ \mu^-)$ 
constraint on $\Delta M_s$}

As has been shown in Ref.~\cite{Buras:2002vd} the chargino box 
diagrams can be neglected if all the squark masses are greater than about $0.5$ 
TeV. We are now interested in setting an upper bound on the FCNC effects induced
by the double penguin contributions. From the form of  Eq.~(\ref{e0j:eq}) and 
Eq.~(\ref{eY:eq}) it is clear that the loop integrals are larger for smaller 
values of the squark masses. The value of $\epsilon_0$ is maximized for large
values of $\mu$ and for values of $M_3$ about twice the overall  squark mass 
value. The value of $\epsilon_Y$ on the other hand, is maximized for large 
values of $A_t$ and values of $\mu$ that are of order two times the overall 
squark mass value. At the same time, large values of $\mu$ and/or $A_t$ may 
induce the presence of color breaking minima~\cite{Kounnas:1983td, 
Casas:1995pd}. Hence, 
values of   $M_3 \sim 2 m_{\tilde{q}} \sim \mu$ maximize $\epsilon_Y$,
while pushing $\epsilon_0$ to large values.
For these values of the parameters, the loop corrections are given by
\begin{eqnarray}
|(\epsilon_0^3)_{MAX}| &\sim& 2.7 \times 10^{-2} \label{epsilon0:eq}\\
|(\epsilon_Y)_{MAX}| &\sim& 1 \times 10^{-2} \label{epsilonY:eq},
\label{maxvalues:eq}
\end{eqnarray}
where we have constrained the trilinear mass parameter $A_t \simlt 3 m_{\tilde{
q}}$, so as not to create color breaking minima~\cite{Kounnas:1983td,Casas:1995pd}. 
Let us stress that the bounds on the parameters coming from 
color breaking minima may be avoided by assuming metastability of the 
electroweak symmetry breaking vacuum.  
However, the somewhat extreme values of the parameters given above  induce additional
anomalies in the low energy spectrum. For instance, values of $A_t \simgt 3.2 
m_{\tilde{q}}$, decrease the physical Higgs mass to values lower than the current 
experimental bound on this quantity~\cite{mhiggsRG1}--\cite{mhiggsEP5}.
It is also important to stress that for negative values of $\mu M_3$, the
coupling $X_{RL}^{JI}$ may be enhanced by taking even larger values of $|\mu|$.
Indeed, $\epsilon_Y$ only falls off  slowly for larger $|\mu|$, while
$\epsilon_0^J$ increases linearly and therefore $X_{RL}^{JI}$ grows with 
increasing $\mu$. We shall comment on the effect of taking larger values of $|\mu|$ below.

In the region of large $\tan \beta$ the heavy CP-even and CP-odd masses are
approximately equal and the Higgs mixing angle $\alpha \sim 1/\tan\beta$, so 
that the dominant contribution to $\mathcal{BR}(B_s\rightarrow \mu^+ \mu^-)$ is
given by
\begin{eqnarray}
\mathcal{BR}(B_s\rightarrow \mu^+ \mu^-) = 4.64 \times 10^{-6} M_{B_s}^2 \left(
\frac{4 \pi^2 m_{\mu} \tan \beta}{\bar{m}_b M_W^2 2^{7/4} G^{3/2} |V_{eff}^{ts}
|}\right)^2 \frac{|(X_{RL}^A)^{32}|^2}{M_A^4}.\label{bstbma:eq}
\end{eqnarray}

Similarly we find the dominant SUSY contribution to $\Delta M_s$ comes from the
$C_2^{LR}$ coefficient. To understand why the $C_2^{LR}$ term is dominant over
the $C_1^{SLL}$ we consider the case when there is no CP violation in the
neutral Higgs sector. In the basis $(H^0,h^0,A)$ we have $x_u^S = (\sin
\alpha,\cos \alpha,-i \cos \beta)$ and $x_d^S=(\cos \alpha,-\sin \alpha,i\sin 
\beta)$, where $\alpha$ is the Higgs mixing angle. Putting these values into 
Eq.(\ref{dp:eq}) for the $(3,2)$ component, we find~\cite{Buras:2002vd}, 
\begin{eqnarray}
C_2^{LR} \propto \bar{m}_b \bar{m}_s \tan^4 \beta \left(\frac{\sin^2(\alpha-
\beta)}{M_{H^0}^2} +\frac{\cos^2 (\alpha-\beta)}{M_{h^0}^2} + \frac{1}{M_A^2}
\right) \\
C_1^{SLL} \propto \bar{m}_b^2 \tan^4 \beta \left(\frac{\sin^2(\alpha-
\beta)}{M_{H^0}^2} +\frac{\cos^2 (\alpha-\beta)}{M_{h^0}^2} - \frac{1}{M_A^2}
\right).
\end{eqnarray}
From a cursory inspection of these two equation it is not clear which term 
is dominant, at large $\tan \beta$, as $C_2^{LR}$ is suppressed by 
a factor of $\bar{m}_s/\bar{m}_b$ 
with respect to $C_1^{SLL}$. However, using the constraint equations that 
relate $M_{h^0}$, $\alpha$ and $\beta$ at tree-level in the 
MSSM~\cite{Carena:2002es} we find
\begin{eqnarray}
& & M_{h^0}^2 \approx M_Z^2 (1- \frac{4}{\tan^2 \beta}) \\
& & \frac{\cos^2(\alpha-\beta)}{M_{h^0}^2} = \frac{M_Z^2-M_{h^0}^2}{M_A^2(M_{
H^0}^2-M_{h^0}^2)} \approx \frac{4M_Z^2}{M_A^4 \tan^2 \beta}
\end{eqnarray} 
where only the lowest order terms in $(M_Z^2/M_A^2)$ have been kept. Using
these tree-level approximations we find that
\begin{eqnarray}
C_2^{LR} \propto \bar{m}_b \bar{m}_s \tan^4 \beta \frac{2}{M_A^2} \\
C_1^{SLL} \propto \bar{m}_b^2 \tan^2 \beta \frac{4M_Z^2}{M_A^4}.
\end{eqnarray}
Thus at large $\tan \beta$ and moderate or large $M_A$, $C_2^{LR}$ clearly dominates 
over $C_1^{SLL}$.\footnote{When the loop factors and phases are included the 
approximation for $C_1^{SLL}$ still holds up to a factor of order 1.} 
The value of $\Delta M_s$, 
including the corrections from new physics, may be represented as 
$(\Delta M_s)^{SM}|1+f_s|$, where $f_s$ is the total SUSY contribution.
Due to 
$C_2^{LR}$ being dominant we find
\begin{eqnarray}
f_s &=& - \frac{16 \pi^2 P_{LR}^2}{G_f^2 M_W^2 S_0(x_t) (V_{eff}^{32})^2} 
\frac{2}{M_A^2} (X_{RL}^A)^{32} (X_{LR}^A)^{32}.
\end{eqnarray}

\begin{figure}
\begin{center}
\resizebox{14cm}{!}{\includegraphics{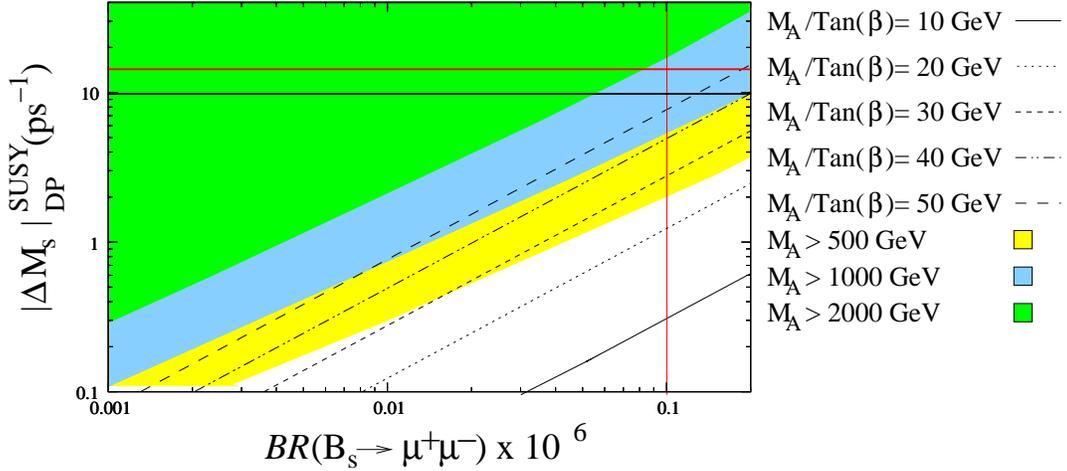}}
\end{center}
\caption{Correlation between $\mathcal{BR}(B_s\rightarrow \mu^+ \mu^-)$ and 
$\Delta M_s$. The squark masses are all uniform and have been set to $2$ TeV.
The rest of the SUSY parameters have been chosen so that $|\epsilon_0|$ and
$|\epsilon_Y|$ have their maximal values. The black lines have 
fixed values of $M_A/\tan \beta$, but varying gluino phase. The contours 
represent $\Delta M_s$ for different ranges of $M_A$ ($M_A \geq 500$, 1000, 2000 GeV) 
for gluino mass and $A_t$ phases equal to $\pi$, and varying $\tan \beta$ values. 
The red (grey) vertical line is the experimental bound on $\mathcal{BR}(B_s\rightarrow 
\mu^+ \mu^-)$. The horizontal black line is the $2\sigma$ upper bound on  the
double penguin contributions to $\Delta M_s$ from the UTFit group while 
red (grey) horizontal line is the same bound from the CKMfitter group.}
\label{dmsbsr1:fig}
\end{figure}
In the limit of universal squark masses, for fixed values of the supersymmetry 
breaking mass parameters, the ratio between $\Delta M_s$ and $\mathcal{BR}(B_s
\rightarrow \mu^+ \mu^-)$ is proportional to $(M_A/\tan \beta)^2$. Furthermore,
$f_s$ is negative~{\cite{Isidori:2001fv,Buras:2002wq,Buras:2002vd}}. Therefore,
unless $|f_s| 
>2$, the double penguin contributions to $\Delta M_s$ always interferes 
destructively with the SM contribution, at large $\tan\beta$. This result, 
showing the suppression of $\Delta M_s$ for enhanced $\mathcal{BR}(B_s 
\rightarrow \mu^+ \mu^-)$, has been known for some time and 
was first shown in Ref.~{\cite{Isidori:2001fv,Buras:2002wq,Buras:2002vd}}.

\begin{figure}
\begin{center}
\resizebox{14cm}{!}{\includegraphics{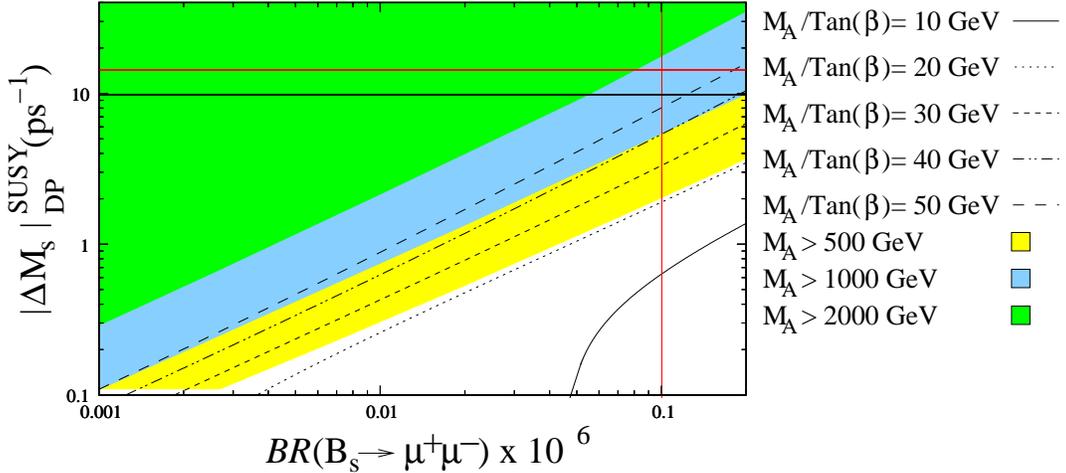}}
\end{center}
\caption{
Same as Fig.~\ref{dmsbsr1:fig}, but for third generation soft
supersymmetry breaking squark masses
equal to 0.5 TeV and first and second generation squark masses equal to
5 TeV.}
\label{dmsbsr10:fig}
\end{figure}

In Figs.~\ref{dmsbsr1:fig} and \ref{dmsbsr10:fig} we show  the correlation
between $\Delta M_s$ and $\mathcal{BR}(B_s \to \mu^+ \mu^-)$ for different
squark spectra and gaugino phases. In 
Fig.~\ref{dmsbsr1:fig} 
the black curves show  the correlation between the double penguin contributions
to $\Delta M_s$ and $\mathcal{BR}(B_s\rightarrow \mu^+ \mu^-)$ for uniform
squark masses $\sim 2$~TeV. We have chosen the uniform squark masses to be 
$\sim 2$~TeV so as to ensure that for $M_A \leq 1$~TeV the effective 
Lagrangian in Eq.(\ref{Leff:eq}) and Eq.(\ref{Lmass:eq}) remains valid.
Had we chosen squark masses of the order of 1 TeV, then the 
low-energy effective theory
would break down for $M_A$ close to 1~TeV, and a more detailed analysis
of the $\epsilon_i$'s momentum dependence would be required for these
large values of $M_A$.  
Each of the black curves have different values of $M_A/\tan\beta$. 
The contours represent the maximal values of $|\Delta M_s|^{DP}$, for a given
value of $BR(B_s \to \mu^+\mu^-)$, and for a given range of values of $M_A$. 
Due to the fact that for fixed $M_A$, the ratio of $|\Delta M_s|^{DP}$ 
to   $\mathcal{BR}(B_s \rightarrow \mu^+\mu^-)$ 
goes like $1/\tan^2\beta$, in order to maximize $|\Delta M_s|$ 
for any given value of $\mathcal{BR}(B_s \rightarrow \mu^+\mu^-)$ we
need to minimize the value of $\tan\beta$.
Inspection of the expressions 
given above shows that this may be achieved by choosing positive values of 
$\mu$, $\arg(M_3)=\arg(A_t)=\pi$ and maximal values of $|\epsilon_0|$ and 
$|\epsilon_Y|$.  In order to define the contours we have taken the values
of the loop corrections given in Eq.~(\ref{maxvalues:eq}). The horizontal 
black and red (grey) line corresponds to an upper bound on the largest possible
contribution to $\Delta M_s$ from new physics using the
$2\sigma$ values obtained by the UTFit and CKMfitter collaborations, 
Eq.~(\ref{deltamsbound:eq}) and Eq.~(\ref{deltamsbounda:eq}), respectively. 
In order to get a precise evaluation
of this bound, a complete fit to the flavor violating processes within
the MSSM should be performed, something that is beyond the scope of this
paper. However, since in this region of parameters the only relevant
new flavor violating contributions are from the double penguin
diagrams, we can make an estimate of this bound in the following way:
From Eq.~(\ref{deltamsbound:eq}) or Eq.~(\ref{deltamsbounda:eq}) we have
a 2-$\sigma$ range that goes from values 
consistent with the experimentally measured value up to values
much larger than the measured values. Therefore the negative
double penguin contribution can be as large as the difference between the
maximum allowed SM value and the smallest allowed experimental value.
This leads to an upper bound on the magnitude of the double penguin 
contributions to $\Delta M_s$ of about $\sim 10$~ps$^{-1}$ for 
the UTFit limits in Eq.~(\ref{deltamsbound:eq}) or $\sim 14.5$~ps$^{-1}$
for the CKMfitter limits in Eq.~(\ref{deltamsbounda:eq}).
From Fig.~\ref{dmsbsr1:fig} it is clear that, for CP-odd Higgs
masses below 1~TeV, this bound  does not lead to any further
constraint beyond the one already obtained by the non-observation
of the branching ratio of the decay $B_s \to \mu^+ \mu^-$.

It is possible to enhance the value of $\Delta M_s$ beyond what we have 
explored, by allowing values of $|\mu| > 2 \; m_{\tilde{q}}$. If, for instance,
we consider values of $\mu \gsim 3 m_{\tilde{q}}$, for the same value of 
$\mathcal{BR}(B_s\rightarrow \mu^+ \mu^-)$ we can enhance 
$\Delta M_s$ by a factor $\sim 1.5$. This suggests that the contours in
Figs.~\ref{dmsbsr1:fig} and \ref{dmsbsr10:fig} are not strict upper bounds,
and can be further enhanced, almost in a linear way, by pushing 
$|\mu|/m_{\tilde{q}}$ to larger values. However, due to the extreme values of 
the mass parameters selected in defining the contours, these are indicative
of the upper bound on the the double penguin contributions to $\Delta M_s$ for 
a given value of $\mathcal{BR}(B_s\rightarrow \mu^+ \mu^-)$ for natural values 
of the mass parameters.

In Fig.~\ref{dmsbsr10:fig} we depart from the limit of universal squark masses,
by setting the third generation squark masses $\sim 0.5$~TeV while the first 
two generation squark masses are $5$ TeV, which leads to $\epsilon_0^3$ having 
its maximal value, but $\epsilon_0^1$ and $\epsilon_0^2$ being $100$ times 
smaller. Hence, this splitting of the squark masses spoils the linear 
correlation between $\Delta M_s$ and $\mathcal{BR}(B_s\rightarrow \mu^+ \mu^-)$
due to the different parametric dependences of $X_{RL}^{32}$ and $X_{RL}^{23}$ 
for split masses. In both  Figs.~\ref{dmsbsr1:fig} and \ref{dmsbsr10:fig} the 
vertical red (grey) line is the experimental bound on $\mathcal{BR}(B_s
\rightarrow \mu^+ \mu^-)$ in Eq. (\ref{bsmumubound:eq}).

\begin{figure}
\begin{center}
\resizebox{14cm}{!}{\includegraphics{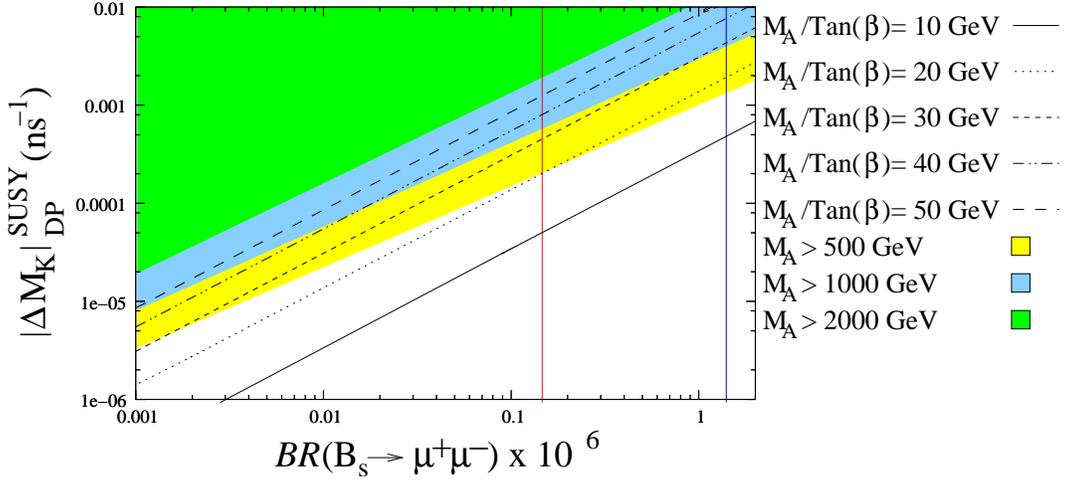}}
\end{center}
\caption{
Correlation between $\mathcal{BR}(B_s\rightarrow \mu^+ \mu^-)$ and 
$\Delta M_K$. The squark masses are all uniform and have been set to $2$ TeV.
The rest of the SUSY parameters have been chosen so that $|\epsilon_0|$ and
$|\epsilon_Y|$ have their maximal values. The black lines have 
fixed values of $M_A/\tan \beta$. The 
contours are the double penguin contributions to $\Delta M_K$ for gluino mass 
and $A_t$ phases equal to $\pi$, but varying $\tan\beta$.
The left red (grey) vertical line is the present experimental bound on 
$\mathcal{BR}(B_s\rightarrow \mu^+ \mu^-)$ while the right blue (black) verticle line 
is the previous limit.} 
\label{dmkbs:fig}
\end{figure}

Figs.~\ref{dmsbsr1:fig} and \ref{dmsbsr10:fig} suggest that large double 
penguin contributions to $|\Delta M_s|$ may not be obtained, for values of 
$\epsilon_0^J$ and $\epsilon_Y$ close to their maximal values in 
Eqs.~(\ref{epsilon0:eq}) and (\ref{epsilonY:eq}), without violating the 
$\mathcal{BR}(B_s\rightarrow \mu^+ \mu^-)$ bound. Due to these bounds, for 
values of $M_A < 1$~TeV, the double penguin 
corrections to $\Delta M_s$ are restricted to be negative and relatively small,
so that $|\Delta M_s|^{\rm SUSY} \simlt 4 \times 10^{-12}$~GeV, or equivalently
$|\Delta M_s|^{\rm SUSY} \simlt 6$~ps$^{-1}$. 

\begin{figure}
\begin{center}
\resizebox{14cm}{!}{\includegraphics{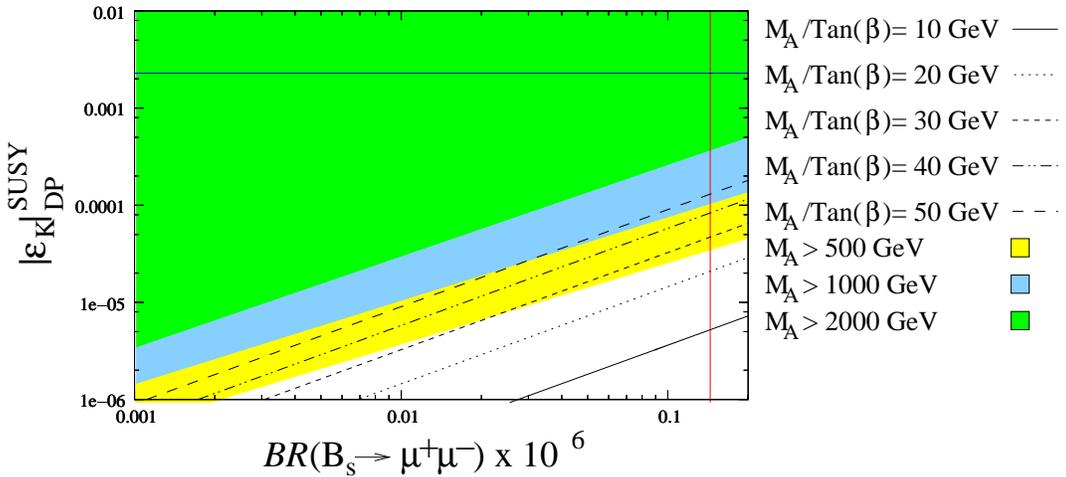}}
\end{center}
\caption{Same as Fig.~\ref{dmkbs:fig}, but for $\epsilon_K$. Only the current 
bound on $\mathcal{BR}(B_s\rightarrow \mu^+ \mu^-)$ is shown, by the vertical 
red (grey) line and the horizontal blue (black) line is the experimentally 
measured value of $\epsilon_K$.} 
\label{ekbs:fig}
\end{figure}

The $\mathcal{BR}(B_s\rightarrow \mu^+ \mu^-)$ bound also constrains 
contributions to $\Delta M_d$ and $\Delta M_K$ to values within experimental 
errors. For example, in Fig.~\ref{dmkbs:fig}, the SUSY contributions to 
$\Delta M_K$ in the Kaon system for uniform squarks masses are below the 
experimental error of $6\times 10^{-18}$~GeV or $0.01 \; ns^{-1}$, even for 
large values of $(M_A/\tan \beta)^2$. These results seem to be at variance with
those obtained in Ref.~\cite{Dedes:2002er}. This is mainly due to the fact that
the authors of Ref.~\cite{Dedes:2002er} represented results in regions of 
parameters where the value of $\mathcal{BR}(B_s \to \mu^+ \mu^-)$ is well 
above the present limit. Observe that, to arrive at this conclusion, the new 
limit on $\mathcal{BR}(B_s \to \mu^+\mu^-)$ is essential. From
Fig.~\ref{dmkbs:fig} we can also see how the improvement in the limit on
$\mathcal{BR}(B_s \to \mu^+\mu^-)$ forces the double penguin contributions
to $|\Delta M_K|$ from SUSY to be small. Finally, Fig.~\ref{ekbs:fig} shows 
similar results for $\epsilon_K$. As happens in the case of $\Delta M_K$, the 
results for values of $M_A < 1$~TeV are far below the current experimental 
value of $2.282 \times 10^{-3}$.

\begin{figure}
\begin{center}
\resizebox{10cm}{!}{\includegraphics{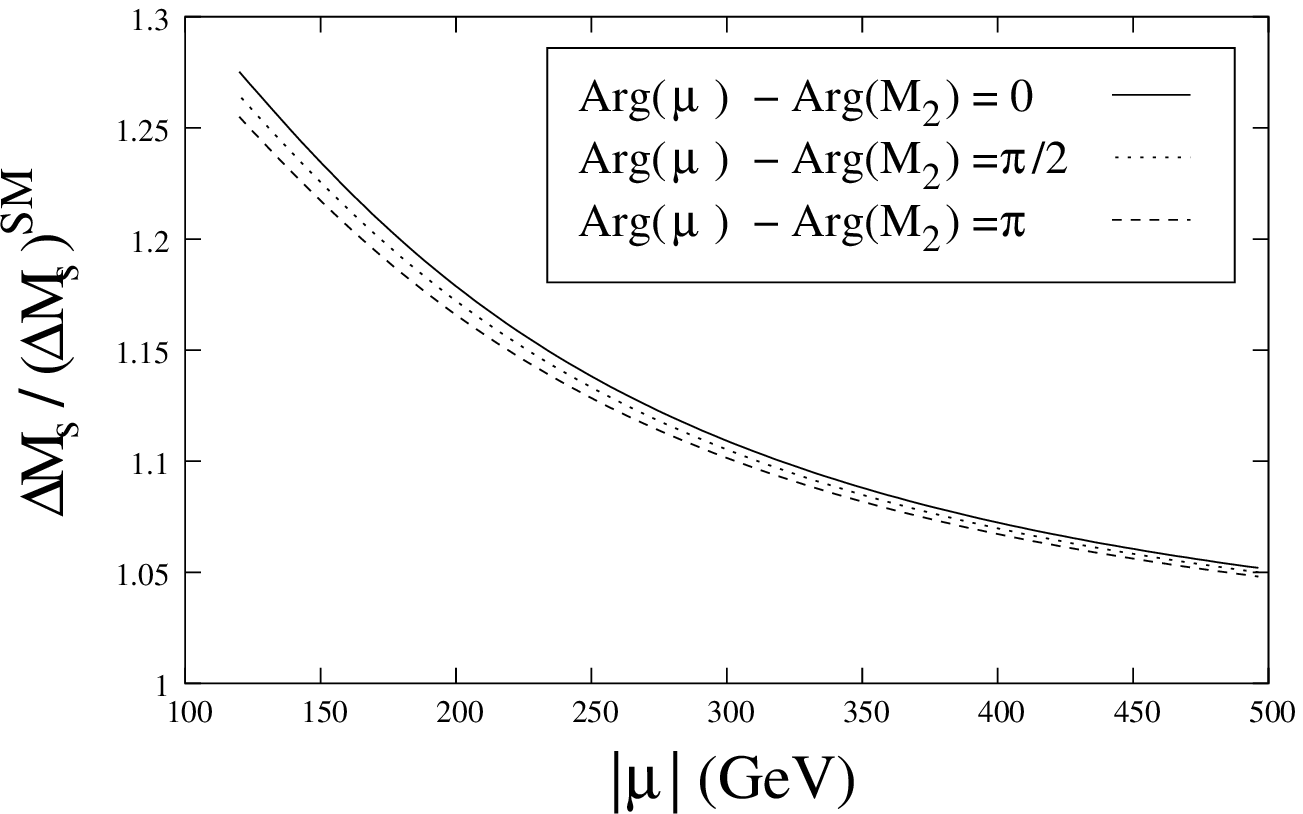}}
\end{center}
\caption{Variation of SUSY contributions to $\Delta M_s$ with input parameters 
$M_A=200$ GeV, $M_3=1000$ GeV, $M_{D_3}= M_{SUSY}=2000$ GeV, $2 M_1=M_2=\mu$, 
$M_{U_3}^2 = -90^2$ GeV$^2$, $A_t=-1000$ GeV and $\tan\beta=10$}
\label{dmsmu:fig}
\end{figure}

\begin{figure}
\begin{center}
\resizebox{10cm}{!}{\includegraphics{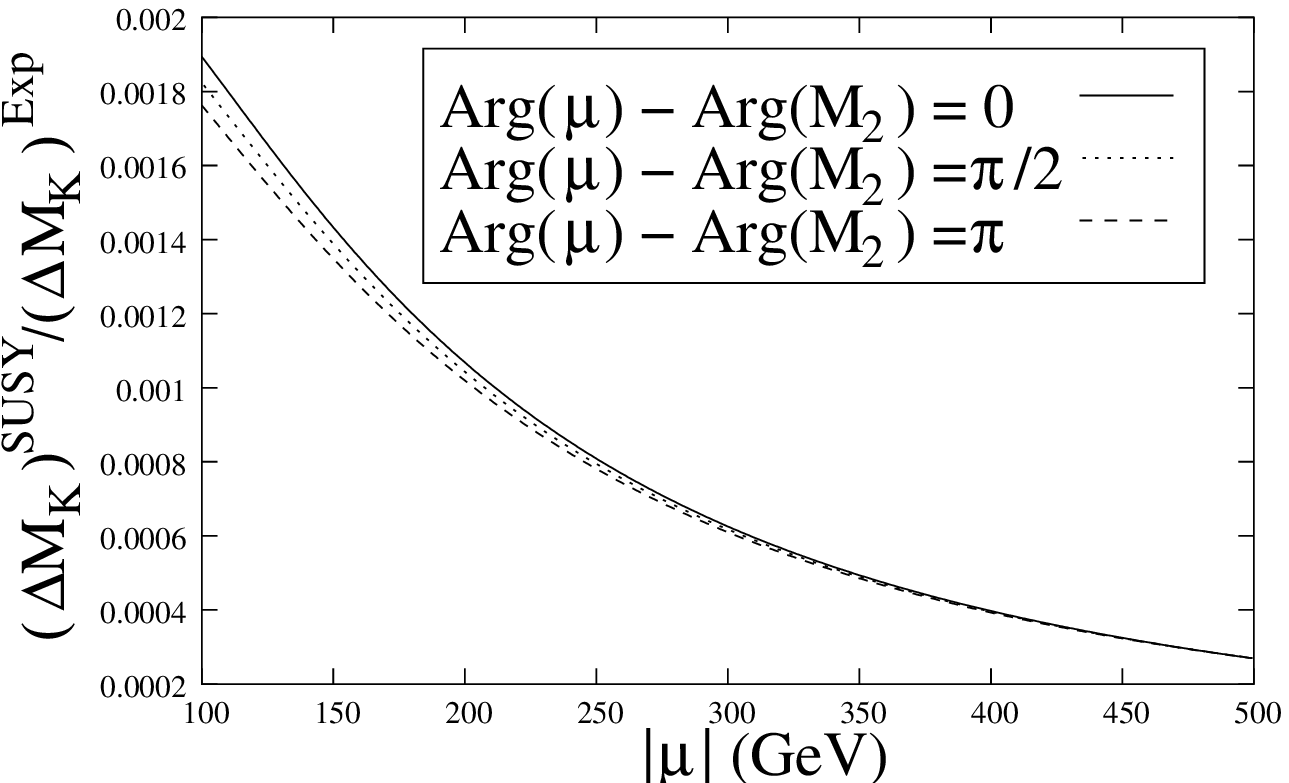}}
\end{center}
\caption{Variation of SUSY contributions to $\Delta M_K$ with input parameters 
$M_A=200$ GeV, $M_3=1000$ GeV, $M_{D_3}= M_{SUSY}=2000$ GeV, $2 M_1=M_2=\mu$, 
$M_{U_3}^2 = -90^2$ GeV$^2$, $A_t=-1000$ GeV and $\tan\beta=10$}
\label{dmkmu:fig}
\end{figure}

\begin{figure}
\begin{center}
\resizebox{10cm}{!}{\includegraphics{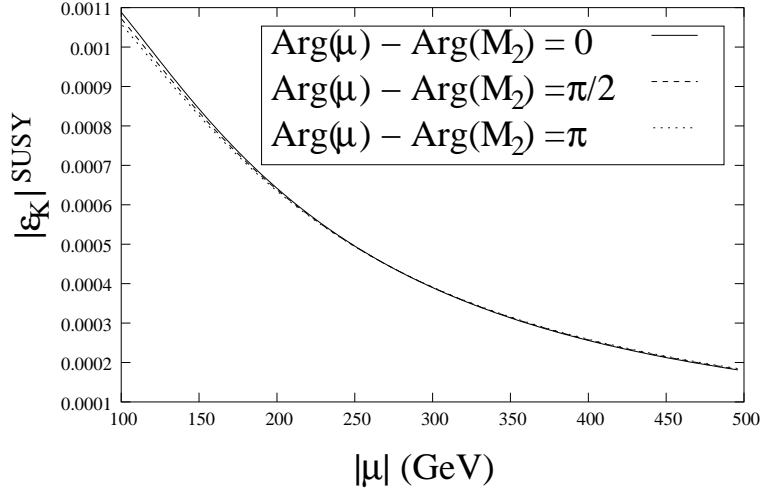}}
\end{center}
\caption{Variation of SUSY contributions to $\epsilon_K$ with input parameters 
$M_A=200$ GeV, $M_3=1000$ GeV, $M_{D_3}= M_{SUSY}=2000$ GeV, $2 M_1=M_2=\mu$, 
$M_{U_3}^2 = -90^2$ GeV$^2$ $A_t=-1000$ GeV and $\tan\beta=10$}
\label{epskmu:fig}
\end{figure}

However, within the Minimal Flavor Violation scheme, large contributions to 
$\Delta M_s$ are possible for scenarios in which the stops and charginos are 
light, so that the chargino-stop box diagrams become larger. Furthermore, the 
bound on $\mathcal{BR}(B_s\rightarrow \mu^+ \mu^-)$ can be satisfied by going 
to regions
of large $M_A$ or low $\tan \beta$ as chargino-stop box contributions are not 
very sensitive to $\tan \beta$. This scenario is similar to that discussed 
in Ref.~\cite{Balazs:2004ae} where low values of $\tan \beta$ satisfy both the 
dark matter and baryogenesis constraints. In Fig.~\ref{dmsmu:fig}, we choose 
SUSY parameters
\begin{eqnarray}
& & M_A=200 \mbox{ GeV, } M_3=1000 \mbox{ GeV, } M_{D_3}= M_{SUSY}=2000 
\mbox{ GeV, } \nonumber \\
& & M_{U_3}^2 = -90^2 \; \mbox{GeV}^2, \; A_t=-1000 \mbox{ GeV } \tan\beta=
10, \nonumber \\
& & \mbox{ and } 100 \mbox{ GeV } \lsim 2 M_1, M_2, \mu \lsim 500 \mbox{ GeV }
\nonumber 
\end{eqnarray}
that agree with  dark matter and baryogenesis constraints and 
produce a value of $\Delta M_s$ that is enhanced with respect to the
SM value. For this kind of particle 
spectrum the double penguin contributions to 
$\Delta M_s$ are small compared to that of chargino stop diagrams. 
Although the enhancement of $\Delta M_s$ is small,  
a comparison of the SM prediction, 
Eq.~(\ref{deltamsbound:eq}) and Eq.~(\ref{deltamsbounda:eq}),
and the experimentally measured value leads to  
disfavor additional positive contributions of $\Delta M_s$, larger than about
3.5~ps$^{-1}$, where we have taken into account the SM allowed range given
by the CKMfitter collaboration Eq.(\ref{deltamsbounda:eq}),
at the 2-$\sigma$ level. Even stronger constraints would be obtained
if the UTfit values in Eq.(\ref{deltamsbounda:eq}) for 
$(\Delta M_s)^{\rm SM}$ were used.  
Therefore the smallest values of $\mu$, smaller than 200~GeV, would be 
disfavored. A global fit to all flavor dependent observables within
this scenario would be necessary in order to determine the precise lower
bound on $\mu$, something that is beyond the scope of this article. Also
observe that for larger values of $\tan\beta$ there may be relevant
double penguin contributions that could cancel the positive box-diagram
contributions and therefore the bound on $\mu$ could be relaxed in this case.

Although this scenario  leads to contributions
to $\Delta M_K$ that are smaller than the present experimental errors on this 
quantity, as can be seen in Fig.~\ref{dmkmu:fig}, it 
leads to interesting corrections to
$\epsilon_K$, as shown in Figure~\ref{epskmu:fig}. The 
results in Fig.~\ref{epskmu:fig} were obtained for a value of the CKM 
phase $\delta = \pi/3$ (the best fit value within
the SM).
Experimentally we know that
\begin{eqnarray}
|\epsilon_K| = (2.282 \pm 0.014) \times 10^{-3}.
\end{eqnarray} 
and therefore the SUSY corrections are significant. For lower values of the 
CKM phase, however, the SUSY contributions to $|\epsilon_K|$ within this 
scenario can be smaller. The experimental value of $\epsilon_K$ is
usually used to put a constraint on the $\bar{\rho}-\bar{\eta}$ 
plane\footnote{$\bar{\rho}$ and $\bar{\eta}$ are the usual corrected 
Wolfenstein parameters of the CKM matrix}. The SM contributions to 
$\epsilon_K$ leads to the constraint equation~\cite{Herrlich:1995hh}
\begin{eqnarray}
5.3 \times 10^{-4} = B_K A^2 \bar{\eta}[(1-\bar{\rho})A^2 \lambda^4 
\eta_2^* S(x_t^*) + \eta_3^* S(x_c^*,x_t^*) - \eta_1^* x_c^*] \label{rhoeta:eq}
\end{eqnarray}
where $B_K = 0.75 \pm 0.10, A \sim 0.85, \lambda = 0.22, \eta_1^* = 
1.32^{+0.21}_{-0.23}, \eta_2^* = 0.57^{+0.00}_{-0.01}, \eta_3^* = 
0.47^{+0.03}_{-0.04}$ and $S(x_t)$ and $S(x_c,x_t)$ are the Inami-Lim 
functions. Because the stops are light
the dominant contributions to $\epsilon_K$ come from the chargino stop 
diagram. Under these approximations we find the $\epsilon_K$ constraint 
equation in $\bar{\rho}-\bar{\eta}$ plane is modified to become
\begin{eqnarray}
5.3 \times 10^{-4} = B_K A^2 \bar{\eta}[(1-\bar{\rho})(1+\zeta)A^2 \lambda^4 
\eta_2^* S(x_t^*) + \eta_3^* S(x_c^*,x_t^*) - \eta_1^* x_c^*].
\label{conststop:eq}
\end{eqnarray}
where $\zeta$ hides all the SUSY dependences. The dominant contribution to 
$\epsilon_K$ from SUSY comes from the $C_{VLL}$ Wilson coefficient. Thus we 
have approximately,
\begin{eqnarray}
\zeta \sim \frac{\bar{P}_{VLL}}{8 G_F^2 M_W^2 S(x_t^*)} D_2(m_{\tilde{t}_2}^2,m_{\tilde{t}_2}^2,
m_{\chi_2}^2,m_{\chi_2}^2) 
\end{eqnarray}
where $m_{\tilde{t}_2}$ is the lightest stop mass, $m_{\chi_2}$ is the lightest
chargino mass and $D_2$ is the Passarino-Veltmann function
\begin{eqnarray}
D_2(x,y,z,t) &=& \frac{y^2}{(y-x)(y-z)(y-t)} \log \left(\frac{y}{x}\right) +
 \frac{z^2}{(z-x)(z-y)(z-t)} \log \left(\frac{z}{x}\right) + \nonumber \\
& & \frac{t^2}{(t-x)(t-y)(t-z)} \log \left(\frac{t}{x}\right).
\end{eqnarray}

Taking the lightest stop mass to be 120~GeV and approximating the
the lightest chargino mass by $|\mu|$ we can 
estimate $\zeta \sim 0.4$ for values of $\mu \sim 100$ GeV. However as $|M_2| \simeq 
|\mu|$ there are also relevant contributions from the heavier chargino.
Including these contributions, we obtain
$\zeta \sim 0.55$. Including this value of $\zeta$ in the theoretical prediction for
$\epsilon_K$ will lead to a modification of the values of $\bar{\rho}$ and $\bar{\eta}$
extracted from the fit to the flavor observables. Although a global fit to these
quantities within the light stop scenario is beyond the scope of this article, we notice
that for $\zeta \simlt 0.55$, the new constraint equation, Eq.~(\ref{conststop:eq}), is 
still consistent with the limits coming from $|V_{ub}|/|V_{cb}|$, $\sin(2\beta)_{\rm eff}$
and $\Delta M_{s,d}$ and therefore this scenario is not ruled out by these considerations.

\subsubsection{The effect of $\mathcal{BR}(B_s\rightarrow \mu^+ \mu^-)$ 
constraint on Higgs physics at the Tevatron and the LHC}

As shown above, in the minimal flavor violating scheme, all dominant FCNC 
effects at large $\tan \beta$ are proportional to $\epsilon_Y$, which is 
directly proportional to  the product of the $\mu$ and $A_t$, but 
inversely proportional to the square of the squark masses. The FCNC effects are
strongly enhanced for large values of $\tan\beta$ and small values of the 
CP-odd Higgs mass. The Tevatron collider is performing searches for 
non-standard Higgs bosons, which become efficient for exactly the same 
conditions. Therefore, in minimal flavor violating models, current bounds on 
the rate $B_s \to \mu^+ \mu^-$ impose strong constraints on the possibility of 
finding non-standard Higgs bosons at the Tevatron collider (for related
studies, see Refs.~\cite{Ibrahim:2002fx}--\cite{Dedes:2004yc}).  
This is 
particularly true for large values of the $A_t$ and $\mu$ parameters, for which
$\epsilon_Y$ is enhanced. 

Low values of the CP-odd Higgs mass are also associated with low values of the 
charged Higgs mass. These values of the charged Higgs 
mass induce large positive corrections to the branching ratio $BR(b \to s 
\gamma)$. Since the measured value of $BR(b \to s \gamma)$ agrees well with 
the SM prediction, these large charged Higgs induced corrections to the rare 
decay rate needs to be cancelled by similarly large corrections induced by 
supersymmetric particles. In minimal flavor violating schemes,
these SUSY corrections are associated with stop-chargino 
loops~\cite{Bertolini:1990if},\cite{Barbieri:1993av}--\cite{Borzumati:2003rr}.
For positive (negative) values of $A_t   \mu$, the
corrections to the amplitude of the decay $b \to s \gamma$ have
the same (opposite) sign to the ones associated with the charged Higgs
corrections, and grow linearly with $\tan\beta$. Therefore, agreement of 
the theoretical predictions with the experimental values of $BR(b \to s 
\gamma)$ for small values of $M_A$  demands negative values of $A_t \mu$.

Additional constraints come from the CP-even Higgs sector.
For a given value of the overall squark masses, the mass of the lightest
CP-even Higgs boson in the large $\tan\beta$ regime depends strongly
on the parameter $A_t$. In particular, this mass is maximized for
a value of $X_t = A_t - \mu/\tan \beta \simeq 2.4 \; M_{SUSY}$ (where 
$M_{SUSY}$ is equal to the  average stop mass) and minimized 
for values of $X_t = 0$~\cite{mhiggsRG1}. Due to the complicated 
dependence  of the Higgs boson 
properties on the supersymmetric mass parameters, searches for Higgs bosons at 
the Tevatron and the LHC are usually interpreted in terms of benchmark 
scenarios~\cite{Carena:2002qg}. For instance, the scenario  with $X_t/M_{SUSY} 
\simeq 2.4$ is named the maximal mixing scenario, since it is associated with 
the values of the stop mixing parameters that maximize the lightest CP-even 
Higgs mass. Similarly, $X_t = 0$ defines the minimal mixing scenario. While for
the maximal mixing scenario the constraints coming from FCNC are particularly 
strong, no constraint from $B_s \to \mu^+ \mu^-$ are expected to be obtained 
in the minimal mixing scenario.

\begin{figure}
\begin{center}
\resizebox{7.cm}{!}{\includegraphics{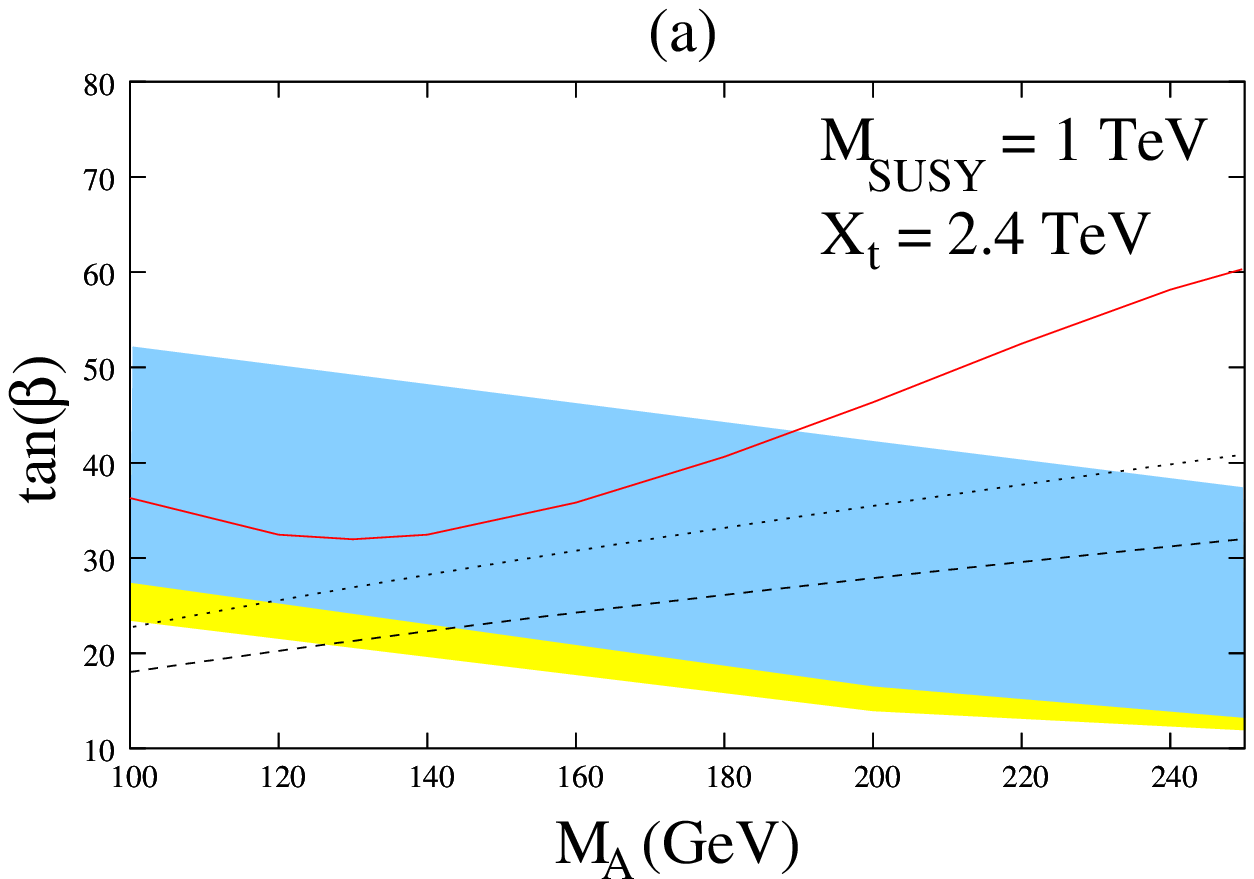}}
\resizebox{7.cm}{!}{\includegraphics{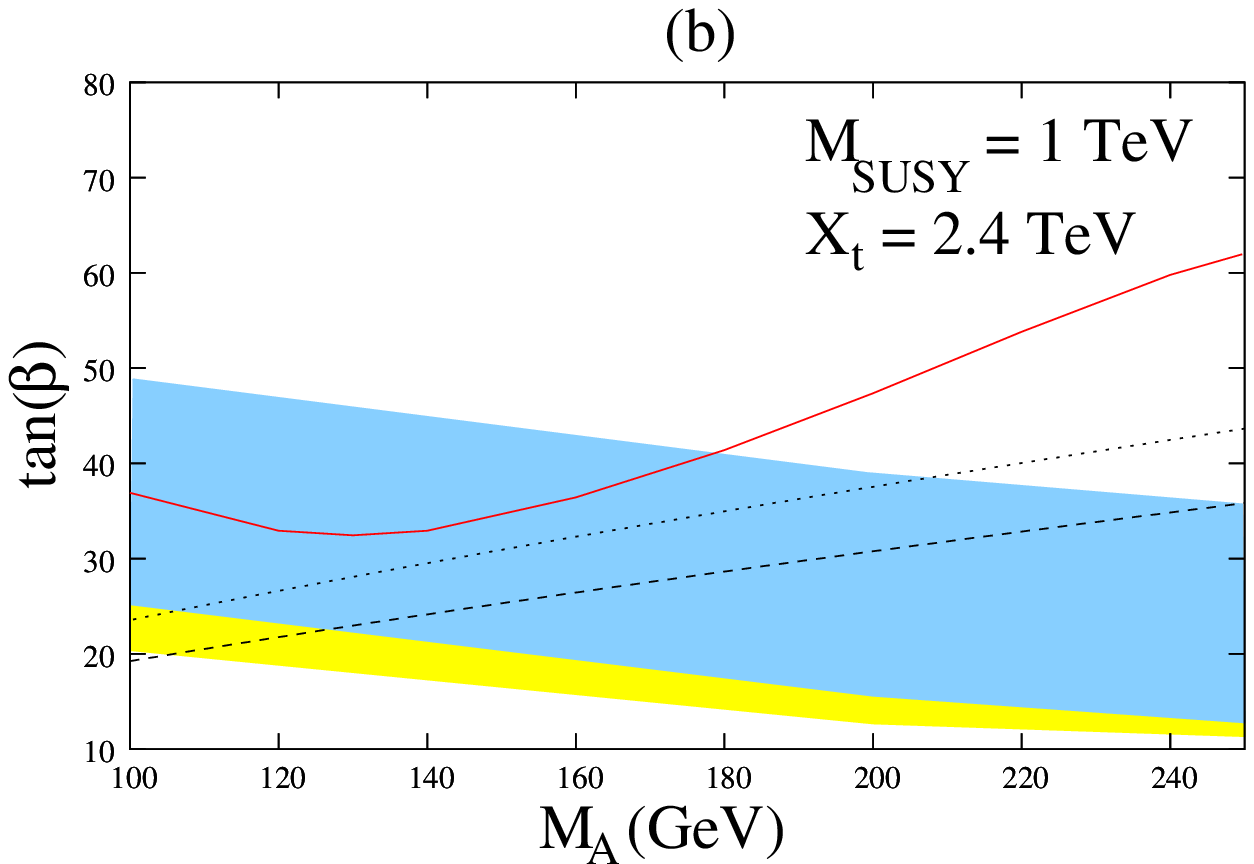}}
\resizebox{7.cm}{!}{\includegraphics{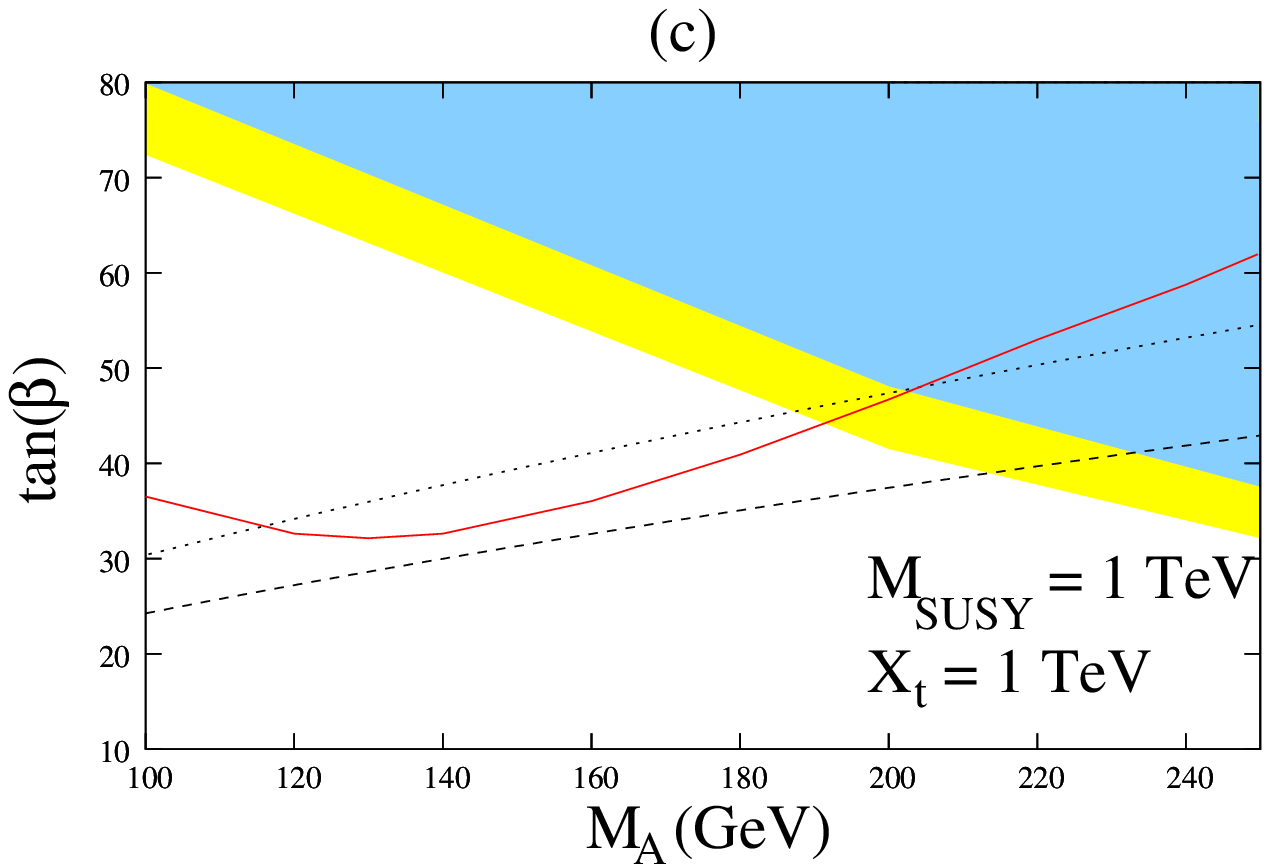}}
\resizebox{7.cm}{!}{\includegraphics{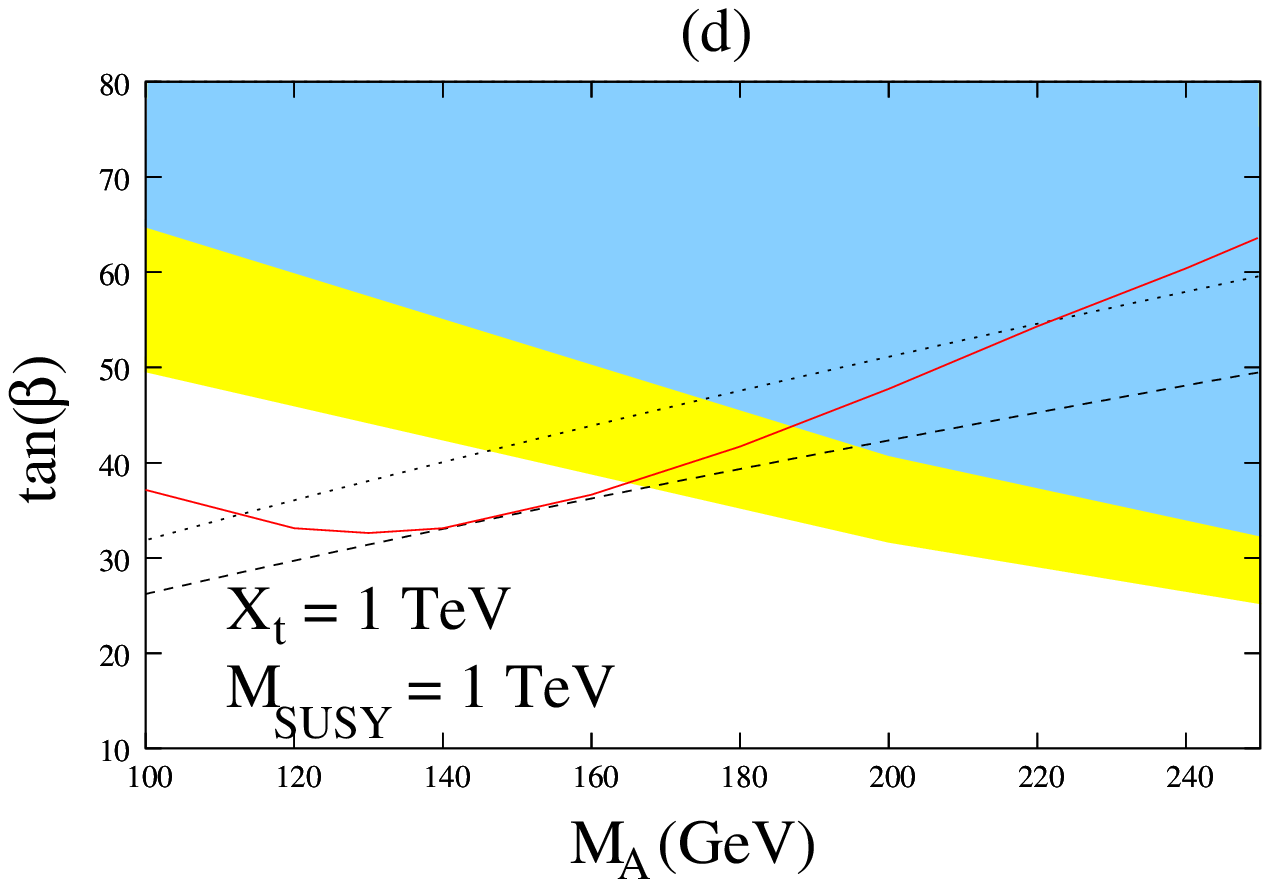}}
\resizebox{7.cm}{!}{\includegraphics{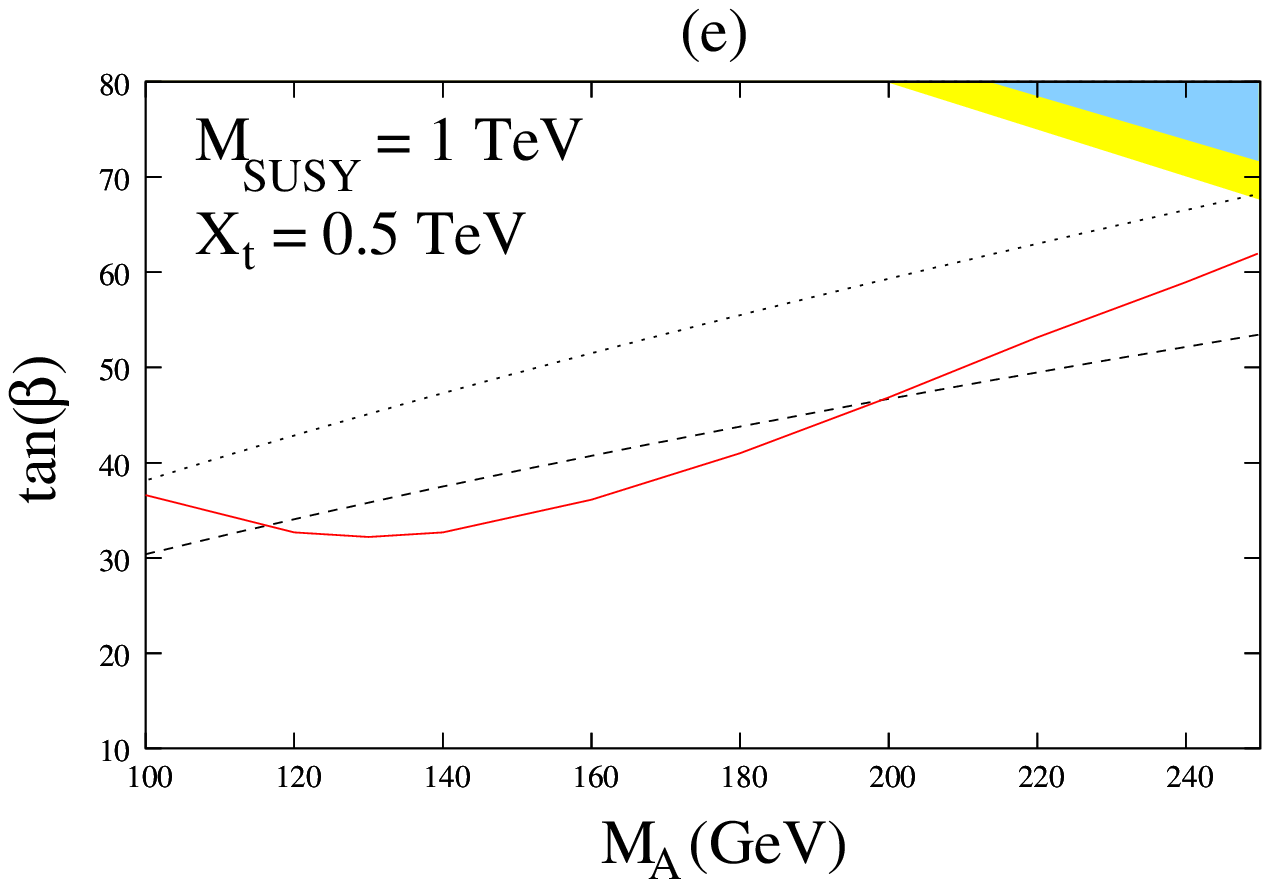}}
\resizebox{7.cm}{!}{\includegraphics{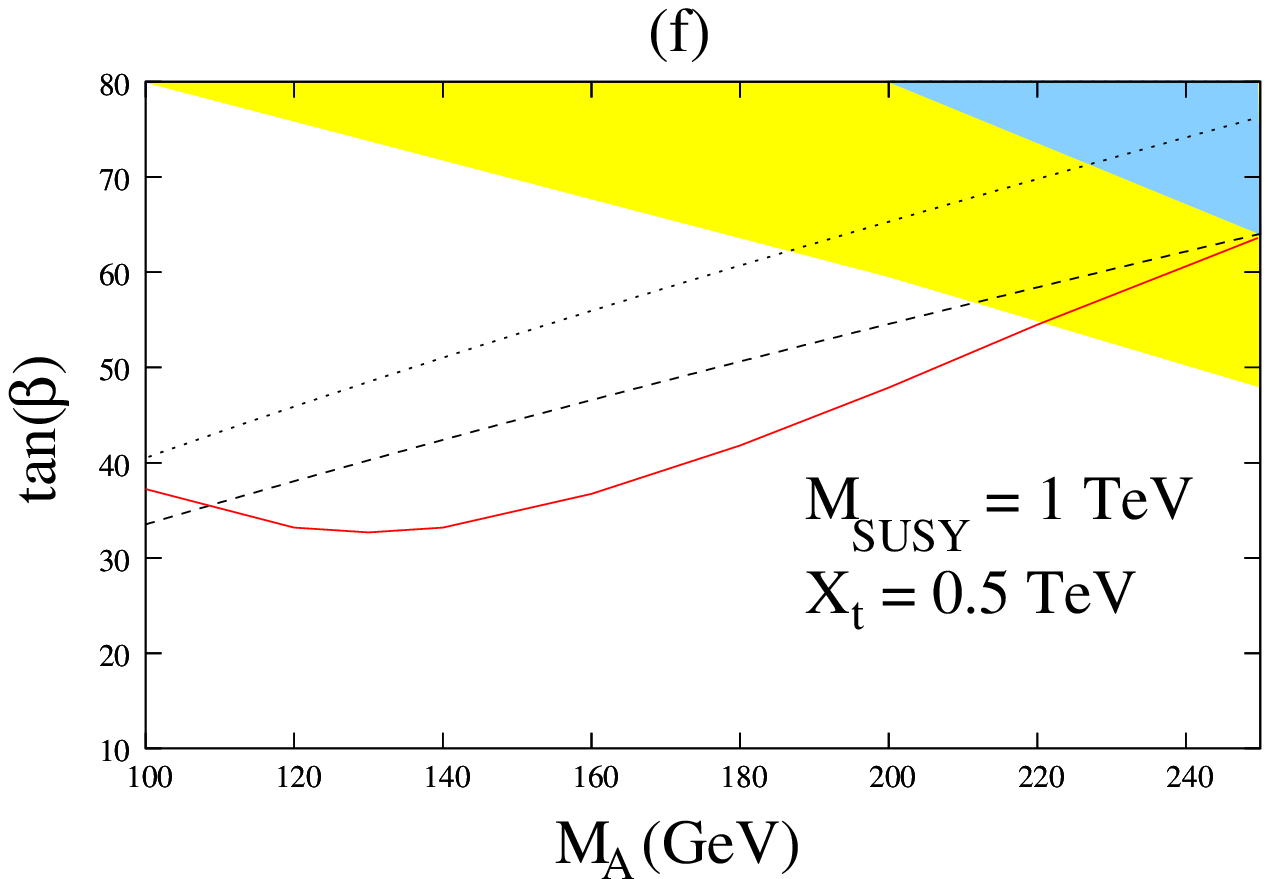}}
\end{center}
\caption{The dashed(dotted) line is the $\mathcal{BR}(B_s\rightarrow 
\mu^+ \mu^-)$ experimental bound in the $M_A-\tan \beta$ plane for 
$\mu = -200(-100)$ GeV and the yellow (light grey) and blue (dark grey) 
bands are the  $b\rightarrow s \gamma$ allowed regions for $\mu = -200$ GeV 
and $-100$ GeV, respectively, in the uniform squark limit with 
$M_{\rm SUSY}=1$~TeV, 
$|M_3|=0.8$ TeV, and $2M_1=M_2=110$ GeV. The red (grey) line is the 
projected CDF limit on $H\rightarrow \tau \tau$ for $1$fb$^{-1}$ luminosity.
Larger luminosities would probe larger $M_A$ and smaller $\tan \beta$.  
Also changing $\mu$ from $-200$~GeV to $-100$~GeV does not affect the CDF limit
significantly. Figures (a),(c) and (e) have different values of $X_t = A_t -\mu
/\tan \beta$ for $\arg(M_3) = 0$ while (b), (d) and (f) have a $\arg(M_3) = 
\pi$}
\label{matbbsxt:fig}
\end{figure}

In Fig.~\ref{matbbsxt:fig}, we display the
constraints in the $M_A$--$\tan\beta$ plane that are induced by
the requirement of obtaining a good agreement with the $BR(b \to s\gamma)$
and the non-observation of $B_s \to \mu^+\mu^-$ at the Tevatron collider.
The results are presented for different values of $X_t$ and
$\mu$ parameters and supersymmetry breaking squark masses equal to 1 TeV. 
The region of parameter space consistent with
$B_s \to \mu^+\mu^-$ for $\mu=-100$~GeV and $\mu=-200$~GeV is below the dotted 
and dashed lines respectively. For each
value of $A_t$, larger values of $|\mu|$ imply consistency with larger
values of $M_A$ and smaller values of $\tan\beta$. On the other hand,
the regions in the $M_A-\tan \beta$ plane that are consistent with the observed
values of 
\begin{eqnarray}
BR(b \to s \gamma)^{Exp} = 3.38^{+0.3}_{-0.28} \times 10^{-4} 
\label{bsgbound:eq}
\end{eqnarray}
and the estimated theoretical uncertainty~\cite{Neubert:2004dd}
\begin{eqnarray}
|BR(b \to s \gamma)^{Exp} - BR(b \to s \gamma)^{SM}| < 1.3 \times 10^{-4}
\end{eqnarray}
are given by the colored bands. For larger values of $|\mu|$ the bands move to 
smaller values of $M_A$ or smaller values of $\tan\beta$. Actually, the 
approximate cancellation of the charged Higgs and chargino stop contributions
implies a correlation between $1/M_A^2$ and $A_t \mu \tan\beta$. We have also 
plotted the projection of the CDF limit for non-standard MSSM Higgs boson 
inclusive searches in the $A,H \rightarrow \tau \tau$ channel for a total 
integrated luminosity of 1~fb$^{-1}$. In order to obtain this limit we have 
used the approximate relation given in Ref.~\cite{Carena:2005ek}
\begin{eqnarray}
\sigma(gg,b\bar{b} \rightarrow A) \times \mathcal{BR}(A\rightarrow \tau^+ \tau^-)
\sim \sigma(gg,b \bar{b} \rightarrow A)_{SM} \frac{\tan^2 \beta}{(1+\epsilon_3 
\tan \beta)^2 + 9},
\end{eqnarray}
along with the Tevatron's reach for scenario of maximal mixing with 
$\mu \sim -200$ GeV and a luminosity of $1$ fb$^{-1}$ shown in 
Ref.~\cite{htautau}. 

The Tevatron collider is only sensitive to values of $M_A$ smaller than
about 300 GeV and values of $\tan\beta$ larger than about 40. For maximal
mixing, Fig.~\ref{matbbsxt:fig}(a) shows that the constraints coming from 
flavor physics are sufficiently strong so as to restrict the  
parameter space consistent with the search for non-standard Higgs bosons 
at the Tevatron collider. On the other hand, for values of $A_t \simeq 1$~TeV, 
Fig.~\ref{matbbsxt:fig}(c) shows that one can obtain borderline consistency with the 
constraints coming from the flavor sector, but only for the smaller values of 
$\mu$ and $M_A\simeq 200$~GeV. Finally, for values of 
$A_t = 500$~GeV or smaller, Fig.~\ref{matbbsxt:fig}(e) shows that the bounds 
coming from $BR(b \rightarrow s \gamma)$ are sufficiently strong as to strongly
restrict the parameter space consistent with non-standard Higgs boson searches 
at the Tevatron collider.
 
The situation is ameliorated for positive values of $\mu M_3$, keeping 
negative values of $\mu A_t$. In Figs.~\ref{matbbsxt:fig}(b),(d) and (f)
we have changed the sign of the gluino mass (the same results
would be obtained by keeping the gluino mass fixed but changing the
sign of $\mu$ and $A_t$). Positive values of $\mu M_3$ diminish
the $\epsilon_0$ contributions and hence make the bound coming from
$\mathcal{BR}(B_s \to \mu^+\mu^-)$ slightly less severe. The bound coming from 
$\mathcal{BR}(b \to s \gamma)$ is also improved, with the colored bands 
being slightly lower. Thus for $X_t \simlt 1$~TeV the region of $M_A \sim 
200$~GeV, small $\mu$ and $\tan \beta \sim 50$, that is not excluded by flavor 
physics, will be probed by the Tevatron Higgs searches in the near future.

Finally, we consider the minimal mixing scenario, $X_t \simeq 0$.  In this
case, the constraints coming from the non-observation of $B_s \to \mu^+\mu^-$
become very weak, even for large values of $|\mu|$. As we will explain
below, this opens up an interesting possibility:
The dominant charged Higgs contribution to the 
$b\rightarrow s \gamma$ amplitude at large $\tan \beta$ is
proportional to the charged Higgs coupling to top and bottom quarks given
in Eq.~(\ref{H+RLj3:eq}). Setting, for simplicity, $A_b=0$ makes the 
$\epsilon_Y' \approx 0$ 
while
\begin{eqnarray}
\epsilon_0^{3'} \approx \frac{2 \alpha_s}{3\pi} \mu M_3
(\cos^2 \theta_{\tilde{t}} C_0(m_{\tilde{s}_L}^2,m_{\tilde{t}_1}^2,M_3^2) 
+ \sin^2 \theta_{\tilde{t}} C_0(m_{\tilde{s}_L}^2,m_{\tilde{t}_2}^2,M_3^2)).
\end{eqnarray}
Therefore, in this case, the charged Higgs contribution to the 
$\mathcal{BR}(b\rightarrow s 
\gamma)$ becomes proportional to~\cite{Degrassi:2000qf,Carena:2000uj}
\begin{eqnarray}
A_{H+} \propto \frac{1 - \frac{2 \alpha_s}{3\pi} \mu M_3 \tan \beta
\left(\cos^2 \theta_{\tilde{t}} C_0(m_{\tilde{s}_L}^2,m_{\tilde{t}_1}^2,M_3^2) 
+ \sin^2 \theta_{\tilde{t}} C_0(m_{\tilde{s}_L}^2,m_{\tilde{t}_2}^2,M_3^2) 
\right)}{1+\epsilon_3 \tan \beta}, \label{bsgh+:eq}
\end{eqnarray}
where $\theta_{\tilde{t}}$ is the stop mixing angle. From Eq.~(\ref{bsgh+:eq}) 
we can clearly see that for large positive values of $M_3 \mu$ and $\tan\beta$,
the charged Higgs amplitude can be strongly reduced. Furthermore when $X_t 
\simeq 0$ the chargino stop contribution to $b\rightarrow s \gamma$ is also 
small. Since, for these parameters, the BSM contributions to 
the $\mathcal{BR}(
b\rightarrow s \gamma)$ are small, the experimental limit in 
Eq.~(\ref{bsgbound:eq}) puts only a weak constraint on the allowed value of 
$M_A$. Moreover, as stressed above,
for this parameter region $B_s \rightarrow \mu^+ \mu^-$ also provides no
constraint because $X_t \sim 0$ implies small values of $\epsilon_Y$.

\begin{figure}[t]
\begin{center}
\resizebox{7.cm}{!}{\includegraphics{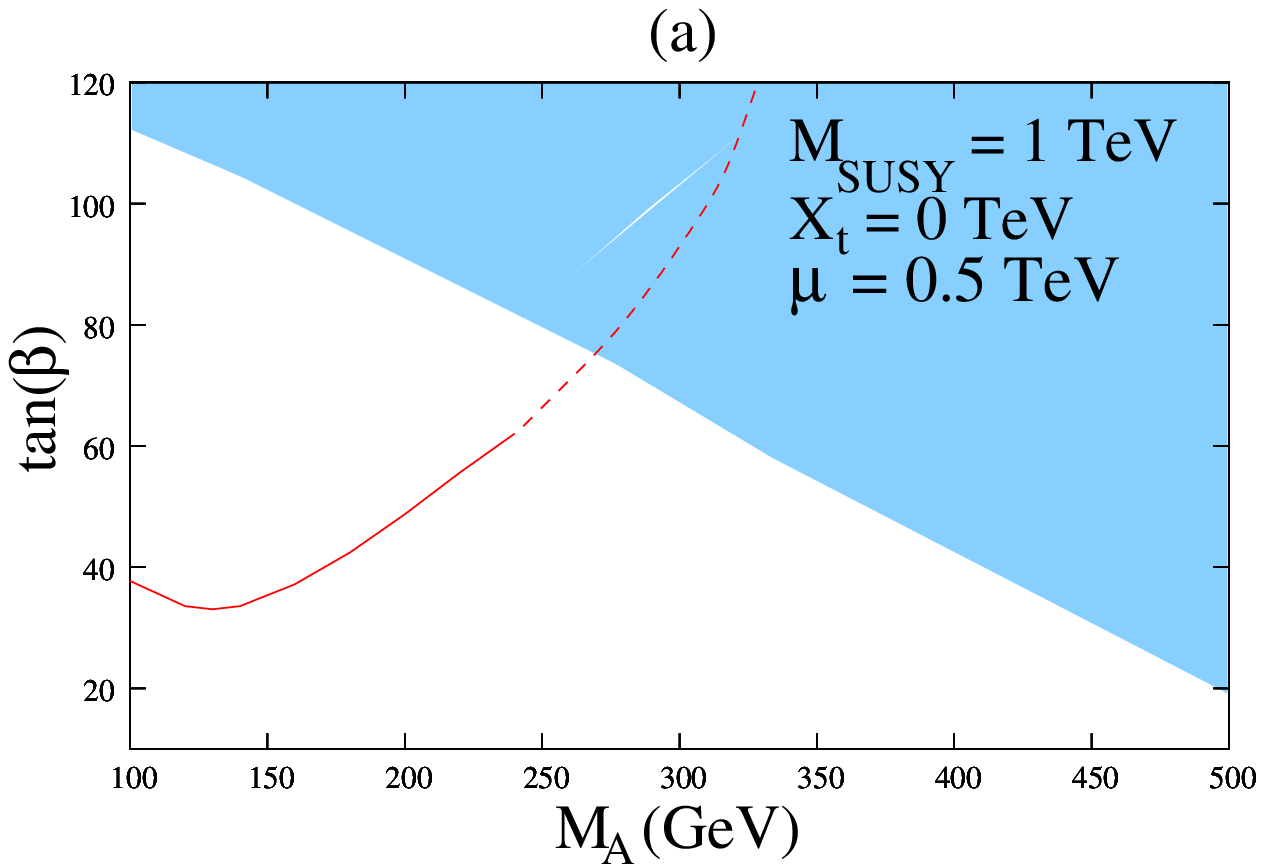}}
\resizebox{7.cm}{!}{\includegraphics{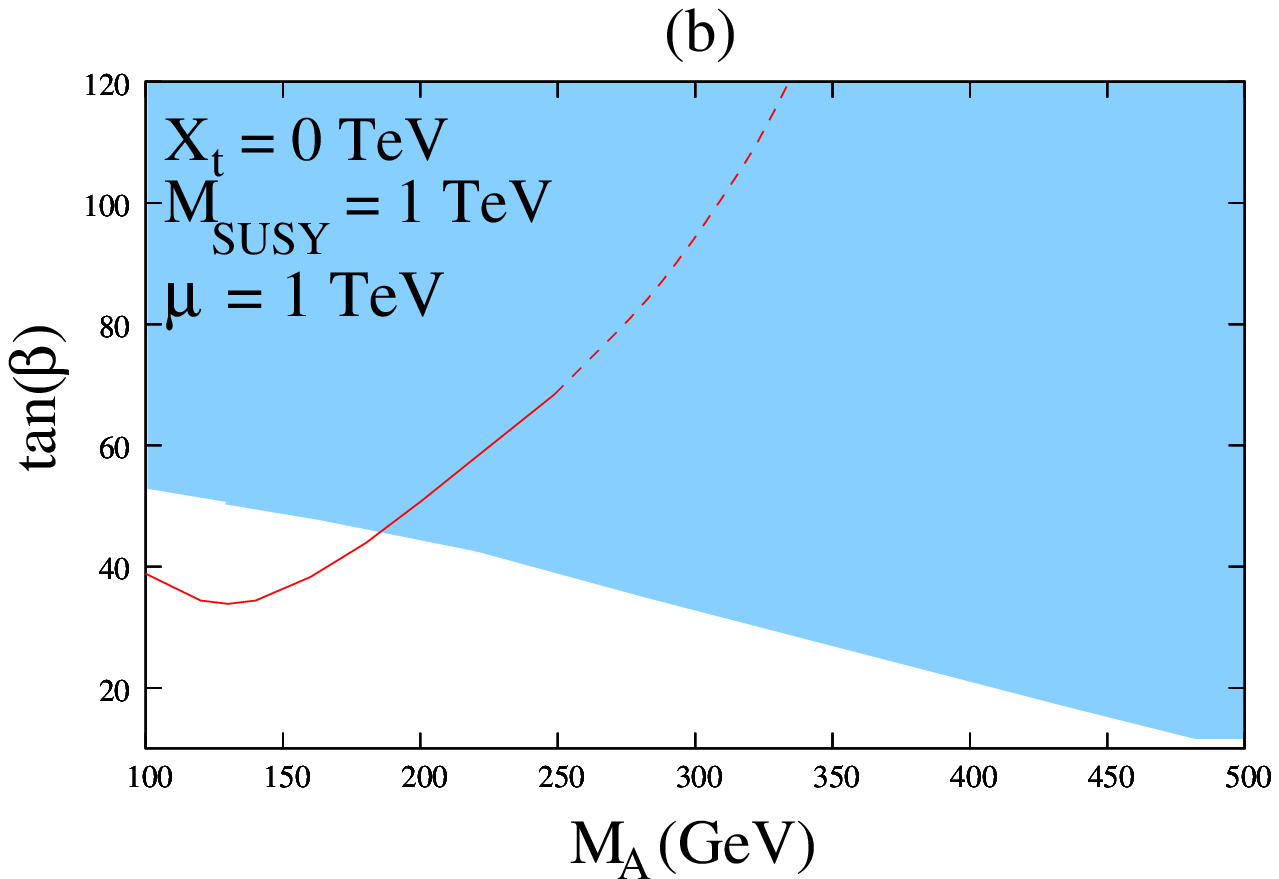}}
\resizebox{7.cm}{!}{\includegraphics{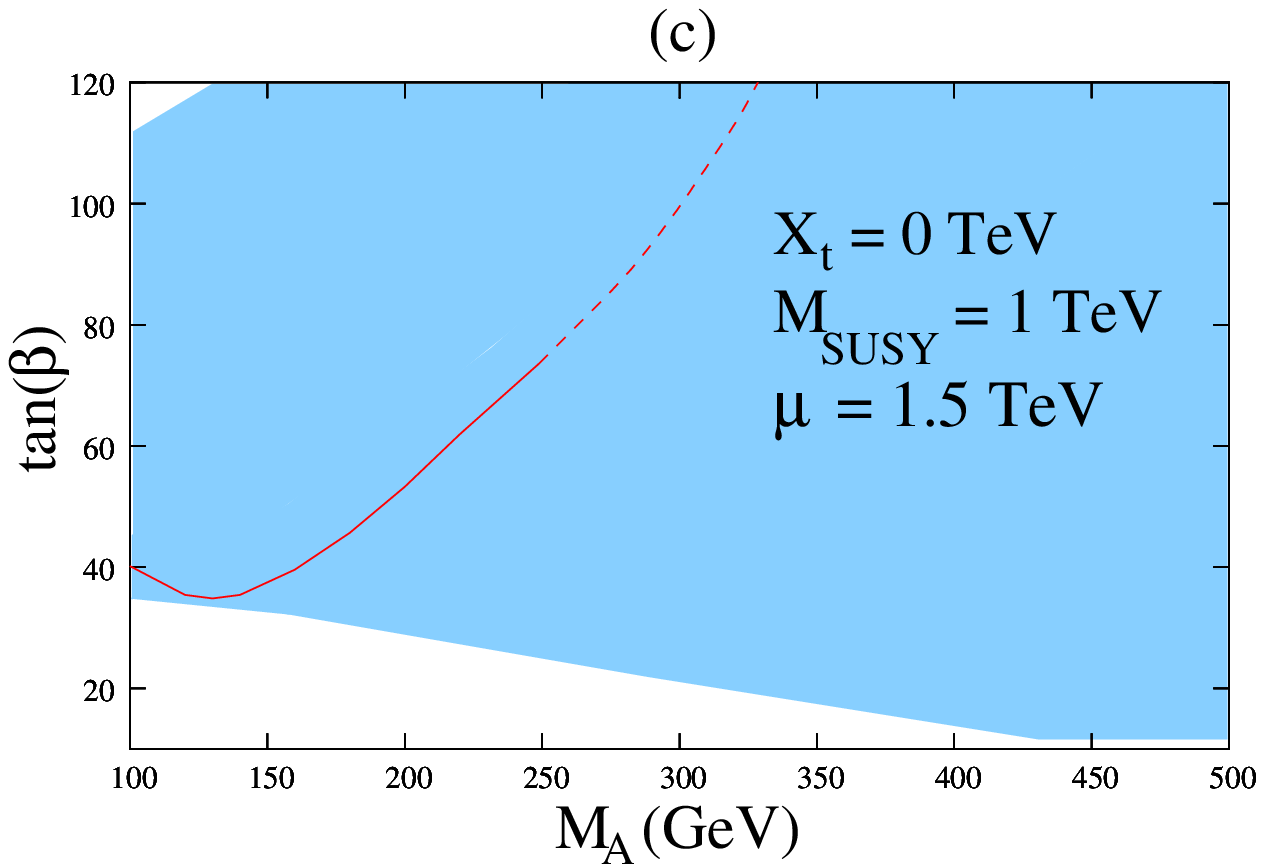}}
\resizebox{7.cm}{!}{\includegraphics{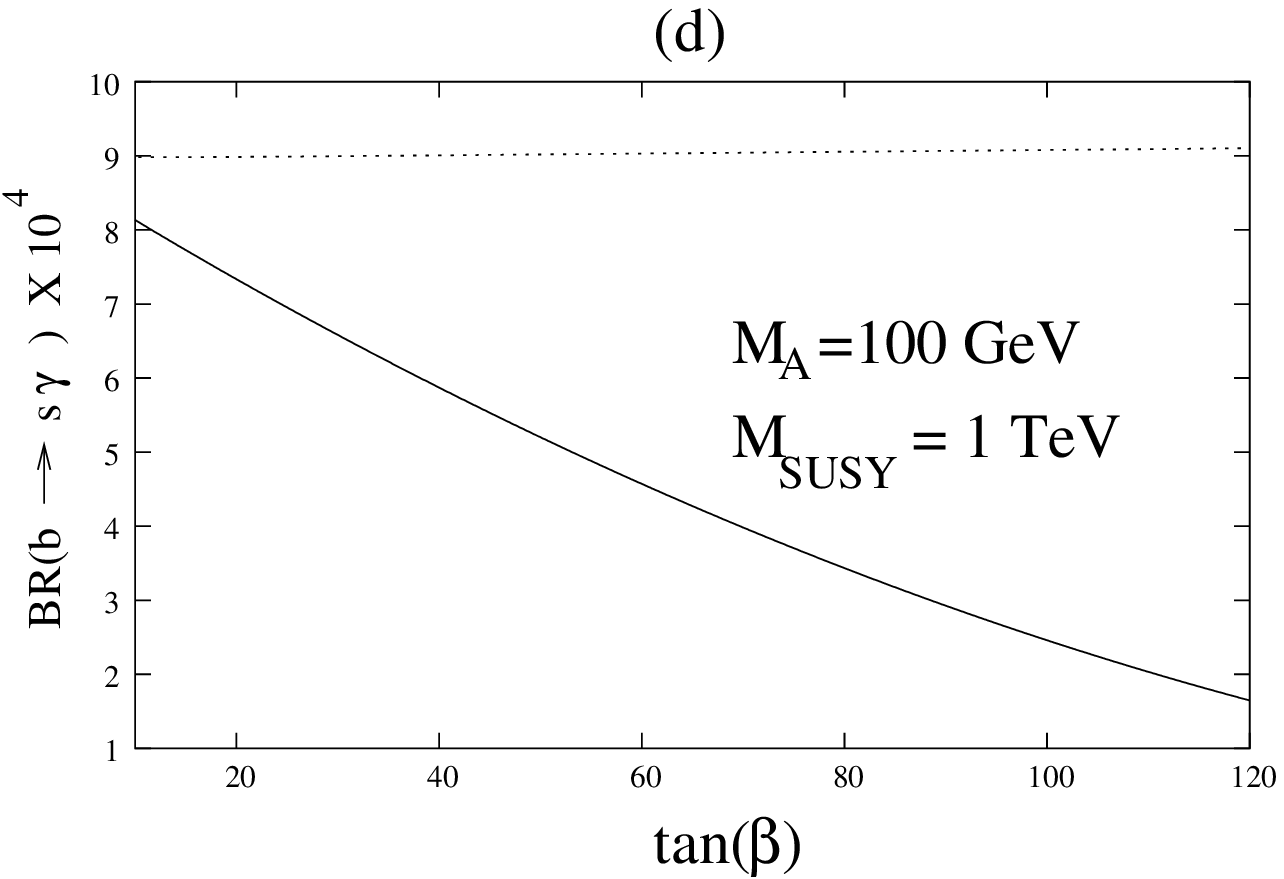}}
\end{center}
\caption{(a)--(c) Corresponds to $\mu = 500 -- 1500$ GeV 
with the blue (dark grey)
band showing $b\rightarrow s\gamma$ allowed regions for these 
values of $\mu$ in the uniform squark mass limit with 
a common value of the squark masses $M_{\rm SUSY}=1$ TeV, 
$M_3=0.8$ TeV, $2M_1=M_2=110$ GeV. The red (grey) line is the projected CDF limit on $H
\rightarrow \tau \tau$ for $1$fb$^{-1}$ luminosity. The dashed part of the
projected Tevatron reach is an extrapolation of the curve. 
(d) Shows the effect of 
including the squark loop correction to $P_{RL}^{H+}$ vertex, 
proportional to $\epsilon_0^{3'}$,
on $b\rightarrow s\gamma$ rate for $\mu = 1$~TeV. 
The dashed line corresponds to the case when corrections are 
not included while the solid line 
corresponds to the case when they are included.}
\label{bsmuext:fig}
\end{figure}


Additionally, for the values of the parameters for which a cancellation
of the charged Higgs contribution to $BR(b \to s \gamma)$ occurs, 
the usual bound on $\tan \beta$ that comes 
from requiring that $y_b$ be perturbative up to the GUT scale may be
relaxed: The bottom 
Yukawa has the form 
\begin{eqnarray}
y_b \simeq \frac{\sqrt{2} m_b \tan\beta}{v (1+\epsilon_3 \tan \beta)}
\end{eqnarray}
and as $\epsilon_3 \tan\beta$ is real and positive, and of order one 
for the cancellation to occur,
the denominator suppresses
the Yukawa for large values of $\tan \beta$. This leads to an 
enhancement of the upper bound on $\tan\beta$ coming from perturbative 
consistency in the bottom quark sector.

In Fig.~\ref{bsmuext:fig} we 
illustrate such a scenario for different values of $|\mu|$. 
Because both the $B_s \rightarrow \mu^+ \mu^-$ and $b\rightarrow s \gamma$ 
constraints allow essentially any value of $M_A \gsim 100$~GeV a large 
region of the $M_A-\tan \beta$ can be probed by the heavy MSSM Higgs searches 
at the Tevatron. Interestingly enough, the lightest Higgs boson mass is also 
close to the experimental bound $m_h \simeq 115$~GeV in this region of 
parameters, and therefore it could be at the reach of the Tevatron collider
searches. 

In conclusion, for minimal flavor violating schemes, the discovery of
a non-standard Higgs signature at the Tevatron collider would point
to a definite region of parameter space, with values of $X_t$ of
order of the squark masses or smaller. Larger values are strongly
restricted by the present Tevatron, CLEO and B-factory experimental 
constraints. It is important to remark that, as the luminosity of the
Tevatron increases, the probability of measuring  $B_s \to \mu^+\mu^-$
increases, and so does the one of measuring a non-standard Higgs
boson signal. However, as it becomes clear from the above discussion,
an improvement of the bound on $B_s \to \mu^+\mu^-$ would put 
strong restrictions on the possibility of measuring a non-standard
Higgs boson signature for moderate or large values of $X_t$. Conversely, 
if a Higgs boson signature were observed, with absence of observation of
$B_s \to \mu^+\mu^-$, 
it would imply either small
values of $X_t$, or a strong departure from minimal flavor violating 
scenarios. 

It is interesting to analyze the constraints 
that the non-observation of $B_s \to \mu^+
\mu^-$ at the LHC, for a total integrated luminosity of order of 10~fb$^{-1}$,
would put on the MSSM parameter space. The projected Atlas bound on 
$BR(B_s \to \mu^+\mu^-)$ in this case would be of 
order $5.5 \; 10^{-9}$~\cite{LHCBstomumu},
and therefore would imply strong constraints on the $M_A$--$\tan\beta$
parameter space (The final Tevatron bound, in case of non-observation
of $B_s \to \mu^+\mu^-$, assuming a total integrated luminosity of
order 8~fb$^{-1}$,
will be close to $2 \; 10^{-8}$~\cite{TeVBstomumu} 
and therefore it will set similarly strong
bounds on the parameter space).
In order to study the possible implications for searches
of non-standard Higgs bosons at the LHC, we have considered the projected
reach of the CMS searches in the inclusive 
$pp \to \Phi + X$, $\Phi \to \tau^+\tau^-$
mode, at a luminosity of 30~fb~$^{-1}$~\cite{LHCtautau}. 

\begin{figure}[t]
\begin{center}
\resizebox{7.cm}{!}{\includegraphics{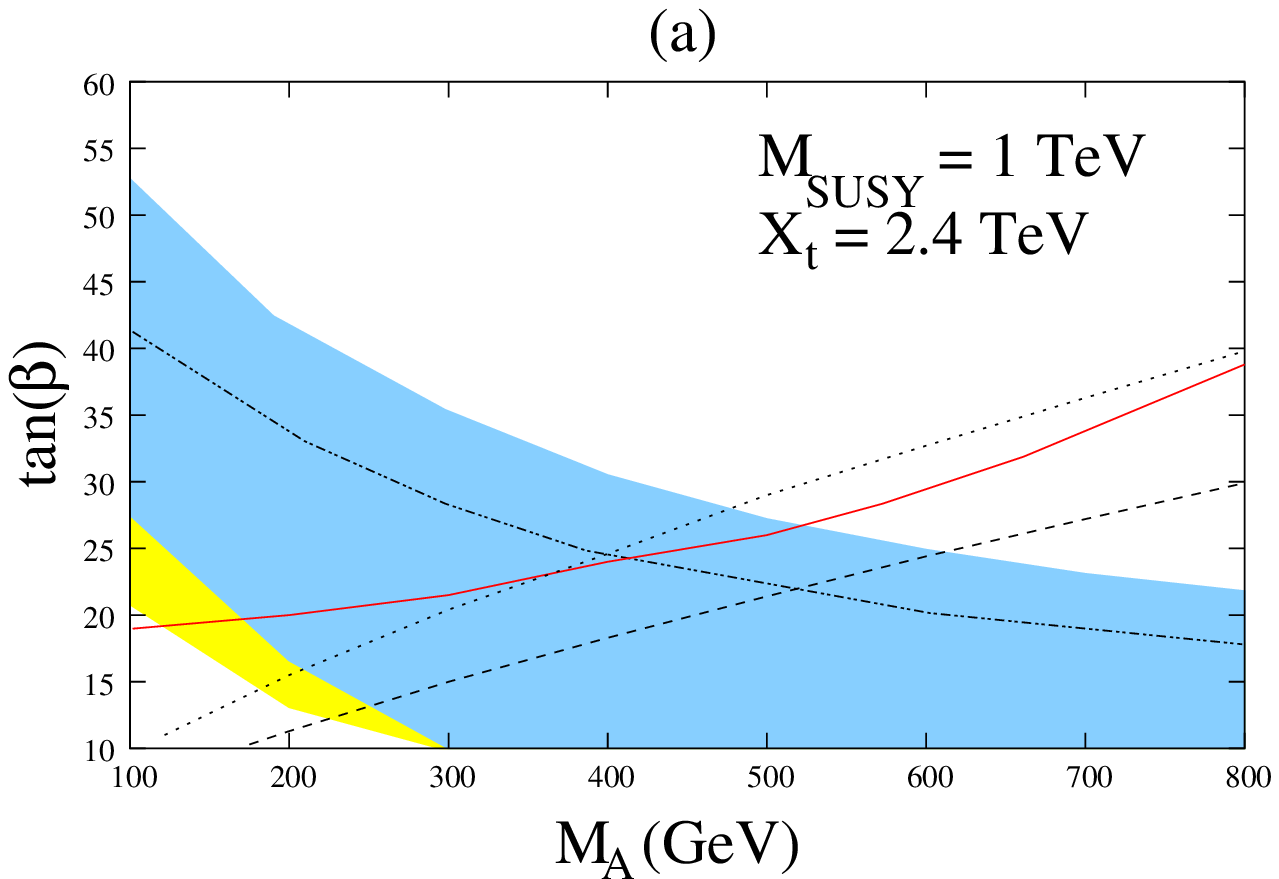}}
\resizebox{7.cm}{!}{\includegraphics{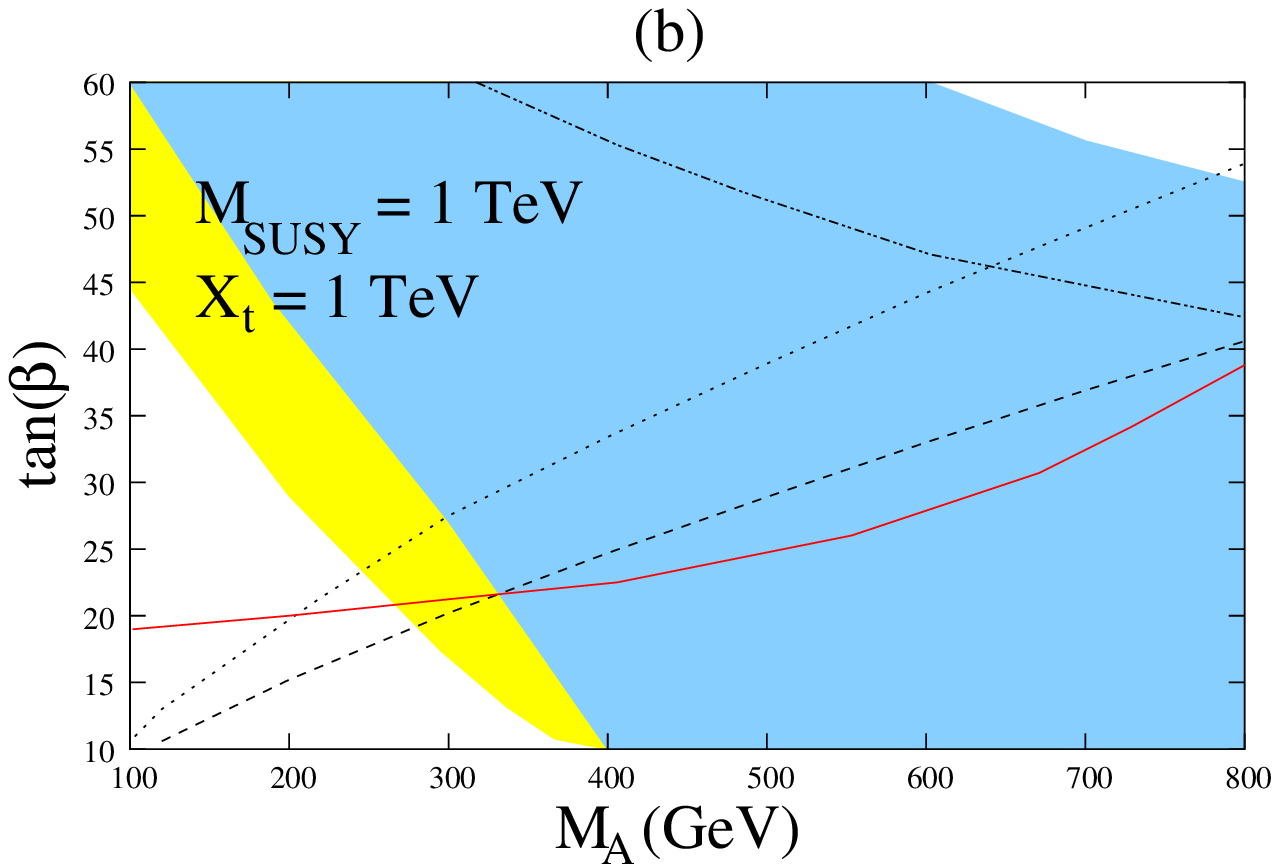}}
\end{center}
\caption{Comparison of the projected reach for non-standard Higgs bosons at the
LHC in the inclusive $pp \to \Phi + X$, $\Phi \to \tau^+\tau^-$ 
mode (red (grey) line) with
the limits that would be obtained in case of non-observation of the
decay mode $B_s \to \mu^+\mu^-$ for an integrated luminosity of 10~fb$^{-1}$
for $\mu = -100$~GeV (dotted line) and $\mu = -300$~GeV (dashed line).
Blue (dark grey) and yellow (light grey) areas correspond to the bounds
coming from $BR(b \to s \gamma)$ for $\mu = -100$~GeV and $\mu = -300$~GeV,
respectively. The upper edge of the $\mu = -300$~GeV area is denoted by 
the dot-dashed line.
We show these results for a common value of the squark
masses $M_{\rm SUSY} = 1$~TeV and (a) $X_t = 2.4$~TeV, (b) $X_t = 1$~TeV,
and positive (negative) values of $\mu M_3$ ($\mu A_t$), and 
$|M_3| \simeq 0.8$~TeV.} 
\label{LHCxt:fig}
\end{figure}


From Fig.~\ref{LHCxt:fig} we can see that 
even for the most restrictive case of maximal
mixing and negative values of $\mu M_3$, the bound
coming from the non-observation of $B_s \to \mu^+\mu^-$ would be consistent
with the observation of a non-standard Higgs boson for small values
of $|\mu| \simeq 100$~GeV and somewhat large values of 
$350 \simlt M_A \simlt 500$~GeV. These bounds are strongly relaxed
for smaller values of $X_t$. For instance, 
for $X_t \simlt 1$~TeV, 
observation of non-standard Higgs bosons would be still allowed for 
any value of $M_A$, provided $|\mu| \simlt 300$~GeV.

\section{Non-minimal Flavor Violation}
\subsection{Gluino Contributions to $\Delta M_s$}

The results in the case of non-minimal flavor violation discussed in 
section~\ref{nmvf:sec} are quite similar to the case of minimal flavor
violation.  
As in the case of MFV for large tan$\beta$, the dominant contribution to 
$\Delta M_s$ comes from the DP diagrams. However, in the non-minimal flavor 
violation scenario introduced here, the effects of gluino boxes can also be 
important and compete with the double penguin contributions. 
The appearence of the gluino-box contributions
is a direct consequence of the 
quark-squark-gluino vertices not being diagonal in the flavor basis. In the 
case of uniform squarks masses these contributions disappear due to the 
CKM matrix being unitary.

The double penguin contributions to $\mathcal{BR}(B_s\rightarrow \mu^+ \mu^-)$ 
in the non-minimal flavor scenario may be significantly larger than in the case
of MVF. For instance, assuming that the third generation left-handed and 
right-handed down squark masses are light implies that the vertices in 
Eq.~(\ref{xrlnmfv:eq}) are proportional to
\begin{eqnarray}
X_{RL}^{JI} \propto V_{eff}^{3J*} V_{eff}^{3I} \left(\left(1-\frac{1}{\rho^2}
\right) \epsilon_0^3 + \epsilon_Y\right)
\end{eqnarray}  
where $\rho = m_{\tilde{q}_{1,2}}/m_{\tilde{q}_3}$. Therefore when the squark 
mass splitting is large these vertices can give large contributions to 
$\Delta M_s$ and $\mathcal{BR}(B_s\rightarrow \mu^+ \mu^-)$. However, the 
linear correlation between $\Delta M_s$ and $\mathcal{BR}(B_s\rightarrow \mu^+ 
\mu^-)$ is not spoiled by the splitting of the squark masses as there is no
flavor dependence in the factor multiplying $m_{d_J} V_{eff}^{3J*} V_{eff}^{3I}
$ in Eq.(\ref{xrlnmfv:eq}). Therefore the $\mathcal{BR}(B_s\rightarrow \mu^+ 
\mu^-)$ bound is still a severe constraint on large double penguin 
contributions to $\Delta M_s$ like in the MFV scenario.

\begin{figure}
\begin{center}
\resizebox{8cm}{!}{\includegraphics{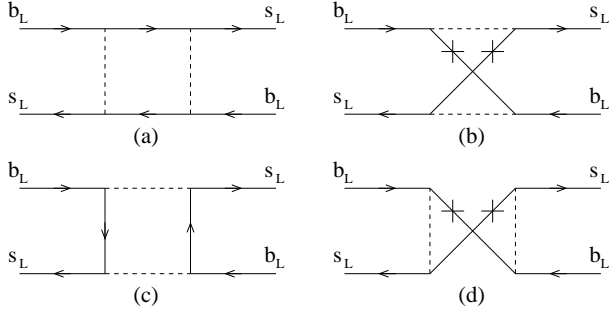}}
\end{center}
\caption{Gluino box diagrams that make contributions to $\Delta M_s$ for the 
Nonminimal flavor violation. Diagrams (b) and (d) are possible because of the
gluinos are Majorana and the lower diagrams have a relative sign difference 
with respect to the upper ones~\cite{Bertolini:1987cw}}.
\label{gluino_box:fig}
\end{figure}

An interesting case is one in which the gluino box diagrams dominate 
over the double penguin contributions to $\Delta M_s$ for moderate values of 
$\rho \sim 2 \mbox{ or }3$. Similar to the light-stop scenario for MFV there 
are situations in which the gluino box diagram contributions are sizeable and 
the other contributions are suppressed. The double penguin contributions are 
suppressed for  low values of $\tan \beta$. On the other hand, large values of 
$\mu$ and $M_2$ suppress the stop-chargino box diagrams. Since the gluino box 
diagram effects are larger for small values of the left-handed squark and 
gluino masses, we shall investigate the case in which the third generation 
left-squark soft supersymmetry breaking parameters are about 100~GeV. To avoid 
the Tevatron bound on sbottoms we also assume that the lightest neutralino is 
within 20 GeV of the sbottom mass~\cite{Demina:1999ty}. We can achieve this 
mass difference by choosing an appropriate value of $M_1$. For larger values of
the soft SUSY breaking sbottom mass parameter, of about $\sim 200$ GeV the 
gluino box contribution becomes negligble. 

\begin{figure}
\begin{center}
\resizebox{8cm}{!}{\includegraphics{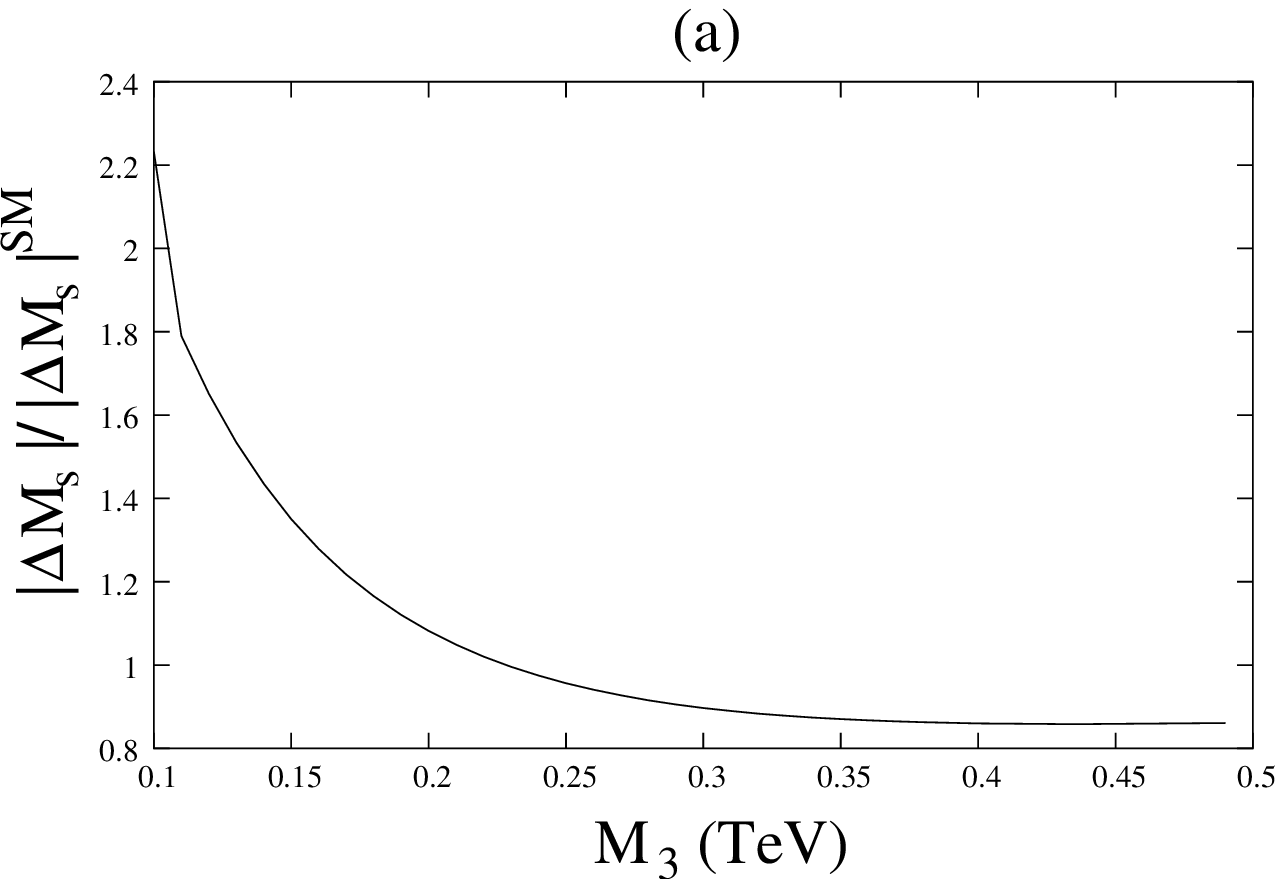}}
\resizebox{8cm}{!}{\includegraphics{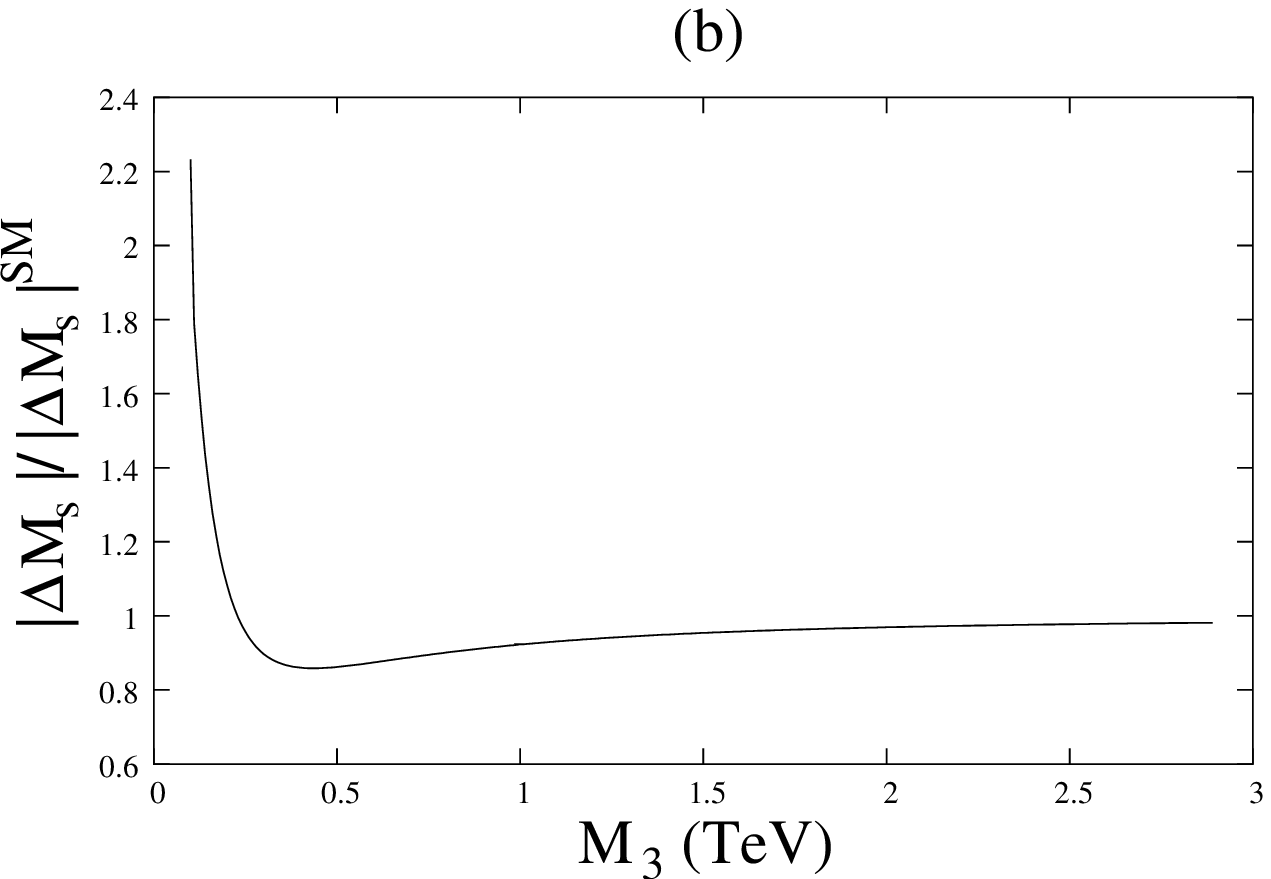}}
\end{center}
\caption{Variation of SUSY contributions to $\Delta M_s$ with input parameters 
$M_A=250$ GeV, $M_{SUSY}=1000$ GeV, $M_1=110$ GeV $M_2=1000$ GeV, $\mu=1100$ 
GeV, $M_{(\tilde{U}_L,\tilde{D}_L)_{12}} = M_{\tilde{U}_R,\tilde{D}_R} = 1000$ 
GeV, $M_{(\tilde{U}_L,\tilde{D}_L)_3}=100$  GeV, $A_t=1110$  GeV, 
$\tan\beta=10$ and all relevant SUSY phases are zero.(a) Shows the variation of
$\Delta M_s$ over small values of gluino mass, while (b) shows that in limit of
large gluino mass we recover the SM value.}
\label{dmsvsmg:fig}
\end{figure}

Light left-handed squarks tend to lead to large values of the $T$-parameter
and hence are constrained by precision electroweak data. These large
contributions to the $T$-parameter are induced by the large difference between
the left-handed sbottom and stop masses and are proportional to the top quark 
mass. However, for some range of values of the right-handed stop mass 
parameter, these large contributions
may be minimized. Indeed, for large values of the right handed stop mass
parameter $M_{\tilde{U}_R}$ and $X_t \simeq M_{\tilde{U}_R}$, 
the lightest stop mass 
becomes mainly left-handed and its mass is given by
\begin{equation}
m_{\tilde{t}_1}^2 \simeq M_{\tilde{U}_L}^2 + m_t^2 \left( 1 - \frac{X_t^2}{M_{\tilde{U}_R}^2}
\right) + D_L^t
\end{equation}
where $D_L^t$ is the small D-term contribution to the left-handed stop mass. 
Observe that for $X_t \simeq M_{\tilde{U}_R}$, the top-quark mass contribution 
is strongly suppressed and hence the contribution to the $T$-parameter becomes 
small~\cite{Carena:1997xb}. In our analysis we have chosen the stop mass 
parameters so that the  relation $X_t = M_{\tilde{U}_R}$ is fulfilled.

\begin{figure}
\begin{center}
\resizebox{8cm}{!}{\includegraphics{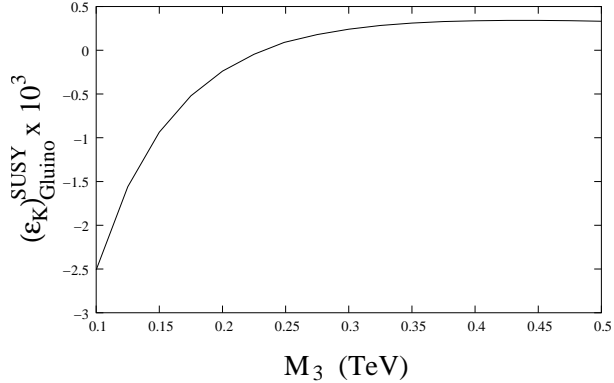}}
\end{center}
\caption{Variation of the gluino contributions to $\epsilon_K$ with the gluino 
mass $M_3$ for the same input parameters as in Fig.~\ref{dmsvsmg:fig}}
\label{eKvsmg:fig}
\end{figure}

In Fig.\ref{dmsvsmg:fig} we see that for 
gluino masses below $200$~GeV, the gluino-sbottom box contribution yields a
value of $\Delta M_s$ that 
is greater than the $1\sigma$ bound coming from the SM. Similarly, in 
Fig.~\ref{eKvsmg:fig} we that there are large negative contributions to 
$\epsilon_K$ from the gluino box-diagrams for $M_3 \lsim 200$~GeV.
The total value $\Delta M_s$ drops below that of the SM, 
for $M_3 \gsim 200$~GeV, because of the interference between the diagrams in 
Fig.~\ref{gluino_box:fig}.  For the region $M_3 \lsim 200$~GeV, where 
$\Delta M_s$ is large, the contributions to $\epsilon_K$ are also larger but 
negative, which seems to predict a total value of $\epsilon_K$ much smaller 
than the experimentally observed one. Therefore, the gluino box 
contributions to $\Delta M_s$, in this non-minimal flavor violating scenario 
with flavor changing effects induced by the CKM matrix elements, are generally 
small and are at most as large as those in the light stop scenario discussed 
above .  In addition this scenario is in general highly contrived as the 
experimental constraints from light gluino and sbottom 
searches~\cite{Demina:1999ty} can be avoided only by going to a small corner of
the MSSM parameter space. 

\section{Conclusions}

In this article, we have studied the constraints on the parameter space
of minimal flavor violating SUSY models coming from the latest constraints
on $B_s \to \mu^+\mu^-$, $\Delta M_s$, $\epsilon_K$ and $BR(b\to s \gamma)$. 
Firstly, we have shown that the analysis of the double penguin contributions 
to observables in the Kaon sector could not be done with the available
formulae in the literature. We derived a new formula that 
describes well the Kaon sector contributions and show that the present
constraints on $B_s \to \mu^+\mu^-$ eliminate the possibility of inducing
relevant double penguin corrections in this sector. Alternative contributions,
coming from chargino and stop loop corrections can produce large contributions
to $\epsilon_K$, which, considering the present theoretical uncertainties, are
consistent with the bounds coming from other flavor observables.

We have also verified that the double penguin contributions to $\Delta M_s$ 
interfere destructively with the SM contribution and are
strongly constrained by the non-observation of $B_s \to \mu^+\mu^-$
at the Tevatron collider. Analyzing the dependence of $\Delta M_s$ on
the supersymmetric loop corrections, we obtained upper bounds on this
quantity for any given value of $B_s \to \mu^+\mu^-$, for natural
values of the supersymmetric mass parameters. We have also shown that for
$M_A <$~1~TeV,
under the current theoretical and experimental uncertainties, this bound
is stronger than the bound on the new physics contributions that is 
obtained from the comparison of the SM predictions and the experimentally
measured values. Finally, if the theoretical errors on $\Delta M_s$ were 
reduced and the SM central value was to remain the same then negative
corrections to $\Delta M_s$, like that of the double penguin contribution, 
would be necessary. However such double penguin corrections to $\Delta M_s$ of 
about a few  ps$^{-1}$'s can be obtained only if $\mathcal{BR}(B_s \to \mu^+ 
\mu^-) \gsim 3 \times 10^{-8}$ for $M_A \leq 1$~TeV, which is within the 
future sensitivity of the Tevatron collider.
 
On the other hand, relevant, positive constributions to 
$\Delta M_s$ may be obtained for light stops and charginos. The contributions 
may be as large as 25 percent of the SM values, almost independently
of the value of $\tan\beta$. Contrary to the double penguin contribution,
the  chargino-stop contributions are positive and they are more
strongly constrained than the negative double penguin ones. Small
values of the Higgsino mass, $\mu < 200$~GeV tend to be disfavored 
for mass parameters consistent with the scenario of electroweak baryogenesis. 
We have also analysed a scenario in which there are flavor violating effects
proportional to CKM matrix elements in the left-handed down squark-gluino
vertices at tree-level. 
Although the box-diagrams may lead to significant contributions
to $\Delta M_s$ for sufficiently small gluino and down squark masses, this
contributions are constrained to be small once the bounds on $\epsilon_K$ 
are taken into account.

We have also analyzed the complementarity of these FCNC constraints
with direct Tevatron searches for heavy MSSM Higgs bosons. We have analyzed
different scenarios and showed
that $\mathcal{BR}(b\rightarrow s\gamma)$ and $\mathcal{BR}(
B_s\rightarrow \mu^+ \mu^-)$ puts strong constraints on the 
$M_A-\tan \beta$ plane.
This study suggests that within minimal flavor violating scenarios, 
the observation of non-standard MSSM Higgs 
bosons at the Tevatron collider would imply either moderate values of 
$|X_t/M_{SUSY}| \simlt 1$ and small values of $|\mu|$, or very
small values of $X_t$ and large values of $|\mu|$.
Interestingly enough, for 
values $X_t \simlt M_{SUSY}$, the lightest CP-even Higgs boson mass is
smaller than 120~GeV and therefore possibly at the reach of Tevatron high
luminosity searches. 

Finally, we have analysed the implications of non-observation of $B_s \to \mu^+
\mu^-$ at the LHC, for a total integrated luminosity of order of 10~fb$^{-1}$, 
on searches for non-standard MSSM Higgs bosons at this collider. Even for the 
most restrictive case of maximal mixing and negative values of $\mu M_3$, this 
situation would be consistent with the observation of a non-standard Higgs boson 
for small values of $|\mu| \simeq 100$~GeV and somewhat large values of 
$350 \simlt M_A \simlt 500$~GeV. For $X_t \simlt 1$~TeV, instead,
observation would be still allowed for any value of $M_A$, provided
$|\mu| \simlt 300$~GeV. 

\vspace{1.5cm}
~\\
\large{\textbf{Acknowledgements:}} \normalsize
We wish to thank Csaba Balasz, Jon Rosner, Ishai Ben-Dov, Thomas Becher, 
Avto Kharchilava, Steve Mrenna, Jae Sik Lee and Ulrich Nierste for helpful 
discussions. R.N.-P and A.S. thank the Fermilab Theory Group for warm 
hospitality and support. Work at ANL is supported in part by the US DOE, 
Div.\ of HEP, Contract W-31-109-ENG-38. Fermilab is operated by Universities 
Research Association Inc. under contract no. DE-AC02-76CH02000 with the DOE. 
This work was also supported in part by the U.S. Department of Energy through 
Grant No. DE-FG02-90ER40560.

\appendix 
\section{Appendix}

\subsection{A corrected perturbative approach for calculating FCNC} 
\label{A1:sec}

We would like to develop a perturbative approach to calculating flavor
changing vertices which in the limit of uniform $\hat{\epsilon}_0$ should 
reproduce the exact result in Eq.\ (\ref{uniformXRL:eq}).

\subsubsection{Basic setup and notation}

As a starting point, we assume the form of the mass matrix
\begin{eqnarray}
(\mathbf{M_d})^{JI} &=& m_{d_J} \left((1+\epsilon_J \tan \beta) 
\delta^{JI} + \epsilon_Y y_t^2 \tan \beta 
\lambda_0^{JI} \right). \label{mass:eq}
\end{eqnarray}
As the off-diagonal elements are suppressed by CKM factors with respect 
to the diagonal elements we expand in terms of the CKM factors. Therefore first 
order terms are proportional to $V_0^{3J}$ for $J\neq3$ and second terms are 
proportional to $V_0^{*32}V_0^{31}$. Strictly speaking we should probably expand
in the Wolfenstein parameter $\lambda$ and not in the CKM elements, however as 
all we want is the leading behaviour, it is sufficient to expand in terms of the
CKM elements. So $\mathbf{M_d}$ has both first and second order terms present 
and can be expanded to be
\begin{eqnarray}
\mathbf{M_d} = (\mathbf{M_d})_0 + \delta \mathbf{M_d} + \delta^2 \mathbf{M_d}.
\end{eqnarray}
where $\delta$ symbolizes terms linear in $V_0^{3J}$ for $J\neq3$ and $\delta^2$
symbolizes terms proportional to $V_0^{*3J}V_0^{3I}$ for $J,I \neq 3$, so that 
\begin{eqnarray}
(\mathbf{M_d})_0^{JI} &=& m_{d_J} (1+\epsilon_J \tan \beta) \\
(\delta \mathbf{M_d})^{JI} &=& \left\{ \begin{array}{cl}
    m_{d_J} \epsilon_Y y_t^2 \tan \beta V_0^{3J*} & J \neq 3 = I \\
    m_b \epsilon_Y y_t^2 \tan \beta V_0^{3I} & J = 3 \neq I \\
    0 & \mbox{otherwise}
\end{array} \right. \\
(\delta^2 \mathbf{M_d})^{JI} &=& \left\{ \begin{array}{cl}
    m_{d_J} \epsilon_Y y_t^2 \tan \beta V_0^{3J*} V_0^{3I}& (J,I) = (1,2),
   (2,1) \\
    0 & \mbox{otherwise}
\end{array}. \right.
\end{eqnarray}

Now as we have second order terms explicitly in the mass matrix we 
need to expand the diagonalization matrices to second order. Additionally they 
have to be unitary to second order and the mass eigenvalues need to be real, 
which leads to the form 
\begin{eqnarray}
(\mathbf{D_L})^{JI} &=& (\mathbf{1+\delta D_L + \delta^2 D_L} + \frac{1}{2} 
\mathbf{\delta D_L\delta D_L})^{JI} \\
(\mathbf{D_L^{\dagger}})^{JI} &=& (\mathbf{1-\delta D_L-\delta^2 D_L} + \frac{1
}{2} \mathbf{\delta D_L \delta D_L})^{JI} \\
(\mathbf{D_R})^{JI} &=& (\mathbf{1+\delta D_R + \delta^2 D_R} + \frac{1}{2} 
\mathbf{\delta D_R \delta D_R})^{JI} e^{i\theta_I} \\
(\mathbf{D_R^{\dagger}})^{JI} &=& (\mathbf{1-\delta D_R-\delta^2 D_R} + \frac{1
}{2} \mathbf{\delta D_R \delta D_R})^{JI} e^{-i\theta_J}.
\end{eqnarray}
where $\mathbf{\delta D_{L,R}^{\dagger}}=-\mathbf{\delta D_{L,R}}$ and $\mathbf{
\delta^2 D_{L,R}^{\dagger}}=-\mathbf{\delta^2 D_{L,R}}$. Now the requirement 
$\mathbf{D_{L,R}}$ diagonalize the mass matrix $\mathbf{M_d}$ for diagonal 
elements gives us the condition
\begin{eqnarray}
\bar{m}_{d_J} &\approx& m_{d_J} |1+\epsilon_J \tan \beta| \\
\theta_J &\approx& \arg(1+\epsilon_J \tan \beta)
\end{eqnarray}
where we have only kept the leading order behaviour (i.e. $\delta^2$ terms have
been neglected). 

All off-diagonal terms automatically vanish at the zeroth order and the first 
order contributions are the same as in Ref.~\cite{Buras:2002vd}
\begin{eqnarray}
e^{-i\theta_J}(\mathbf{-(\delta D_R) (M_d)_0+ \delta M_d+ (M_d)_0 
(\delta D_L)})^{JI} = 0. \label{1stoffdiag:eq}
\end{eqnarray}
Which give us the results 
\begin{eqnarray}
(\mathbf{\delta D_L})^{JI} &=& - \frac{(\mathbf{M_d^{\dagger}}^{JJ})_0 
(\mathbf{\delta M_d})^{JI} + (\mathbf{\delta M_d^{\dagger}})^{JI}
(\mathbf{M_d}^{II})_0}{|(\mathbf{M_d}^{JJ})_0|^2 - |(\mathbf{M_d}^{II})_0|^2 } 
\label{DLgen1:eq} \\
(\mathbf{\delta D_R})^{JI} &=& - \frac{(\mathbf{M_d}^{JJ})_0 (\mathbf{\delta 
M_d^{\dagger}})^{JI} + (\mathbf{\delta M_d})^{JI}(\mathbf{M_d^{\dagger}}^{II}
)_0}{|(\mathbf{M_d}^{JJ})_0|^2 - |(\mathbf{M_d}^{II})_0|^2}. \label{DRgen1:eq}
\end{eqnarray}
As $\mathbf{\delta M_d}=0$ for $(J,I)=(1,2),(2,1)$ these first order corrections
are zero for these elements. To find the leading order contributions to 
$\mathbf{D_{L,R}}$ for these components we need to go to quadratic order
in the expansion parameter. 
Therefore the condition on the leading contributions to $\mathbf{D_{L,R}}$ for
$(J,I)=(1,2),(2,1)$ are
\begin{eqnarray}
e^{-i\theta_J}(\mathbf{-(\delta^2 D_R) (M_d)_0+ \Lambda+ (M_d)_0 
(\delta^2 D_L)})^{JI} = 0 \label{2ndoffdiag:eq}
\end{eqnarray}
where
\begin{eqnarray}
\Lambda^{JI} &=& \mathbf{(\delta D_R)^{J3}  \left((\delta M_d)^{3I} + (\delta 
D_L)^{3I} (M_d)_0^{33} \right)-\frac{1}{2}(\delta D_R)^{J3} (\delta D_R)^{3I}} 
- \nonumber \\ 
& & \mathbf{(\delta^2 M_d)^{JI} -(\delta M_d)^{J3}  (\delta D_L)^{3I} - 
\frac{1}{2} (\delta D_L)^{J3}(\delta D_L)^{3I} (M_d)_0^{33}} \\
&=& \mathbf{\frac{1}{2}(\delta D_R)^{J3} (\delta D_R)^{3I} - (\delta^2 M_d)^{JI}
-(\delta M_d)^{J3}  (\delta D_L)^{3I}} \nonumber \\
& & -\mathbf{\frac{1}{2} (\delta D_L)^{J3} (\delta D_L)^{3I} (M_d)_0^{33}} 
\label{Lambda:eq}.
\end{eqnarray}
To arrive at Eq.(\ref{Lambda:eq}) we used Eq.(\ref{1stoffdiag:eq}) and 
neglected terms of order $\mathcal{O}(m_{d_I}/m_b)$. Using 
Eq.(\ref{2ndoffdiag:eq}) leads to a relation similar to 
the one in Eq.(\ref{DLgen1:eq}) and 
Eq.(\ref{DRgen1:eq}), except that $\mathbf{\delta M_d \rightarrow \Lambda}$
\begin{eqnarray}
(\mathbf{\delta^2 D_L})^{JI} &=& - \frac{(\mathbf{M_d^{\dagger}}^{JJ})_0 
(\mathbf{\Lambda})^{JI} + (\mathbf{\Lambda^{\dagger}})^{JI}
(\mathbf{M_d}^{II})_0}{|(\mathbf{M_d}^{JJ})_0|^2 - |(\mathbf{M_d}^{II})_0|^2} 
\label{DLgen2:eq} \\
(\mathbf{\delta^2 D_R})^{JI} &=& - \frac{(\mathbf{M_d}^{JJ})_0 (\mathbf{
\Lambda^{\dagger}})^{JI} + \mathbf{\Lambda}^{JI}(\mathbf{M_d^{\dagger}}^{II})_0
}{|(\mathbf{M_d}^{JJ})_0|^2 - |(\mathbf{M_d}^{II})_0|^2} \label{DRgen2:eq}.
\end{eqnarray}
Substituting these equations into Eq.~(\ref{DLgen1:eq}) and Eq.(\ref{DRgen1:eq})
and neglecting all terms suppressed by the mass hierarchy we find 
\begin{eqnarray}(\mathbf{\delta D_L})^{JI} = \left\{ \begin{array}{cl}
- \frac{\epsilon_Y y_t^2 \tan \beta}{1+\epsilon_J \tan \beta} V_0^{3I} 
& J = 3 \neq I \\
\frac{\epsilon_Y^* y_t^2 \tan \beta}{1+\epsilon_I^* \tan \beta} V_0^{3J*} 
& J \neq 3 = I \\
0 & \mbox{otherwise}
\end{array} \right. \label{deltaDL1:eq}
\end{eqnarray}
and
\begin{eqnarray}
(\mathbf{\delta D_R})^{JI} = \left\{ \begin{array}{cl}
- \frac{\bar{m}_{d_I}}{\bar{m}_{b}} \left(\frac{\epsilon_Y y_t^2 \tan \beta}{
1+\epsilon_3 \tan \beta} + \frac{\epsilon_Y^* y_t^2 \tan \beta}{1+\epsilon_I^* 
\tan \beta}\right) e^{i(\theta_3-\theta_I)} V_0^{3I} & J = 3 \neq I \\
\frac{\bar{m}_{d_J}}{\bar{m}_{b}} \left(\frac{\epsilon_Y y_t^2 \tan \beta}{
1+\epsilon_J \tan \beta} + \frac{\epsilon_Y^* y_t^2 \tan \beta}{1+\epsilon_3^* 
\tan \beta}\right) e^{i(\theta_J-\theta_3)} V_0^{3J*}  & J \neq 3 = I \\
0 & \mbox{otherwise}
\end{array}. \right.\label{deltaDR1:eq}
\end{eqnarray}
Now to calculate the leading order corrections to the $(J,I)=(2,1),(1,2)$ 
elements we substitute the independent and linear order terms into 
Eq.(\ref{DLgen2:eq}) and Eq.(\ref{DRgen2:eq}) to find
\begin{eqnarray}
(\mathbf{\delta^2 D_L})^{21} &=& V_0^{32*}V_0^{31} \left(-\frac{\epsilon_Y y_t^2
\tan \beta}{1+\epsilon_2 \tan \beta} + \frac{\epsilon_Y^2 y_t^4 \tan^2 \beta}{(1
+\epsilon_2 \tan \beta)(1+\epsilon_3 \tan \beta)} + \right. \nonumber \\
& & \left. \frac{|\epsilon_Y|^2 y_t^4 \tan^2\beta}{2|1+\epsilon_3 \tan \beta|^2}
\right) \label{deltaDL2:eq}\\
(\mathbf{\delta^2 D_R})^{21} &=& V_0^{32*}V_0^{31} \frac{\bar{m}_d}{\bar{m}_s} 
e^{i(\theta_2-\theta_1)} \left[-\left(\frac{\epsilon_Y y_t^2 \tan \beta}{1+
\epsilon_2 \tan \beta} + \frac{\epsilon_Y^* y_t^2 \tan \beta}{1+\epsilon_1^* 
\tan \beta}\right) + \right. \nonumber \\ 
& & \frac{|\epsilon_Y|^2 y_t^4 \tan^2 \beta}{|1+\epsilon_3 \tan
\beta|^2} + \frac{(\epsilon_Y^*)^2 y_t^4 \tan^2 \beta}{(1+\epsilon_1^* \tan 
\beta)(1+\epsilon_3^* \tan \beta)} + \nonumber \\
& & \left. \frac{\epsilon_Y^2 y_t^4 \tan^2 \beta}{(1+
\epsilon_2 \tan \beta)(1+\epsilon_3 \tan \beta)}\right]. \label{deltaDR2:eq}
\end{eqnarray}
Using Eq.(\ref{deltaDL1:eq}), Eq.(\ref{deltaDR1:eq}), Eq.(\ref{deltaDL2:eq}) and
Eq.(\ref{deltaDR2:eq}), we find the same corrections to the effective CKM matrix
to leading order as in Refs.~\cite{Buras:2002vd,Isidori:2001fv,Babu:1999hn,
Hamzaoui:1998nu}
\begin{eqnarray}
V_0^{JI} = \left\{\begin{array}{cl}
V_{eff}^{3I} \frac{1+\epsilon_3 \tan \beta}{1+\epsilon_0^3 \tan \beta} & J = 3 
\neq I \\
V_{eff}^{J3} \frac{1+\epsilon_3^* \tan \beta}{1+\epsilon_0^{3*} \tan \beta} & J 
\neq 3 = I \\
V_{eff}^{JI} & \mbox{otherwise}
\end{array} \right. \label{CKM:eq}
\end{eqnarray}

\subsubsection{Flavor changing effective couplings of the Neutral Higgs Bosons}

Using the relations derived in the previous section,
it is relatively straightforward to calculate the 
coupling of the neutral Higgs bosons to the quarks. The effective lagrangian in 
the initial basis has the form
\begin{eqnarray}
\mathcal{L}_{eff} = - (\bar{d}_J^0)_R \mathbf{F_L^{dS}} (d_I^0)_L S^0 - (\bar{d
}_J^0)_L \mathbf{F_R^{dS}} (d_I^0)_R S^0 
\end{eqnarray}
where $S^0$ can be any of the three neutral scalars which has mixing matrix 
elements $x_d^S$ for the $\Phi_d^{0*}$ Higgs and $x_u^S$  for the $\Phi_u^{0*}$ Higgs.
So if $O^{IJ}$ diagonalizes the neutral Higgs mass matrix, we have
\begin{eqnarray}
x_d^S &=& O^{1S} + i \sin \beta O^{3S} \nonumber \\
x_u^S &=& O^{2S} - i \cos \beta O^{3S} \label{neutHmix:eq}
\end{eqnarray}
Now if we rotate quarks into the physical basis the Lagrangian has the form
\begin{eqnarray}
\mathcal{L}_{eff} = - (\bar{d}_J)_R \left(\mathbf{D_R^{\dagger} F_L^{dS} D_L}
\right) (d_I)_L S^0 - (\bar{d}_J)_L \left(\mathbf{D_L^{\dagger} F_L^{dS} D_R}
\right) (d_I^0)_R S^0
\end{eqnarray}
Therefore, assuming a mass matrix of the form given
in Eq.~(\ref{mass:eq}), we obtain,
\begin{eqnarray}
(\mathbf{F_L^{ds}})^{JI} = \frac{m_{d_J}}{v_d} \left( (x_d^S+ \epsilon_J x_u^S) 
\delta^{JI} + \epsilon_Y y_t^2 x_u^s \lambda_0^{JI} \right).
\end{eqnarray}
which has a dependence up to second order on the CKM elements. Therefore,
we obtain the following expansion in terms of CKM elements
\begin{eqnarray}
\mathbf{F_L^dS} &=& (\mathbf{F_L^{dS}})_0 + \mathbf{\delta F_L^{dS}} + \mathbf{
\delta^2 F_L^{dS}}
\end{eqnarray}
where
\begin{eqnarray}
(\mathbf{F_L^{dS}})_0^{JI} &=& \frac{\bar{m}_{d_J} e^{i \theta_J}}{v_d (1+
\epsilon_J \tan \beta)} (x_d^S + \epsilon_J x_u^S) \delta^{JI} \label{FdS0:eq}\\
(\mathbf{\delta F_L^{dS}})^{JI} &=& \left\{ \begin{array}{cl}
\frac{\bar{m}_{d_J} e^{i \theta_J} \epsilon_Y y_t^2 x_u^S}{v_d (1+\epsilon_J 
\tan \beta)} V_0^{3J*} & J \neq 3 = I \\
\frac{\bar{m}_{d_3} e^{i \theta_3} \epsilon_Y y_t^2 x_u^S}{v_d (1+\epsilon_3 
\tan \beta)} V_0^{3I} & J = 3 \neq I \\
0 & \mbox{otherwise}
\end{array} \right.\label{deltaFdS:eq}\\
(\mathbf{\delta^2 F_L^{dS}})^{JI} &=& \left\{ \begin{array}{cc}
\frac{\bar{m}_{d_J} e^{i \theta_J} \epsilon_Y y_t^2 x_u^S}{v_d (1+\epsilon_J 
\tan \beta)} V_0^{3J*} V_0^{3I* }& (J,I) = (1,2),(2,1) \\
0 & \mbox{otherwise}
\end{array}. \right.
\end{eqnarray}
Therefore the leading order contribution to the diagonal terms of 
$ddS^0$ coupling is just Eq.~(\ref{FdS0:eq}). Again the zeroth term makes no
contribution to the off diagonal elements of the $ddS$ couplings. Hence, 
at linear 
order we have for $J \neq I$ 
\begin{eqnarray}
\delta \left(\mathbf{D_R^{\dagger} F_L^{dS} D_L}\right)^{JI} = e^{-i\theta_J} 
\left(\mathbf{-(\delta D_R)^{JI} (F_L^{dS})_0^{II}+ (\delta F_L^{dS})^{JI} +}
\right. \nonumber \\ 
\left. \mathbf{(F_L^{dS})_0^{JJ} (\delta D_L)^{JI}} \right)
\end{eqnarray}
which also disappears for $(J,I)=(1,2),(2,1)$. So the only contributions that 
are none zero at this order are when either $J=3$ or $I=3$. Using 
Eq.~(\ref{deltaDL1:eq}), Eq.~(\ref{deltaDR1:eq}), Eq.~(\ref{CKM:eq}) and 
Eq.~(\ref{deltaFdS:eq}) and neglecting terms suppressed by the mass hierarchy we
find that
\begin{eqnarray}
& &(X_{RL}^S)^{JI}=\delta \left(\mathbf{D_R^{\dagger} F_L^{dS} D_L}\right)^{JI}
= \nonumber \\
& & \left\{ \begin{array}{cc}
\frac{\bar{m}_b \epsilon_Y y_t^2}{v_d (1+\epsilon_3 \tan \beta)(1+\epsilon_0^3 
\tan \beta)} V_{eff}^{3I} (x_u^S - x_d^S \tan \beta) & J = 3 \neq I \\
\frac{\bar{m}_{d_J} y_t^2 \Gamma^{J3}}{v_d (1+\epsilon_3 \tan \beta)(1+
\epsilon_J \tan \beta)} V_{eff}^{3J*} (x_u^S - x_d^S \tan \beta) & J \neq 3 = I 
\\
0 & \mbox{otherwise}
\end{array} \right.
\end{eqnarray}
where
\begin{eqnarray}
\Gamma^{J3} = \frac{\epsilon_Y(1+\epsilon_3^* \tan \beta)-\epsilon_Y^*(
\epsilon_3 - \epsilon_J) \tan \beta}{1+\epsilon_0^{3*} \tan \beta}
\end{eqnarray}

Finally to find the leading corrections to qqH coupling for 
$(J,I)=(2,1),(1,2)$ we need to go to quadratic order in which case we have
\begin{eqnarray}
(X_{RL}^S)^{21} &=& \delta^2 \left(\mathbf{D_R^{\dagger} F_L^{dS} D_L}
\right)^{21}, \nonumber \\
&=& \frac{\bar{m}_s y_t^2 \Gamma^{21} (x_u^S - x_d^S \tan \beta)}{v_d (1+
\epsilon_2 \tan \beta) (1+\epsilon_3 \tan \beta)} V_{eff}^{32*} V_{eff}^{31} \\
(X_{RL}^S)^{12} &=& \delta^2 \left(\mathbf{D_R^{\dagger} F_L^{dS} D_L}
\right)^{12} \nonumber \\
&=& 
\frac{\bar{m}_d y_t^2 \Gamma^{12} (x_u^S - x_d^S \tan \beta)}{v_d (1+\epsilon_1 
\tan \beta)(1+\epsilon_3 \tan \beta)} V_{eff}^{31*} V_{eff}^{32},
\end{eqnarray}
where
\begin{eqnarray}
\Gamma^{21} &=& \frac{\epsilon_Y}{(1+\epsilon_2 \tan \beta)|1+\epsilon_0^3 \tan 
\beta|^2} \left[(1+\epsilon_0^3\tan \beta)|1+\epsilon_3 \tan 
\beta|^2 - \right. \nonumber\\
& & \left. \epsilon_Y y_t^2 \tan \beta (1+\epsilon_3^*\tan \beta)(1+\epsilon_2
\tan \beta)-  \right. \nonumber \\ 
& & \left.
\epsilon_Y^* y_t^2 \tan \beta(1+\epsilon_2 \tan \beta)^2 \right], \\
\Gamma^{12} &=& \frac{\epsilon_Y}{(1+\epsilon_2 \tan \beta) |1+\epsilon_0^3 
\tan \beta|^2}\left\{(1+\epsilon_0^3\tan \beta)|1+\epsilon_3 \tan \beta|^2 -
 \right.\nonumber\\
& & \left. \epsilon_Y y_t^2 \tan \beta (1+\epsilon_3^*\tan \beta)(1+\epsilon_2
\tan \beta)-\epsilon_Y^* y_t^2
\tan \beta(1+\epsilon_2 \tan \beta)  \right. \nonumber \\ 
& & \left.
(1+\epsilon_1 \tan \beta) + \frac{\epsilon_1-\epsilon_2}{\epsilon_Y} \left[ 
\frac{\epsilon_Y^* \tan \beta}{1+\epsilon_2^* \tan \beta}  \right. \right. 
\nonumber \\ 
& & \left. \left.
- \frac{(\epsilon_Y^*)^2 \tan^2 \beta 
y_t^2}{(1+\epsilon_2^* \tan \beta)(1+\epsilon_3^* \tan \beta)} - \frac{|
\epsilon_Y|^2 \tan^2 \beta y_t^2}{|1+\epsilon_3 \tan \beta|^2} - \right] 
\right\}.
\end{eqnarray}
In the limit that $\epsilon_0^J$'s are uniform then of leading order 
contributions will collapse to Eq.~(\ref{uniformXRL:eq}) as each of 
$\Gamma^{IJ}$ elements go to $\epsilon_Y$. 
As the effective lagrangian is real, the $LR$ 
couplings are related to $RL$, so that
\begin{eqnarray}
X_{LR}^S = (X_{RL}^S)^{\dagger}.
\end{eqnarray}

\subsection{Calculation of Loop factors} \label{A2:sec}

The assumption that the squark mass matrices are block diagonal in the tree 
level CKM basis gives us
\begin{eqnarray}
\mathcal{M}_D^2 &=& \left(\begin{array}{cc}
(M_Q^2)_J \delta^{JI} & \frac{1}{\sqrt{2}} y_{d_J} \hat{\mu}_J^* v_u 
\delta^{JI} \\
\frac{1}{\sqrt{2}} y_{d_J} \hat{\mu}_J v_u \delta^{JI} &  (M_D^2)_J 
\delta^{JI} 
\end{array} \right) \\
\mathcal{M}_U^2 &=& \left(\begin{array}{cc}
(M_Q^2)_J \delta^{JI} +m_t^2 \delta^{J3} \delta^{I3} & - \frac{1}{\sqrt{2}} 
y_{u_J}\tilde{\mu}_J^* v_u \delta^{JI} \\
- \frac{1}{\sqrt{2}} y_{u_J} \tilde{\mu}_J v_u \delta^{JI} & 
(M_U^2)_J \delta^{JI} + m_t^2 \delta^{J3} \delta^{I3}
\end{array} \right) 
\end{eqnarray}
where $\tilde{\mu}_J = \frac{\mu}{\tan \beta} - A_{u_J}$ and $\hat{\mu}_J = 
\mu - \frac{A_{d_J}}{\tan \beta}$. Therefore the diagonalization matrices have 
the simple form
\begin{eqnarray}
Z_{(U,D)} = \left(\begin{array}{cc}
\delta^{IJ} \cos \alpha_I^{(U,D)} & \delta^{IJ} e^{-i\phi_I^{(U,D)}}\sin 
\alpha_I^{(U,D)} \\
-\delta^{IJ} e^{i\phi_I^{(U,D)}}\sin \alpha_I^{(U,D)} &\delta^{IJ} \cos 
\alpha_I^{(U,D)} \end{array} \right)
\end{eqnarray}
where $\phi_I^D$ ($\phi_I^U$) is the phase of $\hat{\mu}$ ($\tilde{\mu}$) and 
\begin{eqnarray}
\cot 2\alpha_J^D &=& -\frac{(m_Q^2)_J - (m_D^2)_J}{\sqrt{2} y_{d_J} |
\hat{\mu}_J| v_u}\\
\cot 2\alpha_J^U &=& \frac{(m_Q^2)_J - (m_U^2)_J}{\sqrt{2} y_{u_J} |\tilde{\mu
}_J| v_u}
\end{eqnarray}
Following the 
notation of Ref.~\cite{Rosiek:1995kg} $Z_+$ and $Z_-$ diagonalize the
chargino mass matrix and $Z_N$ and $Z_N^T$ diagonalize the neutralino mass 
matrix. Additionally, if there is a splitting in the mass spectrum so that 
the squarks of the first two generation have uniform masses (i.e. $m_{D_1}=
m_{D_2}=m_{D_4}= m_{D_5}=m_{U_1}=m_{U_2}=m_{U_4}=m_{U_5}=M_{SUSY}$) we find  
\begin{eqnarray}
\epsilon_0^J &=& \frac{1}{16\pi^2 v_u} \left(\frac{32 \pi \alpha_s}{3} M_3 
\mu^* v_u C_0(|M_3|^2,m_{D_J}^2, m_{D_{J+3}}^2) + \right. \nonumber \\
& & \sum_{l=1}^4 m_{N_l} \left(P_D^{lJ} C_2(m_{N_l}^2,m_{D_J}^2,m_{D_{J+3}}^2) 
+Q_D^{lJ} C_0(m_{N_l}^2,m_{D_J}^2,m_{D_{J+3}}^2) \right) - \label{e0form:eq}\\
& & \left. \sqrt{2} \sum_{l=1}^2 m_{C_l} Z_-^{1l} Z_+^{2l} \left(C_2(m_{C_l}^2,
M_{SUSY}^2,M_{SUSY}^2) + M_{SUSY}^2 C_0(m_{N_l}^2,M_{SUSY}^2,M_{SUSY}^2) \right) \right) \nonumber \\
\epsilon_Y &=& \frac{1}{16\pi^2 v_u} \sum_{l=1}^2 m_{C_l} \left[-\sqrt{2} 
\frac{g_2}{y_t^2} Z_-^{2l} Z_+^{1l} \left(C_2(m_{C_l}^2, m_{U_3}^2, m_{U_6}^2) +
\right.\right. \nonumber \\
& & (M_{U_3}^2 + m_t^2) C_0(m_{C_l}^2, m_{U_3}^2, m_{U_6}^2) - 
C_2(m_{C_l}^2, M_{SUSY}^2, M_{SUSY}^2)- \label{eyform:eq} \\
& &\left. \left. M_{SUSY}^2  C_0(m_{C_l}^2, M_{SUSY}^2, M_{SUSY}^2)\right) - Z_-^{2l} Z_+^{2l} 
\tilde{\mu}_3 v_u C_0(m_{C_l}^2, m_{U_3}^2, m_{U_6}^2)\right] \nonumber
\end{eqnarray}
where $C_i$ are the Passarino-Veltman functions, $m_i$ are the physical squark 
masses, $M_i$ are the squark soft mass parameters and
\begin{eqnarray}
P_D^{lJ} &=& Z_N^{3l} \left(g_1Z_N^{1l}-g_2 Z_N^{2l} \right) \\
Q_D^{lJ} &=& -\frac{g_1 Z_N^{1l}}{3} \hat{\mu}_J^*v_u\left(
\frac{g_1Z_N^{1l}}{3} -g_2 Z_N^{2l} \right) + \frac{2 g_1 Z_N^{1l}}{3} Z_N^{3l} 
\left((M_Q^2)_J +m_{d_J}^2\right) \nonumber \\
& & + Z_N^{3l} \left(\frac{g_1Z_N^{1l}}{3} -g_2 Z_N^{2l} \right) \left((M_D^2)_J
+(m_d^2)_J\right) - y_{d_J}^2 (Z_N^{3l})^2 \hat{\mu}_J v_u.
\end{eqnarray}
Similarly for the antiholomorphic corrections to the up Yukawas have the form
\begin{eqnarray}
\epsilon_0^{'J} &=& \frac{1}{16\pi^2 v_u} \left(-\frac{32 \pi \alpha_s}{3} M_3 
\mu^* v_u C_0(|M_3|^2,m_{U_J}^2, m_{U_{J+3}}^2) + \right. \nonumber \\
& & \sum_{l=1}^4 m_{N_l} \left(P_U^{lJ} C_2(m_{N_l}^2,m_{U_J}^2,m_{U_{J+3}}^2) 
+ Q_U^{lJ} C_0(m_{N_l}^2,m_{U_J}^2,m_{U_{J+3}}^2) \right) - \\
& & \left. \sqrt{2} \sum_{l=1}^2 m_{C_l} Z_-^{1l} Z_+^{2l} \left(C_2(m_{C_l}^2,
M_{SUSY}^2,M_{SUSY}^2) + M_{SUSY}^2 C_0(m_{N_l}^2,M_{SUSY}^2,M_{SUSY}^2) \right) \right) \nonumber \\
\epsilon_Y^{'} &=& \frac{1}{16\pi^2 v_u} \sum_{l=1}^2 m_{C_l} \left[- \sqrt{2} 
\frac{g_2}{y_b^2} Z_+^{2l} Z_-^{1l} \left(C_2(m_{C_l}^2, m_{D_3}^2, m_{D_6}^2) +
\right.\right. \nonumber \\
& & (M_{D_3}^2 + m_b^2) C_0(m_{C_l}^2, m_{D_3}^2, m_{D_6}^2) - 
C_2(m_{C_l}^2, M_{SUSY}^2, M_{SUSY}^2)- \nonumber \\
& &\left. \left. M_{SUSY}^2  C_0(m_{C_l}^2, M_{SUSY}^2, M_{SUSY}^2)\right) - Z_-^{2l} Z_+^{2l} 
\hat{\mu}_3 v_u C_0(m_{C_l}^2, m_{D_3}^2, m_{D_6}^2)\right]
\end{eqnarray}
where
\begin{eqnarray}
P_U^{lJ} &=&- Z_N^{4l} \left(g_1Z_N^{1l} -g_2 Z_N^{2l} \right) \\
Q_U^{lJ} &=& -\frac{2 g_1 Z_N^{1l}}{3} \tilde{\mu}_J^*v_u\left(
\frac{g_1Z_N^{1l}}{3} +g_2 Z_N^{2l} \right) - \frac{4 g_1 Z_N^{1l}}{3} Z_N^{4l} 
\left((M_Q^2)_J +m_{u_J}^2\right) \nonumber \\
& & + Z_N^{4l} \left(\frac{g_1Z_N^{1l}}{3} +g_2 Z_N^{2l} \right) \left((M_U^2)_J
+ m_{d_J}^2 \right) + y_{u_J}^2 (Z_N^{4l})^2 \tilde{\mu}_J v_u.
\end{eqnarray}
The infinities present in $C_2$ in $\epsilon_Y$'s clearly cancel, however the 
infinities in $\epsilon_0$'s need to be absorbed by counter terms in the 
effective lagrangian. So that the $C_2$ contributions to the $\epsilon)$'s in 
the above formulae are purely the finite pieces.

\end{document}